\documentclass{article}

\usepackage{amssymb}

\usepackage{latexsym,amsmath,amsfonts,amscd}
\usepackage{xfrac}
\usepackage{graphics}
\usepackage{caption}
\usepackage{subcaption}
\usepackage{epstopdf}
\usepackage{color}
\usepackage{changebar}
\usepackage{indentfirst}
\usepackage{verbatim}
\usepackage{xcolor}
\usepackage{hyperref}
\usepackage{amssymb}
\usepackage{lscape}
\usepackage{array,multirow}

\renewcommand{\d}{{\rm d}}			
\newcommand{\CFL}{\mathsf{CFL}}		

\usepackage{authblk}
\usepackage[margin=3.0cm]{geometry}

\title{Modeling blood flow in viscoelastic vessels: \\ the 1D augmented fluid-structure interaction system}

\author[$\dagger$]{Giulia Bertaglia \footnote{Corresponding author. Email address: \textit{giulia.bertaglia@unife.it}}}
\author[$\dagger$]{Valerio Caleffi}
\author[$\dagger$]{Alessandro Valiani}

\affil[$\dagger$]{\small Department of Engineering, University of Ferrara, Via G. Saragat 1, 44122 Ferrara, Italy}

\begin{document}

\maketitle

\begin{abstract}
Nowadays mathematical models and numerical simulations are widely used in the field of hemodynamics, representing a valuable resource to better understand physiological and pathological processes in different medical sectors. The theory behind blood flow modeling is closely related to the study of incompressible flow through compliant thin-walled tubes, starting from the incompressible Navier-Stokes equations. Furthermore, the mechanical interaction between blood flow and vessels wall must be properly described by the model. Recent works showed the benefits of characterizing the rheology of the vessel wall through a viscoelastic law. Taking into account the viscous contribution of the wall material and not simply the elastic one leads to a more realistic representation of the vessel behavior, which manifests not only an instantaneous elastic strain but also a viscous damping effect on pulse pressure waves, coupled to energy losses. In this context, the aim of this work is to propose an easily extensible one-dimensional mathematical model able to accurately capture fluid-structure interactions. The originality of the model lies in the introduction of a viscoelastic tube law in PDE form, valid for both arterial and venous networks, leading to an augmented fluid-structure interaction system. In contrast to well established mathematical models, the proposed one is natively hyperbolic. The model is solved with an efficient and robust second-order numerical scheme; the time integration is based on an Implicit-Explicit Runge-Kutta scheme conceived for applications to hyperbolic systems with stiff relaxation terms. The validation of the proposed model is performed on several different test cases. Results obtained in Riemann problems, adopting a simple elastic tube law for the characterization of the vessel wall, are compared with available exact solutions. To validate the contribution given by the viscoelastic term, the Method of Manufactured Solutions has been applied. Specific tests have been designed to verify the well-balancing with respect to fluid-at-rest condition and the accuracy-preserving property of the scheme. Finally, a specific test case with an inlet pulse pressure wave has been designed to assess the effects of viscoelasticity with respect to a simple elastic behavior of the vessel wall. The complete code, written in MATLAB (MathWorks Inc.) language, with the implemented test cases, is made available in Mendeley Data repository.
\end{abstract}



\section{Introduction}
The availability of robust and efficient mathematical instruments, together with the engineering know-how in the fluid mechanics sector, represents an invaluable tool for a consistent support in hemodynamics studies. It is a proven fact that computational modeling can provide efficient approaches for the quantification of fluid dynamics phenomena in the cardiovascular network, supplying meaningful data that otherwise would require invasive techniques or simply would not be available with general clinical measurements \cite{formaggia2009,ambrosi2012,willemet2016}. Mathematical models and numerical simulations can also help the prediction of the possible onset of diseases and development of pathologies \cite{toro2016a,liang2018,muller2019}.\par
In recent years, mathematical models have been consistently developed, focusing on different aspects and fundamental issues that need to be addressed to successfully model the circulatory system. Among these, it has to be considered that blood flow mechanically interacts with vessel walls and tissue, giving rise to complex fluid-structure interactions whose mathematical analysis is difficult to properly describe and numerically simulate in an efficient manner \cite{holenstein1980,leguy2019}. The wall of a blood vessel consists of three layers of different tissues: an epithelial inner lining, a middle layer consisting of a smooth muscle and elastic connective tissue, and a connective tissue outer covering. These three structural layers, from innermost to outermost, are called tunica interna (intima), tunica media and tunica externa \cite{tortora2013}. At a macroscopic level, the arterial wall can be seen as a complex multi-layer viscoelastic structure which deforms under the action of blood pressure \cite{nichols2011,wang2016}, even collapsing, in the case of veins, under certain circumstances \cite{shapiro1977,toro2013,spiller2017}. The modeling of the interaction between blood flow and vessel wall mechanics requires the definition of a constitutive law which has to correctly describe the energy transfer between the two means, to accurately represent wave propagation phenomena \cite{fung1997,leguy2019}. Even though frequently in hemodynamics models the viscosity of vessels is neglected for simplicity, there is an increasing number of contributions showing the benefits of modeling the mechanical behavior of the vessel wall using a viscoelastic rheological characterization \cite{valdez-jasso2009,reymond2009b,alastruey2011,alastruey2012a}. In viscoelastic materials the energy put into the system during strain is not totally recovered during relaxation, causing a viscous damping of the pulse waves. This phenomenon can be found in many biological tissues and it is visible when plotting pressure against area variations in time: the presence of a widening pressure-area loop (hysteresis) represents the effective energy dissipated during dilatation and contraction cycles \cite{valdez-jasso2009,vignon-clementel2011,battista2015}. When modeling the vessel wall mechanics simply by means of an elastic law, the whole information related to the loss of energy of the phenomenon vanishes and pressure peaks levels could be overestimated \cite{holenstein1980,alastruey2011,montecinos2014,battista2015}. For this reason, many attempts have recently been made to improve the rheological characterization of vessels wall on the basis of linear, quasilinear viscoelasticity or more complex non-linear models \cite{reymond2009b,valdez-jasso2009,battista2015,ghigo2016}. \par
In the present work, a novel one-dimensional augmented fluid-structure interaction (FSI) system, able to capture viscoelastic wall effects, is presented for the blood flow modeling of both arterial and venous network and solved with a Finite Volume Method (FVM) with the implementation of an IMEX-SSP2 Runge-Kutta scheme \cite{pareschi2005}. Usually, the viscoelastic behavior of vessels is described by analogous mechanical models consisting of springs and dashpots. In this work, the use of the Standard Linear Solid Model (SLSM) is proposed, represented by one spring in series with the so-called Kelvin-Voigt (KV) unit, composed by a spring and a dashpot in parallel. This model is yet able to exhibit all the three primary features of a viscoelastic material: creep, stress relaxation and hysteresis \cite{Lakes2009}. The same viscoelastic model is indeed already used by Bessems et al. \cite{bessems2008} and by Valdez-Jasso et al. \cite{valdez-jasso2009}, with successful comparisons of numerical results with experimental data. Moreover, Bessems et al. \cite{bessems2008} uses a mathematical approach similar to the one presented in this work, based on an augmented system of equations, but with a different formulation of the final SLSM constitutive law and solely for the arterial network, associated to a spectral element method discretization. Nevertheless, the most commonly used viscoelastic model merely considers a Kelvin-Voigt unit, due to its simplicity \cite{alastruey2011,montecinos2014,wang2014,mynard2015,liang2018}. This model formulates pressure as a function of cross-sectional area, making it straightforward the incorporation into a fluid dynamics model. However, the KV model has the lack of being able to describe an exponential relaxation of the stress (pressure) over time that is one of the main attributes of viscoelastic materials \cite{Lakes2009}. Furthermore, when inserted into the system of equations in its general formulation, the KV viscoelastic law give rise to a second-order derivative and consequent numerical issues related to the parabolic term. In literature this complication is treated in different manners. Alastruey et al. \cite{alastruey2011} adopts a discontinous Galerkin scheme with a spectral/$hp$ spatial discretisation. Montecinos et al. \cite{montecinos2014} passes through a hyperbolic reformulation of the system introducing a numerical relaxation parameter (applying Cattaneo's law) and solving it with an ADER scheme. While, Formaggia et al. \cite{formaggia2003}, Mynard and Smolich \cite{mynard2015} and Wang et al. \cite{wang2014} employ a specific operator-splitting procedure.  The SLSM, together with the augmented FSI system proposed in the present work, directly overcome these issues, still maintaining ease of implementation and usage, with very good efficiency and robustness granted by the IMEX Runge-Kutta algorithm chosen.\par
The paper is structured as follows: in section \ref{section_mathematicalmodel} the mathematical model is presented first in its general formulation and successively in the augmented form, with the description of the chosen constitutive tube laws, whether elastic or viscoelastic. In section \ref{section_numericalmodel} the IMEX Runge-Kutta scheme for applications to hyperbolic systems with stiff relaxation terms is presented, with an analytical proof of the well-balancing of the model, intended as exact conservation of a rest initial condition (C-property), as defined in \cite{bermudez1994}. All the numerical results, including Riemann problems, C-property tests, problems for the validation of the viscoelastic contribution through the Method of Manufactured Solutions, accuracy analysis and a pulse wave test case designed for a common carotid artery, are presented and discussed in section \ref{section_numericalresults}. Finally, in section \ref{section_conclusions}, the advantages of the model proposed are summarized together with some concluding remarks.
\section{Mathematical model}
\label{section_mathematicalmodel}
\subsection{General one-dimensional formulation}
\label{section_general_formulation}
The standard one-dimensional mathematical model for blood flow, valid for medium to large-size vessels, is obtained averaging the incompressible Navier-Stokes equations over the cross-section, under the assumption of axial symmetry of
the vessel and of the flow, obtaining the well established equations of conservation of mass and momentum \cite{formaggia2009}:
\begin{subequations}
\begin{align}
	&\partial_t A + \partial_x (Au) = 0 \label{eq.contST}\\ 
	&\partial_t (Au) + \partial_x (Au^2)  + \dfrac{A}{\rho} \partial_x p = 0 . \label{eq.momST}
\end{align}
\label{eq.cont&mom}	
\end{subequations}
Here \(A(x,t)\) is the cross-sectional area of the vessel, \(u(x,t)\) is the averaged fluid velocity, \(p(x,t)\) is the averaged fluid pressure, \(\rho\) is the cross-sectional averaged density of the fluid and \(x\) and \(t\) are respectively space and time. To close the governing partial differential equation (PDE) system \eqref{eq.cont&mom}, a tube law, representative of the interaction between vessel wall displacement (through the cross-sectional area \(A\)) and blood pressure \(p\), is required. This can be performed via a constitutive model, relating strain and stress.
\subsection{Elastic constitutive tube law}
\label{section_elastictubelaw}
In the simplest case, the pressure-area relationship is defined considering a perfectly elastic behavior of the vessel wall, with the widely adopted elastic constitutive tube law \cite{formaggia2003,matthys2007,muller2013}:
\begin{equation}
p = p_{ext} +\psi_{el} .
\label{elastictubelaw}
\end{equation}
In this equation, \(p_{ext}(x)\) is the external pressure and $\psi_{el}(A,A_0,E_0)$ is the elastic contribution of the transmural pressure, assumed as:
\begin{equation}
\psi_{el} = K \left(\alpha^m - \alpha^n\right) ,
\label{transmuralpressure}
\end{equation}
where \(\alpha = \sfrac{A}{A_0}\) is the non-dimensional cross-sectional area rescaled with respect to \(A_0(x)\), equilibrium cross-sectional area, \(K(x)\) represents the stiffness coefficient of the material, which accounts for the instantaneous Young (elastic) modulus, $E_0(x)$, and the wall thickness, $h_0$, and finally \(m\) and \(n\) are specific parameters related to the behavior of the vessel wall, whether artery or vein. If dealing with arteries, this tube law corresponds to the well known Laplace law \cite{wylie1978}. Indeed, defining
\begin{equation}
W = \frac{h_0}{R_0} ,
\label{W}
\end{equation}
as the ratio between the thickness $h_0$ and the equilibrium radius of the wall, $R_0 = \sfrac{\sqrt{A_0}}{\sqrt{\pi}}$, for an artery we have:
\begin{equation}
K = \frac{E_0}{W}, \quad m = 1/2, \quad n = 0 .
\label{Ka}
\end{equation}
When dealing with veins, their possible collapse in case of large negative transmural pressures has to be considered \cite{carpenter2001,toro2013,murillo2019}. The collapsed state for veins is identified by a cross-sectional area assuming a buckled, dumbbell shape configuration, in which opposite sides of the interior wall touch each other, still leaving some fluid in the two extremes \cite{carpenter2001,spiller2017}. This particular aspect leads to the assumption of different parameters for the mechanical characterization of the wall behavior \cite{shapiro1977}:
\begin{equation}
K = \frac{E_0}{12W^3}, \quad m = 10, \quad n = -3/2 .
\label{Kv}
\end{equation}
Following what presented in \cite{bertaglia2018,bertaglia2018a}, it is also possible to derive with respect to time eq.~\eqref{elastictubelaw} and to use the continuity eq.~\eqref{eq.contST} to obtain a PDE representative of the elastic behavior of the vessel wall:
\begin{equation}
\partial_t p + \frac{K}{A} \left(m \alpha^m - n \alpha^n\right) \partial_x (Au) = 0 ,
\label{elasticPDEtubelaw}
\end{equation}
with the same set of parameters $K, m$ and $n$ discussed above for the characterization of arteries \eqref{Ka} or veins \eqref{Kv}.
\subsection{Viscoelastic constitutive tube law}
\label{section_SLSM}
Even though mathematical models representing the blood circulation frequently neglect the viscous component of the vessel wall, it is well known that blood vessels (and living tissues in general) exhibit viscoelastic properties \cite{fung1997,nichols2011,salvi2012}. Viscoelastic effects are simulated in literature using different more or less complex rheological models, whether linear or not \cite{holenstein1980,bessems2008,valdez-jasso2009,wang2016,ghigo2016,mitsotakis2018a}. A constitutive relation of linear viscoelasticity is built up considering the material as a sum of linear elastic springs, each one defined by a Young modulus $E$, and linear viscous dashpots, characterized by a viscosity coefficient $\eta$, to take into account also the time dependent relaxation of the wall and its damping effect on pressure waves. \par
If a single spring and a dashpot are connected in parallel, so that they both experience the same deformation or strain and the total stress is the sum of the stresses in each element, we have what is known as Kelvin-Voigt (KV) model \cite{Lakes2009}. The constitutive law which describes the behavior of a KV unit relates strain $\epsilon(t)$ and stress $\sigma(t)$ as follows:
\begin{equation}
\sigma = E\epsilon + \eta \dot{\epsilon} .
\label{eq.constitutiveKV}
\end{equation}
Considering initially a generic artery, the deformation of the wall is geometrically related to the cross-sectional area through equation $\epsilon = \alpha^m - \alpha^n$, with $m=\sfrac{1}{2}$ and $n=0$ as values in eq.~\eqref{Ka}. Involving Barlow's formula, for which $\sigma = W(p-p_{ext})$, and the continuity eq.~\eqref{eq.contST}, it is possible to obtain the following KV viscoelastic tube law:
\begin{equation}
p = p_{ext} + K\left(\alpha^m - \alpha^n\right) - \frac{\Gamma}{A_0\sqrt{A}}\frac{\partial(Au)}{\partial x} ,
\label{eq.KV}
\end{equation}
with
\begin{equation}
\Gamma = \frac{\eta h_0 \sqrt{\pi}}{2}
\label{gamma}
\end{equation}
representing the viscous contribution of the material. With the same approach, considering $m=10$ and $n=-\sfrac{3}{2}$ as in eq.~\eqref{Kv} for $\epsilon = \alpha^m - \alpha^n$ and assuming a relationship between stress and pressure defined as $\sigma = 12W^3(p-p_{ext})$, to take into account the potential collapsibility of the vessel wall, it is possible to obtain exactly the same equation \eqref{eq.KV}, valid also for veins.\\
This viscoelastic law, which is widely adopted among literature's well recognized blood flow models \cite{alastruey2011,montecinos2014,wang2014,mynard2015}, has the deficiency of defining a relaxation response that is a constant plus a Dirac delta function. Setting the strain to be a constant, indeed, the constitutive eq.~\eqref{eq.constitutiveKV} reduces to the simple Hooke's law: $\sigma = E\epsilon$. In this way, the stress is taken up by the spring and is constant; so in fact there is no stress relaxation over time \cite{Lakes2009}.\par
\begin{figure}[t]
\centering
\includegraphics[width=0.4\linewidth]{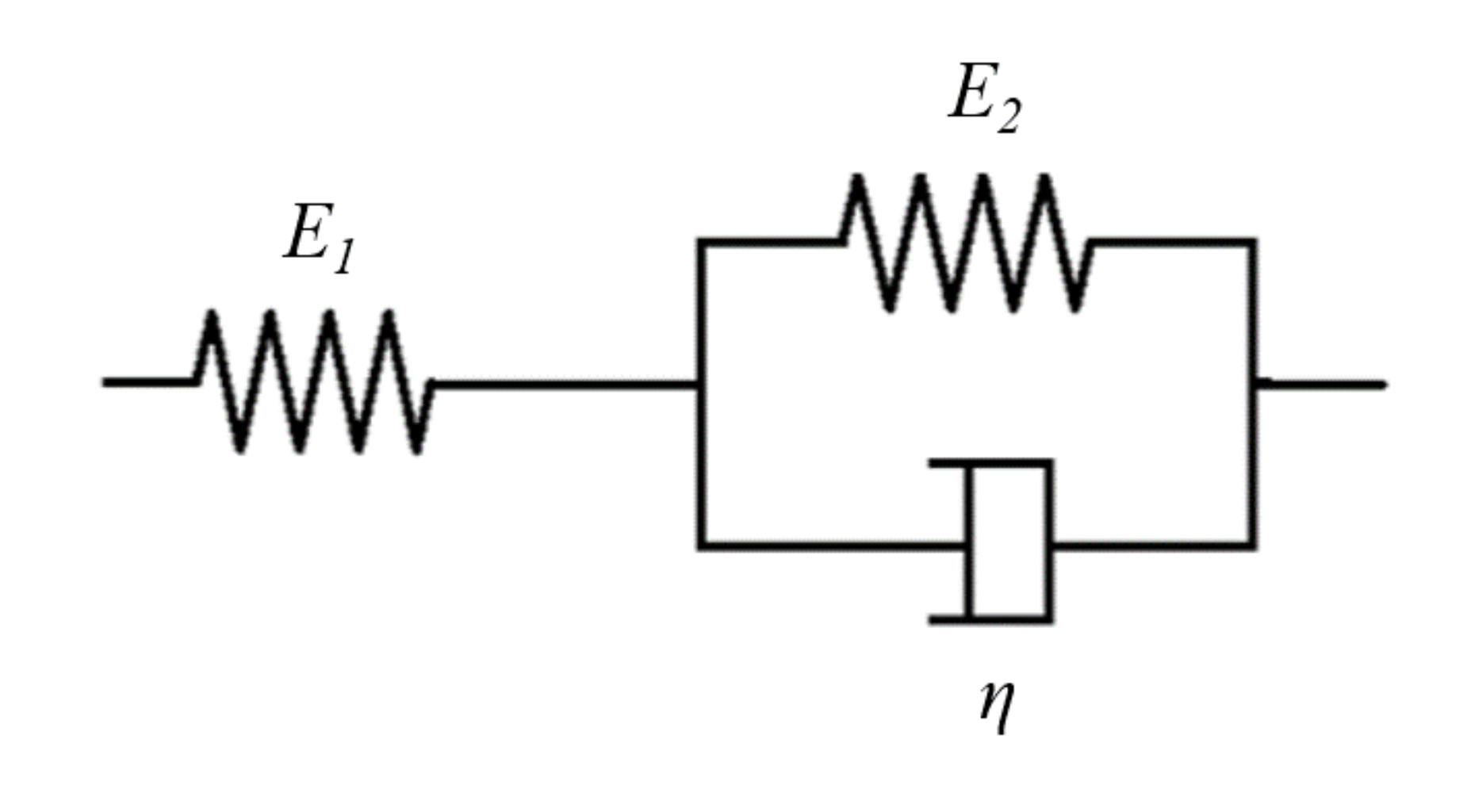}
\caption{Scheme of the Standard Linear Solid Model. $E_1$ is the Young modulus of the first spring (in series with the Kelvin-Voigt unit); $E_2$ is the Young modulus of the elastic spring of the Kelvin-Voigt element itself; $\eta$ is the viscosity coefficient of the dashpot.}
\label{fig.SLSM}
\end{figure}

A more realistic behavior can be modeled by the Standard Linear Solid Model (SLSM), represented in fig.~\ref{fig.SLSM} in its version with a Kelvin-Voigt unit, in series with an additional elastic spring.
Evaluating the constitutive equation of the model, expressed in terms of stress $\sigma$ and strain $\epsilon$,
\begin{equation}
\label{constitutiveEq.}
	\dot{\sigma} = E_0 \dot{\epsilon} - \frac{1}{\tau_r}(\sigma - E_\infty \epsilon),
\end{equation}
we have that the instantaneous Young modulus \(E_0\), the asymptotic Young modulus \(E_\infty\) and the relaxation time \(\tau_r\) are respectively:
\[E_0 = E_1 , \qquad E_\infty = \frac{E_1E_2}{E_1 + E_2}, \qquad \tau_r = \frac{\eta}{E_1 + E_2} ,\] 
with $E_1$ Young modulus of the additional spring, in series with the KV unit, and $E_2$ Young modulus of the elastic spring of the KV element itself, as shown in fig.~\ref{fig.SLSM}. The SLSM is yet able to exhibit all the three primary features of a viscoelastic material: creep, stress relaxation and hysteresis \cite{battista2015}. Concerning the latter aspect, indeed, it is possible to see vessels pressure-area loops representing the energy dissipated during expansion and contraction cycles \cite{nichols2011,salvi2012}. Moreover, in the case of the SLSM, it is possible to define the so-called relaxation function, which describes, through the relaxation time $\tau_r$, how the stiffness of the material changes in time, starting from the instantaneous value, $E_0$, and reaching the asymptotic one, $E_\infty$:
\begin{equation}
E(t) = E_0 e^{-\frac{t}{\tau_r}} + E_{\infty} \left(1- e^{-\frac{t}{\tau_r}}\right) .
\label{relaxationfunction}
\end{equation}
As previously presented for the KV model and in \cite{bertaglia2018}, it is possible to write the SLSM constitutive equation in terms of pressure and area.  Taking into account a generic artery and parameters $K, m$ and $n$ as presented in eq.~\eqref{Ka}, differentiating with respect to time equation $\epsilon = \alpha^m - \alpha^n$ and Barlow's formula, it follows that: $\dot{\epsilon} = A^{-1} \left(m \alpha^m - n \alpha^n\right) \dot{A} \quad \mathrm{and} \quad \dot{\sigma} = W \dot{p}$.
Introducing these equations in the rheological law \eqref{constitutiveEq.} and using the continuity equation \eqref{eq.contST}, the sought PDE is obtained:
\begin{equation}
\partial_t p + \frac{K}{A} \left(m \alpha^m - n \alpha^n\right) \partial_x (Au) = \frac{1}{\tau_r}\left[\frac{E_{\infty}}{E_0}\psi_{el} - (p-p_{ext})\right] .
\label{viscoPDEtubelaw}
\end{equation}
Also with the SLSM it is possible to extend the derivation of the equation to the case of veins with the same procedure discussed for the KV model, considering in this case the set of parameters $K, m$ and $n$ presented in eq.~\eqref{Kv}. It is here highlighted that the viscosity coefficient $\eta$ of the SLSM can be re-conducted to the KV model's viscosity parameter $\Gamma$ directly through eq.~\eqref{gamma}. 
\subsection{The augmented FSI system}
\label{section_FSIsystem}
Based on the choice of how to simulate the mechanical behavior of the vessel wall, whether elastic or viscoelastic, adding respectively eq.~\eqref{elasticPDEtubelaw} or eq.~\eqref{viscoPDEtubelaw} inside system \eqref{eq.cont&mom}, it is possible to obtain a novel 1D augmented fluid-structure interaction system valid for the blood flow modeling. Furthermore, to accommodate longitudinal discontinuities of geometrical and material properties, such as equilibrium cross-sectional area, instantaneous Young modulus \(E_0\) (or eventually the asymptotic one, $E_{\infty}$) and external pressure \(p_{ext}\), it is necessary to introduce additional equations to the system, to allow a formally correct treatment \cite{muller2013}. Considering that these variables are constant in time, the additional equations result: \(\partial_t A_0=0\), \(\partial_t E_0=0\) and \(\partial_t p_{ext}=0\). \\
The complete coupled system of the FSI problem finally results:
\begin{subequations}
\begin{align}
	&\partial_t A + \partial_x (Au) = 0 \label{eq.cont}\\
	&\partial_t (Au) + \partial_x (Au^2)  + \frac{A}{\rho} \hspace{0.5mm} \partial_x p = 0 		 \label{eq.mom}\\
	&\partial_t p + d \hspace{1mm}\partial_x (Au) = S \label{eq.PDE}\\
	&\partial_t A_0 = 0 \\
	&\partial_t E_0 = 0 \\
	&\partial_t p_{ext} = 0 
\end{align}
\label{completesyst}
\end{subequations}
with the choice, in the present work, to neglect friction losses in favor of a better analysis of the energy losses attributable only to viscoelastic effects. In eq.~\eqref{eq.PDE}, the parameter $d(A,A_0,E_0)$ represents the elastic contribution of the tube law, having indeed the same formulation if choosing an elastic or a viscoelastic characterization of the wall:
\begin{equation}
d = \frac{K}{A} \left(m \alpha^m - n \alpha^n\right) .
\end{equation}
In the same equation, the source term $S(x,t)$ takes into account the viscous information of the vessel wall behavior, hence simply be $S = 0$, as in eq.~\eqref{elasticPDEtubelaw}, if an elastic wall behavior is considered, or 
\begin{equation}
S = \frac{1}{\tau_r}\left[\frac{E_{\infty}}{E_0}\psi_{el} - (p-p_{ext})\right] ,
\label{source}
\end{equation}
as in eq.~\eqref{viscoPDEtubelaw}, if a more realistic viscoelastic behavior is chosen. \par
The reader is invited to observe how the formulation of the source term in eq.~\eqref{source} is coherent to the wall's mechanical behavior assumed. If we consider that the viscosity coefficient tends to zero, $\eta~\rightarrow~0$, and so also the relaxation time $\tau_r~\rightarrow~0$, we are looking at asymptotically achieve a perfectly elastic behavior of the material, which automatically leads to $S~\rightarrow~0$ through eq.~\eqref{eq.PDE}. On the other hand, a relaxation time that tends to zero means, referring to the relaxation function presented in eq.~\eqref{relaxationfunction}, that the Young modulus is constant in time, being $E_{\infty} \equiv E_0$, and thus $\sfrac{E_{\infty}}{E_0}~\rightarrow~1$. This is, moreover, in accordance with the fact that, when we are tending to a perfectly elastic behavior, one direct consequence is that $\psi_{el}~\rightarrow~(p - p_{ext})$. The two latter aspects, together, lead again to the result $S~\rightarrow~0$. \par
It is also worth to underline that the choice of inserting the tube law in the form of a PDE straight inside the system of equations is even advantageous when selecting a viscoelastic rheological characterization of the vessel wall. Indeed, if the classical formulation is followed, together with the choice of the KV viscoelastic model, eq.~\eqref{eq.KV} enters inside the system through eq.~\eqref{eq.momST}, and in particular through the derivative in space of the pressure. This procedure gives rise to a second order derivative in space of the flow rate $Au$, which leads to deal with a non-hyperbolic system and also to substantial numerical issues. In literature this problem is treated differently, e.g. resorting a hyperbolic reformulation of the system, introducing a numerical relaxation parameter mimic Cattaneo's law \cite{montecinos2014}, using a discontinuous Galerkin scheme \cite{alastruey2011}, or considering a specific operator-splitting procedure \cite{formaggia2003,mynard2015}. When considering the proposed augmented FSI system, instead, the system persists to be naturally hyperbolic.\par
Writing the non-linear non-conservative system \eqref{completesyst} in the general compact form leads to:
\begin{equation}
\partial_t \boldsymbol{Q} + \partial_x \boldsymbol{f}(\boldsymbol{Q}) + \boldsymbol{B}(\boldsymbol{Q}) \partial_x \boldsymbol{Q} = \boldsymbol{S}(\boldsymbol{Q}) ,
\label{systcompactform}
\end{equation}
in which
	\[\boldsymbol{Q} =
\begin{pmatrix} 
  	A \\ A u \\ p \\ A_0 \\ E_0 \\ p_{ext}
\end{pmatrix}, \quad
	\boldsymbol{f}(\boldsymbol{Q}) = 
\begin{pmatrix} 
 	A u \\ A u^2 \\ 0 \\ 0 \\ 0 \\ 0
\end{pmatrix}, \quad
	\boldsymbol{S}(\boldsymbol{Q}) = 
\begin{pmatrix} 
  	0 \\ 0 \\ S \\ 0 \\ 0 \\ 0
\end{pmatrix}, \quad
\boldsymbol{B}(\boldsymbol{Q}) =
\begin{pmatrix} 
  	0 &0 &0 &0 &0 &0 \\ 0 &0 &\frac{A}{\rho} &0 &0 &0 \\ 0 &d &0 &0 &0 &0 \\ 0 &0 &0 &0 &0 &0 \\ 0 &0 &0 &0 &0 &0 \\ 0 &0 &0 &0 &0 &0
\end{pmatrix} .\]
The system can also be written in the quasi-linear form,
\begin{equation}
\partial_t \boldsymbol{Q} + \boldsymbol{A}(\boldsymbol{Q})\partial_x \boldsymbol{Q} = \boldsymbol{S}(\boldsymbol{Q}) ,
\label{systqlinearform}
\end{equation}
considering \(\boldsymbol{A}(\boldsymbol{Q}) = \partial \boldsymbol{f}/\partial \boldsymbol{Q} + \boldsymbol{B}(\boldsymbol{Q})\), with
\[\frac{\partial \boldsymbol{f}}{\partial \boldsymbol{Q}} =
\begin{pmatrix} 
  	0 &1 &0 &0 &0 &0 \\ -u^2 &2u &0 &0 &0 &0 \\ 0 &0 &0 &0 &0 &0 \\ 0 &0 &0 &0 &0 &0 \\ 0 &0 &0 &0 &0 &0 \\ 0 &0 &0 &0 &0 &0
\end{pmatrix} .\]
It can be demonstrated that the system is hyperbolic, being the matrix \(\boldsymbol{A}(\boldsymbol{Q})\) diagonalizable, with a diagonal matrix \(\boldsymbol{\Lambda}(\boldsymbol{Q})\) containing all real eigenvalues \(\lambda_l\), with $l = 1,\cdots, N$ and $N$ number of unknowns of the system (in this work $N = 6$), and a complete set of linearly independent eigenvectors represented by the columns of the matrix \(\boldsymbol{R}(\boldsymbol{Q})\):\\
	\[\boldsymbol{\Lambda}(\boldsymbol{Q}) =
\begin{pmatrix} 
  	0 &0 &0 &0 &0 &0 \\ 0 &0 &0 &0 &0 &0 \\ 0 &0 &0 &0 &0 &0 \\ 0 &0 &0 &0 &0 &0 \\ 0 &0 &0 &0 &u+c &0 \\ 0 &0 &0 &0 &0 &u-c 
\end{pmatrix}, \quad
	\boldsymbol{R}(\boldsymbol{Q}) =
\begin{pmatrix} 
  	\frac{A}{\rho u^2} &0 &0 &0 &\frac{1}{d} &\frac{1}{d} \\ 0 &0 &0 &0 &\frac{u+c}{d} &\frac{u-c}{d} \\ 1 &0 &0 &0 &1 &1 \\ 0 &1 &0 &0 &0 &0 \\ 0 &0 &1 &0 &0 &0 \\ 0 &0 &0 &1 &0 &0
\end{pmatrix}, \]
with $c$ the wave speed:
\begin{equation}
c = \sqrt{\frac{A}{\rho} \hspace{0.5mm} d } = \sqrt{\frac{A}{\rho} \hspace{0.5mm} \frac{\partial p}{\partial A} } = \sqrt{\frac{K}{\rho} \left(m\alpha^m - n\alpha^n\right)} \quad .
\label{eq:cel}
\end{equation} 
Concerning the eigenvectors, the first, the fifth and the sixth characteristic fields are genuinely non-linear and are associated with shocks and rarefactions, whereas the remaining fields are linearly degenerate (LD) and are associated with stationary contact discontinuities. Evaluating the Riemann invariants (RI) of the system,  those associated with the genuinely non-linear fields are:
\begin{equation*}
\Gamma_1 = u + \int\frac{c}{A}\d A, \quad \Gamma_2 = u - \int\frac{c}{A}\d A, \quad \Gamma_3 = p - \int d \hspace{0.5mm} \d A \hspace{0.5mm} ;
\end{equation*}
while those associated with the linearly degenerate fields result:
\begin{equation*}
\Gamma_1^{LD} = p +\frac{1}{2}\rho u^2, \quad \Gamma_2^{LD} = Au \hspace{0.5mm} .
\end{equation*}
It has to be noticed that when dealing with arteries, integrals in the RI associated with the genuinely non-linear fields can be analytically resolved, resulting:
\[\Gamma_1 = u + 4c , \qquad  \Gamma_2 = u - 4c , \qquad  \Gamma_3 = p - K\sqrt{\alpha} .\]
Finally, it is here mentioned that, under physiological conditions, the source term of system \eqref{completesyst} may become stiff, depending on the spatial discretization applied. Referring to \cite{muller2012}, indeed, a source term can be considered stiff if 
\begin{equation}
\Delta x \frac{\max \{ \mid\beta_l\mid \}}{\max \{\mid\lambda_l\mid \}} > 1 , \quad l = 1,\cdots, N ,
\label{eq:def.stiff}
\end{equation}
with $\beta_l$ representing the $l$-$th$ eigenvalue of the Jacobian of $\boldsymbol{S}(\boldsymbol{Q})$ and $\Delta x$ mesh size. It can be evaluated that $\max \{ \mid\beta_l\mid \} = \sfrac{1}{\tau_r}$, which could reach values up to 5 orders of magnitude more than $\max \{\mid\lambda_l\mid \} = u+c$. Additionally, it is evident that the relaxation time of the material $\tau_r$, related to viscoelasticity, is the key parameter leading to stiffness.
\section{Numerical model}
\label{section_numericalmodel}
To solve system \eqref{completesyst}, which behaves exactly like a hyperbolic systems with stiff relaxation terms, the Implicit-Explicit (IMEX) Runge-Kutta schemes proposed by Pareschi and Russo \cite{pareschi2005} for applications to these kind of systems have been considered. This scheme is asymptotic preserving and asymptotic accurate in the zero relaxation limit (i.e. the consistency of the scheme with the equilibrium system is guaranteed and the order of accuracy is maintained in the stiff limit), with an elevated robustness given by the use of an implicit Runge-Kutta scheme for the treatment of the stiff part. Usually, simpler splitting techniques are preferred to solve these kind of problems; for example Strang splitting provides second order of accuracy if each step is at least second order accurate in space \cite{strang1968}. However, as shown in different references \cite{leveque1990,descombes2004,pareschi2005,duarte2011} and confirmed here for the specific application presented in section \ref{section_numericalresults_accuracy}, this technique reduces to first order of accuracy when the problem becomes highly stiff. In these situations, indeed, the fastest time scales play a leading role in the global physics of the phenomenon and the composed solution of the splitting technique fails to capture the proper dynamics of the event. Recently developed Runge-Kutta schemes overcome this issue. Thus, a formally implicit finite volume discretization is adopted, applying a second-order L-stable diagonally implicit Runge-Kutta method (DIRK) to the stiff part and a second-order explicit strong-stability-preserving (SSP) method to the non-stiff terms, with the addition of the path-conservative Dumbser-Osher-Toro (DOT) Riemann solver, as applied in \cite{bertaglia2018} for compressible flows in polymer tubes. 
\subsection{IMEX Runge-Kutta scheme with DOT solver}
The second-order IMEX Runge-Kutta finite volume discretization of system \eqref{systcompactform} takes the form:
\begin{subequations}
\begin{align}
	&\boldsymbol{Q}^{(k)}_i = \boldsymbol{Q}^n_i -  \frac{\Delta t}{\Delta x} \sum_{j=1}^{k-1} \tilde{a}_{kj} \left[ \left( \boldsymbol{F}_{i + \frac{1}{2}}^{(j)}  - \boldsymbol{F}_{i - \frac{1}{2}}^{(j)} \right)  + \left(\boldsymbol{D}_{i+\frac{1}{2}}^{(j)}  + \boldsymbol{D}_{i-\frac{1}{2}}^{(j)}\right) + \boldsymbol{B}\left( \boldsymbol{Q}_i^{(j)}\right) \Delta \boldsymbol{Q}_i^{(j)} \right] + \Delta t \sum_{j=1}^{s} a_{kj} \boldsymbol{S}\left(\boldsymbol{Q}_i^{(j)}\right)
	\label{eq.iterIMEX} \\
	& \boldsymbol{Q}^{n+1}_i = \boldsymbol{Q}^n_i -  \frac{\Delta t}{\Delta x} \sum_{k=1}^{s} \tilde{\omega}_{k} \left[ \left( \boldsymbol{F}_{i + \frac{1}{2}}^{(k)}  - \boldsymbol{F}_{i - \frac{1}{2}}^{(k)} \right)  + \left(\boldsymbol{D}_{i+\frac{1}{2}}^{(k)}  + \boldsymbol{D}_{i-\frac{1}{2}}^{(k)}\right) + \boldsymbol{B}\left( \boldsymbol{Q}_i^{(k)}\right) \Delta \boldsymbol{Q}_i^{(k)} \right] + \Delta t \sum_{k=1}^{s} \omega_{k} \boldsymbol{S}\left(\boldsymbol{Q}_i^{(k)}\right)  \label{eq.finalIMEX}
\end{align}
\label{eq.IMEX}
\end{subequations}
using a uniform grid of length $L$ and \(N_x\) elements with mesh spacing \(\Delta x = x_{i+\frac{1}{2}}-x_{i-\frac{1}{2}} = L/N_x\) and a time step size \(\Delta t = t^{n+1}-t^{n}\) that follows the \(\CFL\) condition, with $\boldsymbol{Q}^n_i$ vector of the averaged variables on the $i$-$th$ cell of the domain at time $t^n$. Matrices $\tilde A = (\tilde a_{kj})$, with $\tilde a_{kj} = 0 $ for $ j\geq k$ and $A = (a_{kj})$ are $s \times s$ matrices such that the resulting scheme is implicit in $\boldsymbol{S}(\boldsymbol{Q})$ and explicit for all the rest, with $s$ number of stages. Moreover, being a DIRK scheme, $a_{kj} = 0 $ for $ j > k$. An IMEX Runge-Kutta scheme is characterized by these two matrices and by the coefficient vectors $\tilde \omega = (\tilde \omega_1, ...,\tilde \omega_s)^T$, $\omega = (\omega_1, ...,\omega_s)^T$, which can be easily represented by a double tableau in the usual Butcher notation \cite{pareschi2005}:
\begin{center}
\begin{tabular}{c | c}
$\tilde c_k$ & $\tilde a_{kj}$ \\  
\hline \\[-1.0em] 
 &  $\tilde \omega_k^T$ \\ 
\end{tabular}
\hspace{2.0cm}
\begin{tabular}{c | c}
$c_k$ & $a_{kj}$ \\  
\hline \\[-1.0em] 
 &  $\omega_k^T$ \\ 
\end{tabular}
\end{center}
where coefficient vectors $\tilde c$ and $c$ are given by:
\begin{equation*}
\tilde c_k = \sum^{k-1}_{j=1} \tilde a_{kj} , \qquad \qquad c_k = \sum^{k}_{j=1} a_{kj} .
\end{equation*}
In particular, in the present work it has been chosen the stiffly accurate IMEX-SSP2(3,3,2) Runge-Kutta scheme, characterized by $s=3$ stages for the implicit part, 3 stages for the explicit part and 2nd order of accuracy, which can be defined by the following tableau (explicit part on the left and implicit part on the right):
\begin{center}
\begin{tabular}{c | c c c}
0 & 0 & 0 & 0 \\
1/2 & 1/2 & 0 & 0 \\
1 & 1/2 & 1/2 & 0 \\ \hline
 & 1/3  & 1/3 & 1/3 \\
\end{tabular}
\hspace{1.0cm}
\begin{tabular}{c | c c c}
1/4 & 1/4 & 0 & 0 \\
1/4 & 0 & 1/4 & 0 \\
1 & 1/3 & 1/3 & 1/3 \\ \hline
 & 1/3  & 1/3 & 1/3 \\
\end{tabular}
\end{center}
noticing that the explicit discretization coincides with an improved Euler method (Heun's method). The explicit Runge-Kutta methods are precisely those for which the only non-zero entries in the $\tilde A$ matrix of the table lie strictly below the diagonal. Entries at or above the diagonal will cause the right hand side of eq.~\eqref{eq.iterIMEX} to involve $\boldsymbol Q_i^{(j)}$, giving a formally implicit method. Nevertheless, it is worth to highlight that, for the system of equations presented in this work, it is possible to obtain a totally explicit algorithm, as discussed in the Appendix, which leads to a consistent reduction of the computational cost. More details concerning the implemented IMEX algorithm can be found in the just mentioned Appendix, at the end of the paper. \par
For each step of the method, the numerical fluxes are obtained applying the DOT solver as defined in \cite{dumbser2011a}:
\begin{equation}
\label{eq:flux}
	\boldsymbol{F}_{i\pm\frac{1}{2}} = \frac{1}{2} \left[ \boldsymbol{f}\left(\boldsymbol{Q}_{i\pm\frac{1}{2}}^{+}\right) + \boldsymbol{f}\left(\boldsymbol{Q}_{i\pm\frac{1}{2}}^{-}\right)\right] - \frac{1}{2} \int_{0}^{1} \left \lvert \boldsymbol{A}\left(\Psi\left(\boldsymbol{Q}_{i\pm\frac{1}{2}}^{-},\boldsymbol{Q}_{i\pm\frac{1}{2}}^{+},s\right)\right)\right \rvert \frac{\partial \Psi}{\partial s}\d s ,
\end{equation}
with a numerical dissipation related to matrix \(\boldsymbol{A}(\boldsymbol{Q})\) that includes both conservative and non-conservative terms. The fluctuations given by the non-conservative part then read \cite{dumbser2011}:
\begin{equation}
\label{eq:fluct}
	\boldsymbol{D}_{i\pm \frac{1}{2}} = \frac{1}{2} \int_{0}^{1} \boldsymbol{B}\left(\Psi\left(\boldsymbol{Q}_{i\pm \frac{1}{2}}^{-},\boldsymbol{Q}_{i\pm\frac{1}{2}}^{+},s\right)\right)\frac{\partial \Psi}{\partial s}\d s .
\end{equation}
The boundary-extrapolated values within cell \(i\) are given by:
\begin{equation*}
\label{eq:Qbound}
	\boldsymbol{Q}_{i+\frac{1}{2}}^{-} = \boldsymbol{Q}_i + \frac{1}{2}\Delta	\boldsymbol{Q}_i,
	\qquad \qquad
\boldsymbol{Q}_{i-\frac{1}{2}}^{+} = \boldsymbol{Q}_i - \frac{1}{2}\Delta\boldsymbol{Q}_i .
\end{equation*}
The slope \( \Delta \boldsymbol{Q}_i \) is evaluated using the classical minmod slope limiter to achieve second-order of accuracy also in space \cite{Toro2009}, avoiding spurious oscillations near discontinuities. The symbol \(\Psi\) stands for the path connecting left to right boundary values in the phase-space; in this work a simple linear segment has been chosen \cite{Pares2006}, hence:
\begin{equation}
\label{eq:path}
	\Psi = \Psi\left(\boldsymbol{Q}_{i+\frac{1}{2}}^{-},\boldsymbol{Q}_{i+\frac{1}{2}}^{+},s\right) = \boldsymbol{Q}_{i+\frac{1}{2}}^{-} + s\left(\boldsymbol{Q}_{i+\frac{1}{2}}^{+} - \boldsymbol{Q}_{i+\frac{1}{2}}^{-}\right).
\end{equation}
The integrals in relations \eqref{eq:flux} and \eqref{eq:fluct} are approximated by a simple 3-points Gauss-Legendre quadrature formula.
\subsection{Well-balancing proof}
\label{section_wellbalancing}
A numerical scheme is here defined exactly well-balanced (or, with the same connotation, satisfies the exact conservation property, i.e. C-property \cite{bermudez1994}) if it is exact when applied to the stationary case at zero flow rate. For the particular case of a scheme of the form \eqref{eq.IMEX} for system \eqref{completesyst}, this means that, for a given initial state of rest $\boldsymbol{Q}^n_i,  \forall i \in [1,\cdots,N_x]$, for each Runge-Kutta stage it turns out that:
\begin{equation*}
\boldsymbol{F}_{i + \frac{1}{2}}  - \boldsymbol{F}_{i - \frac{1}{2}} = 0 , \qquad
\boldsymbol{D}_{i+\frac{1}{2}}  + \boldsymbol{D}_{i-\frac{1}{2}} = 0 ,\qquad
\boldsymbol{B}\left( \boldsymbol{Q}_i\right) \Delta \boldsymbol{Q}_i = 0 , \qquad
\forall i \in [1,\cdots,N_x].
\end{equation*}
Considering a condition of zero velocity ($u=0$, $Au=0$), observing system \eqref{completesyst}, it must be that $p$ is constant in $x$. Thus, both the second and the third component in $\Delta \boldsymbol{Q}_i$ will necessarily be zero. Evaluating then the product $\boldsymbol{B}\left( \boldsymbol{Q}_i\right) \Delta \boldsymbol{Q}_i$, it can be noticed that the only non-zero columns in $\boldsymbol{B}\left( \boldsymbol{Q}_i\right)$ multiply exactly the two zero components in $\Delta \boldsymbol{Q}_i$, obtaining a null vector as result. Choosing a simple linear path $\Psi$, the same applies also to the term $\boldsymbol{B}\left(\Psi\right)\frac{\partial \Psi}{\partial s}$ when evaluating the fluctuations $\boldsymbol{D}_{i\pm \frac{1}{2}}$ through eq.~\eqref{eq:fluct} and to the product $\left \lvert \boldsymbol{A}\left(\Psi\right)\right \rvert \frac{\partial \Psi}{\partial s}$ when evaluating the numerical fluxes $\boldsymbol{F}_{i\pm\frac{1}{2}}$ with eq.~\eqref{eq:flux}. Finally, the analytical fluxes $\boldsymbol{f}\left(\boldsymbol{Q}_{i\pm\frac{1}{2}}^{\pm}\right)$ are automatically null when considering zero flow rate. It is therefore proved that the numerical model here discussed is well-balanced for the augmented FSI system of blood flow equations. It is also underlined that the C-property arises in a straightforward way from the construction of the explicit part of the scheme and not from the implicit one, which leads to have more freedom in choosing the particular set of coefficients of the IMEX method.
\subsection{Pressure update analysis}
Evaluating the Rankine-Hugoniot condition related to the continuity equation \eqref{eq.cont} and to the additional elastic constitutive PDE \eqref{eq.PDE}, it is observed that the celerity of the shock wave results respectively \cite{Toro2009}:
\begin{equation*}
\xi = \frac{\Delta(Au)}{\Delta A},
\qquad \qquad
\xi = \frac{\Delta(Au)}{\Delta p} \int_0^1 d \hspace{0.5mm} \d s .
\end{equation*}
Since the celerity of the discontinuity is unique, it is concluded that it must be necessarily:
\begin{equation*}
\frac{\Delta p}{\Delta A} = \int_0^1 d \hspace{0.5mm} \d s = \int_0^1 \frac{\partial p}{\partial A} \hspace{0.5mm} \d s .
\end{equation*}
This expression is always true if dealing with linear systems and can be considered a good approximation in case of mildly non-linear systems, but it is certainly not valid when working with highly deformable and even collapsible veins \cite{carpenter2001}. In fact, adding the tube law inside the system of equations leads, in the elastic case, to the establishment of an over-abundant system and a consequent inconsistency.\\
Hence, to correctly update through the tube law the non-conservative variable $p$, in accordance with the updating of the related conservative variable $A$ even in case of non-linear systems, an alternative evolution of the pressure has been introduced in the scheme. Defining for each $k$-$th$ Runge-Kutta step
\begin{equation*}
	\tilde{p}^{(k)}_i = p_{ext,i} + \psi_{el,i}^{(k)} ,
	\qquad \qquad
	\tilde{p}^{(k+1)}_i = p_{ext,i} + \psi_{el,i}^{(k+1)}
\end{equation*}
the pressure with physical sense if an elastic behavior of the vessel wall is considered, the contribution for the pressure variation in time exclusively linked to the area variation, both in the elastic and the viscoelastic case, results:
\begin{equation*}
	\Delta \tilde{p}^{(k)}_i = \tilde{p}^{(k+1)}_i - \tilde{p}^{(k)}_i .
\end{equation*}
Thus, the additional pressure update, containing all the information of the time evolution, even regarding the viscoelastic contribution enclosed in the source term, results:
\begin{equation}
p^{(k+1)}_i = p^{(k)}_i + \Delta \tilde{p}^{(k)}_i + \Delta t \sum_{j=1}^{s} a_{kj} \boldsymbol{S}\left(\boldsymbol{Q}_i^{(j)}\right) .
\end{equation}
In the same way, the final update between time $t^n$ and $t^{n+1}$ will be:
\begin{equation}
p^{n+1}_i = p^n_i + \Delta \tilde{p}^n_i + \Delta t \sum_{k=1}^{s} \omega_{k} \boldsymbol{S}\left(\boldsymbol{Q}_i^{(k)}\right) .
\end{equation}
Involving this additional evaluation, $p$ is properly updated following the time evolution of the variable $A$ through the constitutive law, with the due phase-synchronization of the two variables requested by the elastic contribution of the rheological model.\par
\begin{table}[b!]
\centering
\begin{tabular}{c | c c c c c}
\hline
\\[-1.1em]
	 Variable &RP1 &RP2 &RP3 &RP4 &RP5 \\
\\[-1.1em]
\hline
	\(A_L\)~[mm$^2$] &6.4138 &2.5082 &0.9900 &4.7030 &2.0000 \\
	\(A_R\)~[mm$^2$] &3.1282 &3.2921 &2.0800 &2.1947 &0.2222 \\
	\(u_L\)~[m/s] &0.00 &1.00 &0.00 &0.00 &0.00 \\
	\(u_R\)~[m/s] &0.00 &0.00 &0.00 &0.00 &0.00 \\
	\(p_L\)~[mmHg] &80.00 &146.67 &9.97 &178.99 &43.38 \\
	\(p_R\)~[mmHg] &80.00 &108.78 &46.05 &8.05 &4.58 \\
	\(A_{0,L}\)~[mm$^2$] &6.2706 &1.5677 &1.1000 &3.1353 &1.0000 \\
	\(A_{0,R}\)~[mm$^2$] &3.1353 &3.1353 &1.3000 &3.1353 &1.0000 \\
	\(E_{0,L}\)~[MPa] &2.7655 &1.3828 &0.4604 &1.9555 &0.3991 \\
	\(E_{0,R}\)~[MPa] &1.9555 &1.9555 &5.9153 &1.9555 &0.3991 \\
	\(p_{ext,L}\)~[mmHg] &75.00 &30.00 &10.00 &80.00 &5.00 \\
	\(p_{ext,R}\)~[mmHg] &85.00 &0.00  &5.00  &80.00 &5.00 \\
	\(x_0\)~[m] &0.10 &0.05 &0.05 &0.10 &0.15 \\
\hline
\end{tabular}
\caption{Initial states for the Riemann problems. Subscripts \(L\) and \(R\) stand respectively for left and right state of the piece-wise constant initial values, while $x_0$ indicates the position of the initial discontinuity. For all the tests the length of the domain $L = 0.2$ m and the vessel wall thickness $h_0 = 0.3$ mm.}
\label{tab.RPdata}
\end{table}

\begin{figure}[t!]
\begin{subfigure}{0.5\textwidth}
\centering
\includegraphics[width=1\linewidth]{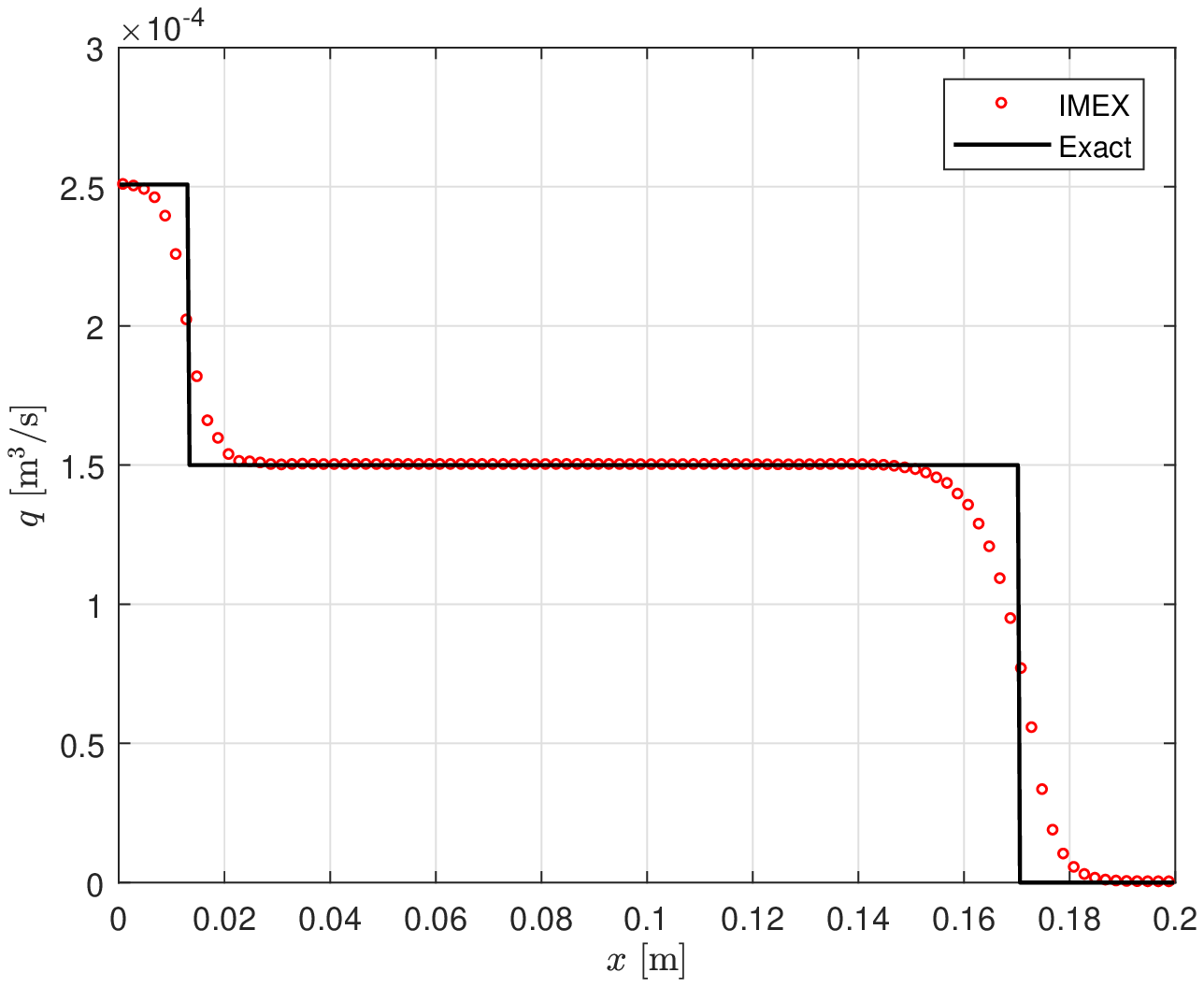}
\vspace*{-5mm}
\caption{}
\label{fig.RP2q}
\end{subfigure}
\begin{subfigure}{0.5\textwidth}
\centering
\includegraphics[width=1\linewidth]{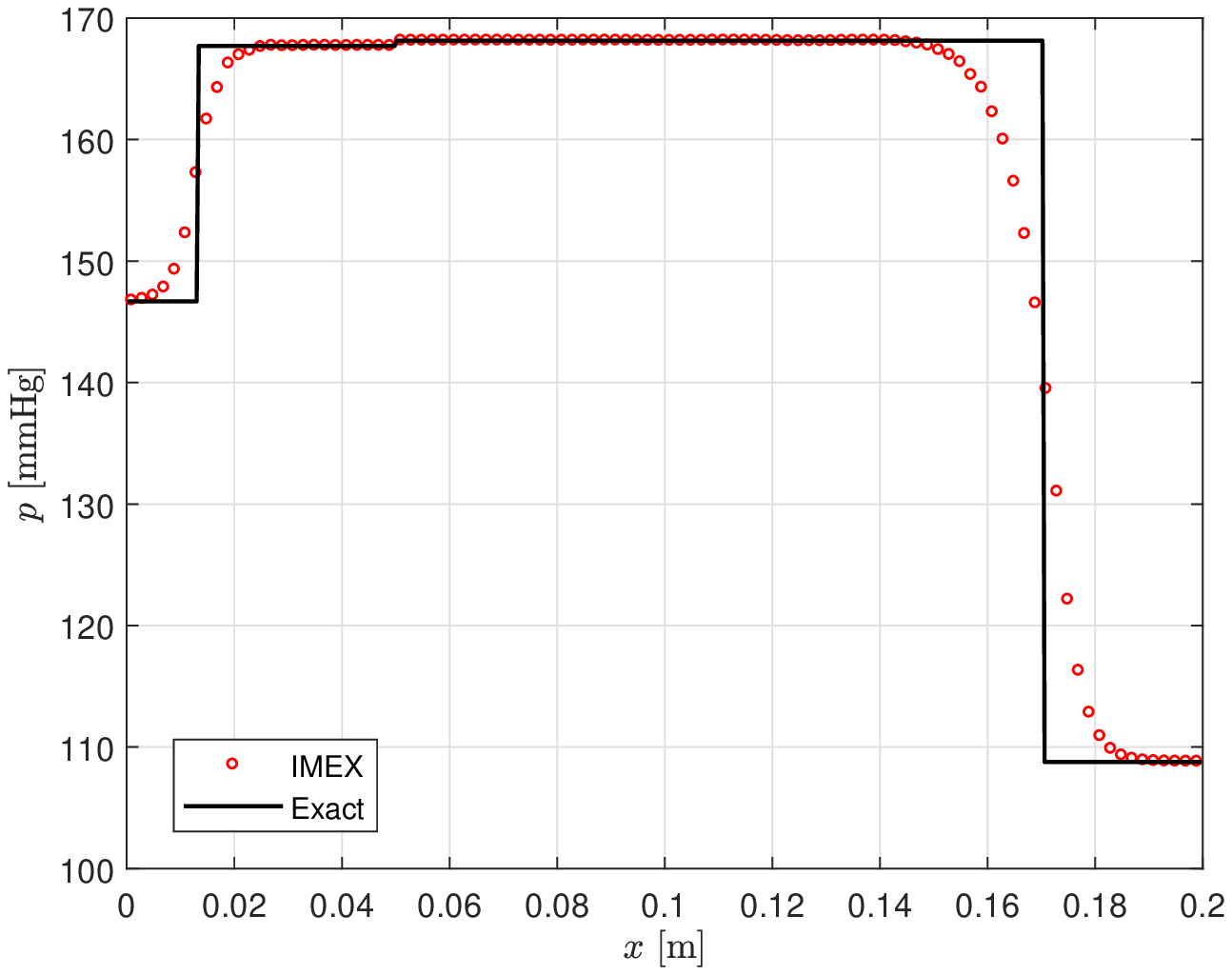}
\vspace*{-5mm}
\caption{}
\label{fig.RP2p}
\end{subfigure}
\begin{subfigure}{0.5\textwidth}
\centering
\includegraphics[width=1\linewidth]{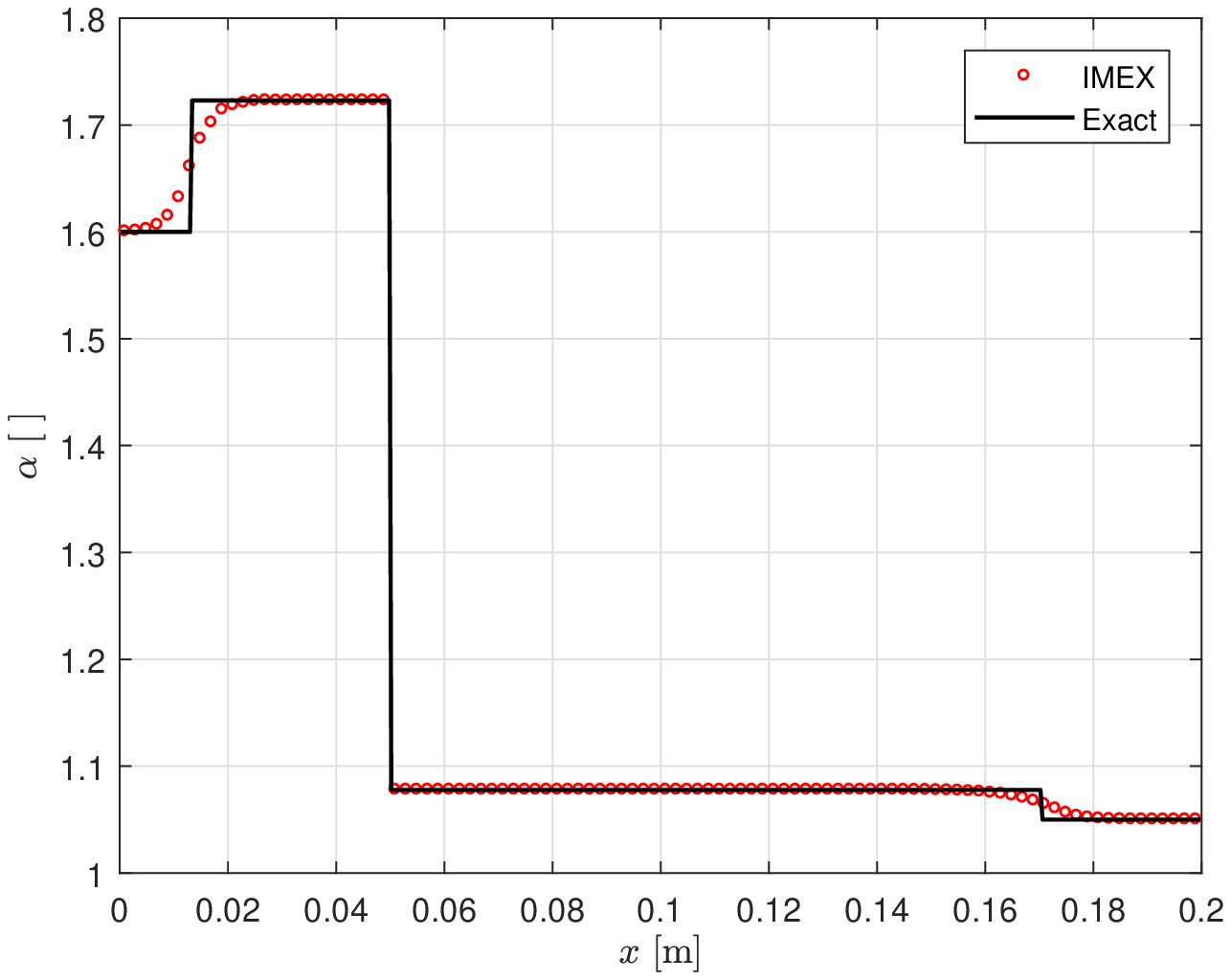}
\vspace*{-5mm}
\caption{}
\label{fig.RP2alpha}
\end{subfigure}
\begin{subfigure}{0.5\textwidth}
\centering
\includegraphics[width=1\linewidth]{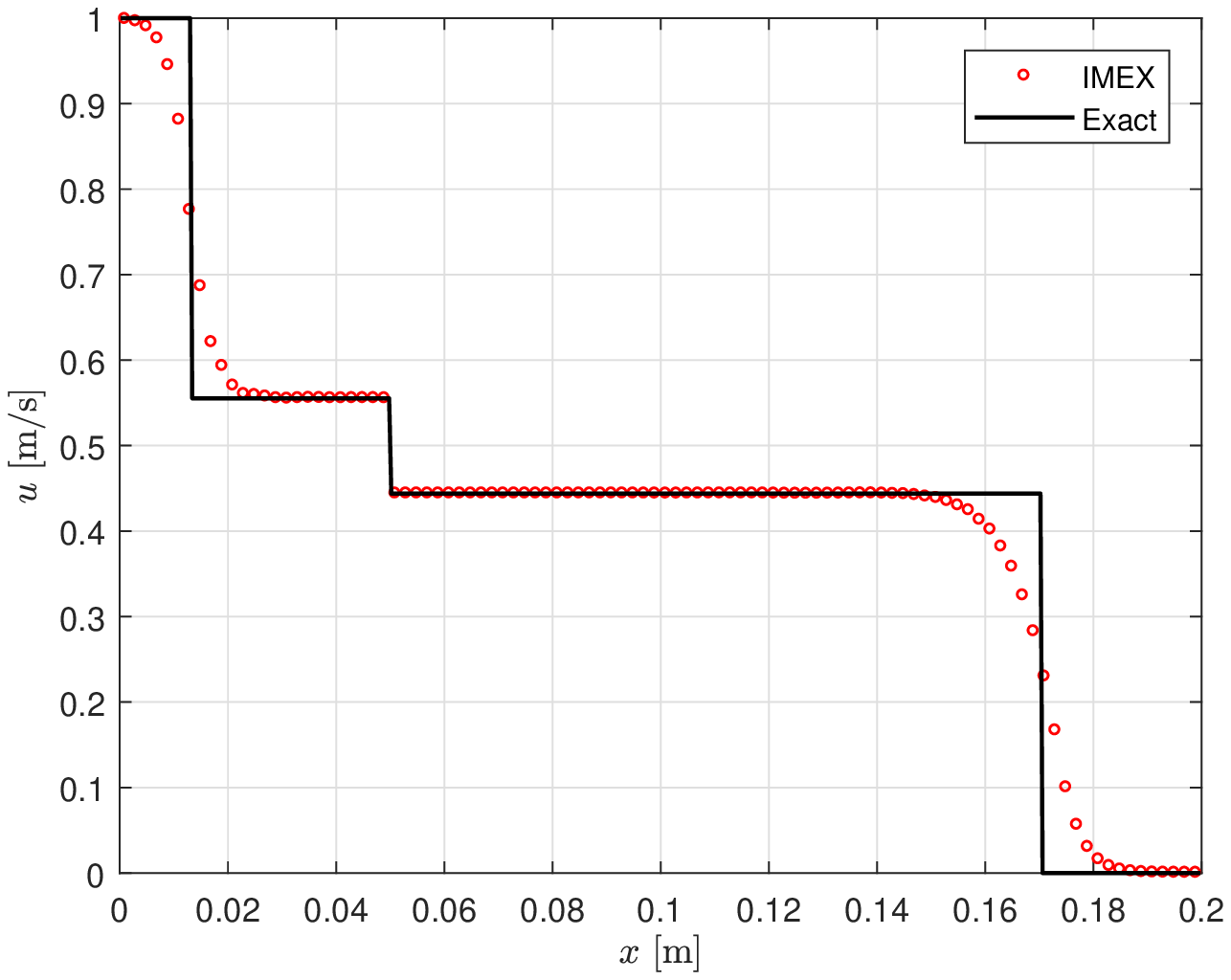}
\vspace*{-5mm}
\caption{}
\label{fig.RP2u}
\end{subfigure}
\caption{Results obtained in test RP2, at time \(t_{end}\) = 0.007 s, solving the augmented FSI system with the IMEX Runge-Kutta scheme in terms of (a) flow rate, (b) pressure, (c) non-dimensional cross-sectional area and (d) velocity, with respect to the exact solution.}
\label{fig.RP2}
\end{figure}

\begin{figure}[t!]
\begin{subfigure}{0.5\textwidth}
\centering
\includegraphics[width=1\linewidth]{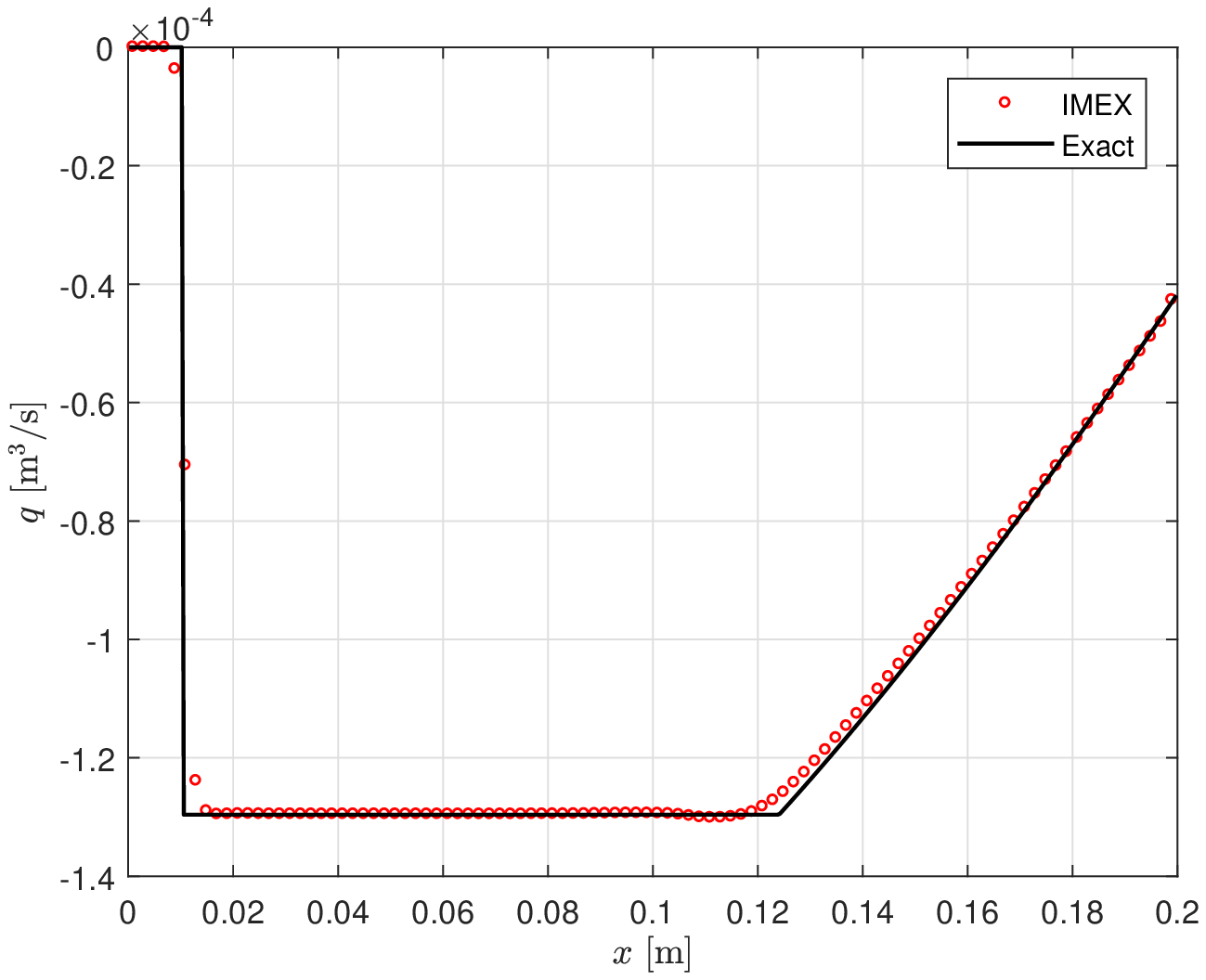}
\vspace*{-5mm}
\caption{}
\label{fig.RP3q}
\end{subfigure}
\begin{subfigure}{0.5\textwidth}
\centering
\includegraphics[width=1\linewidth]{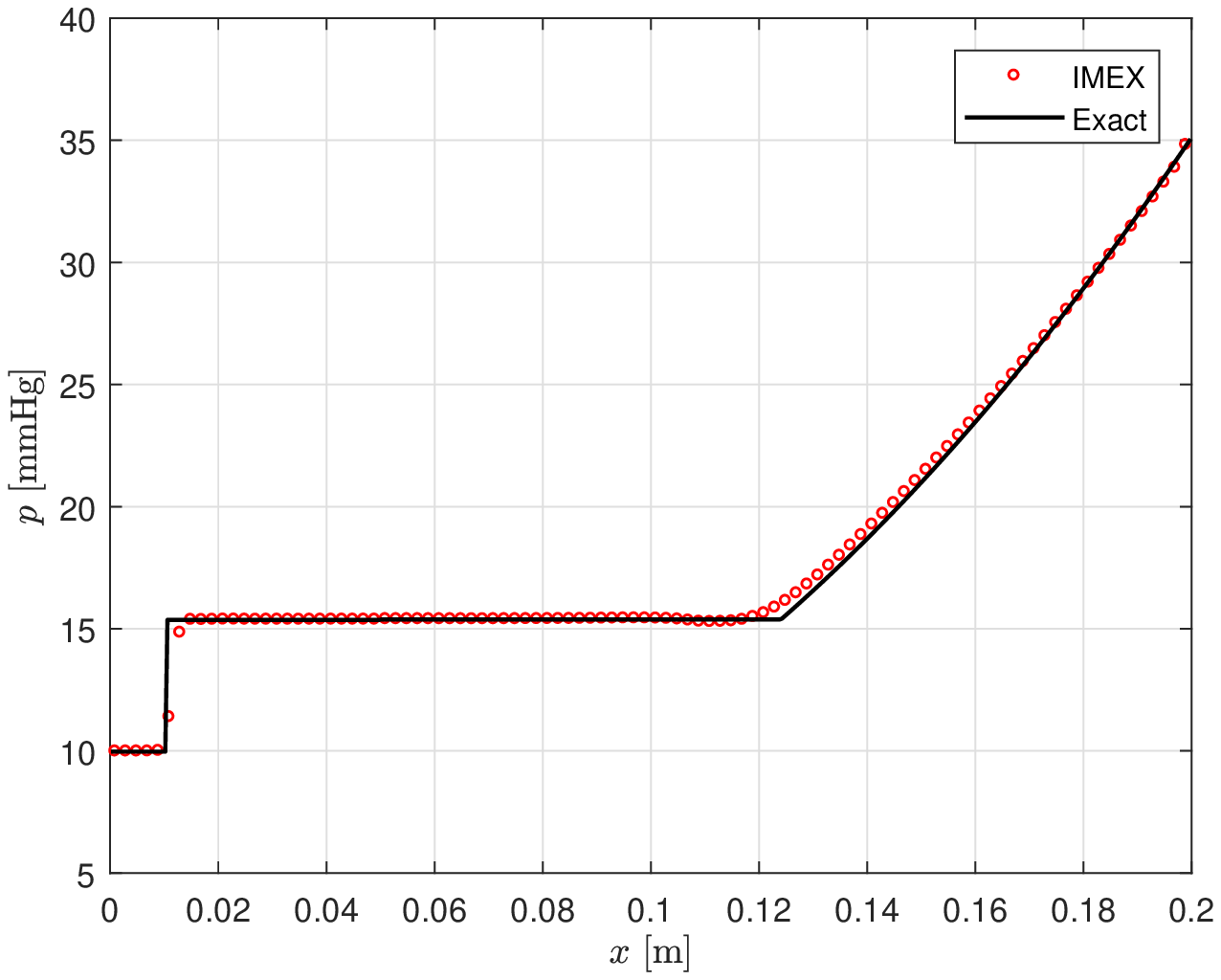}
\vspace*{-5mm}
\caption{}
\label{fig.RP3p}
\end{subfigure}
\begin{subfigure}{0.5\textwidth}
\centering
\includegraphics[width=1\linewidth]{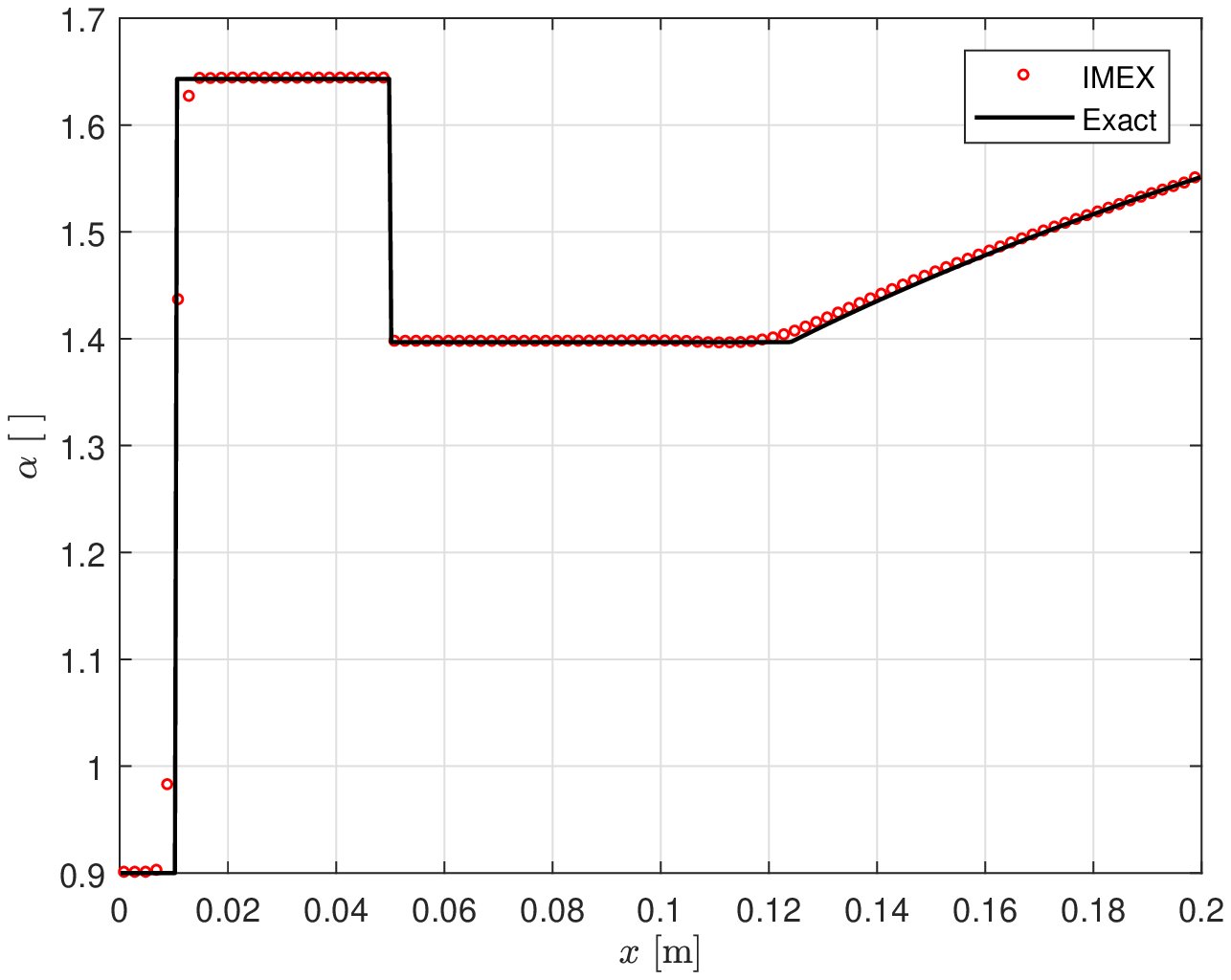}
\vspace*{-5mm}
\caption{}
\label{fig.RP3alpha}
\end{subfigure}
\begin{subfigure}{0.5\textwidth}
\centering
\includegraphics[width=1\linewidth]{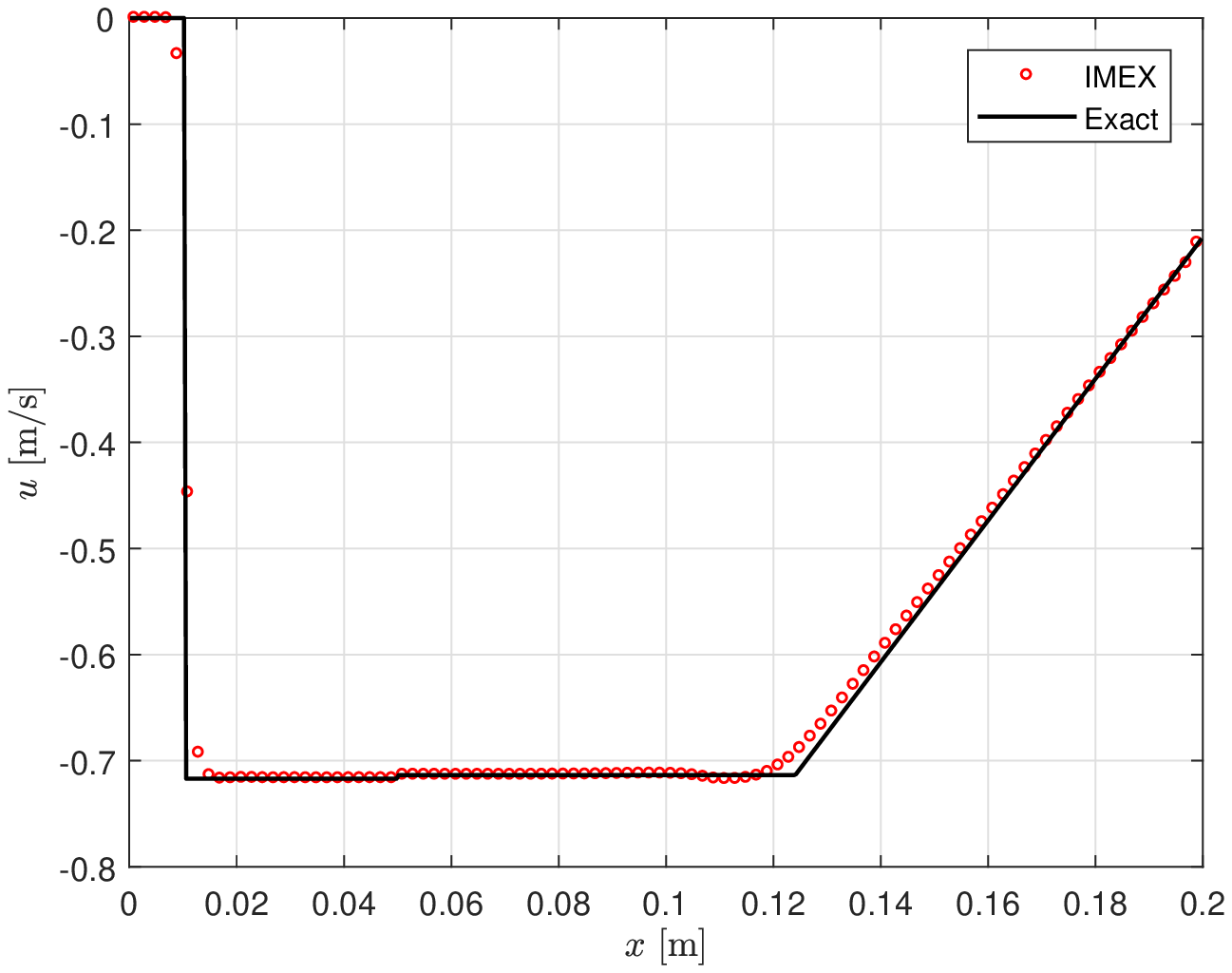}
\vspace*{-5mm}
\caption{}
\label{fig.RP3u}
\end{subfigure}
\caption{Results obtained in test RP3, at time \(t_{end}\) = 0.025 s, solving the augmented FSI system with the IMEX Runge-Kutta scheme in terms of (a) flow rate, (b) pressure, (c) non-dimensional cross-sectional area and (d) velocity, with respect to the exact solution.}
\label{fig.RP3}
\end{figure}

\begin{figure}[t!]
\begin{subfigure}{0.5\textwidth}
\centering
\includegraphics[width=1\linewidth]{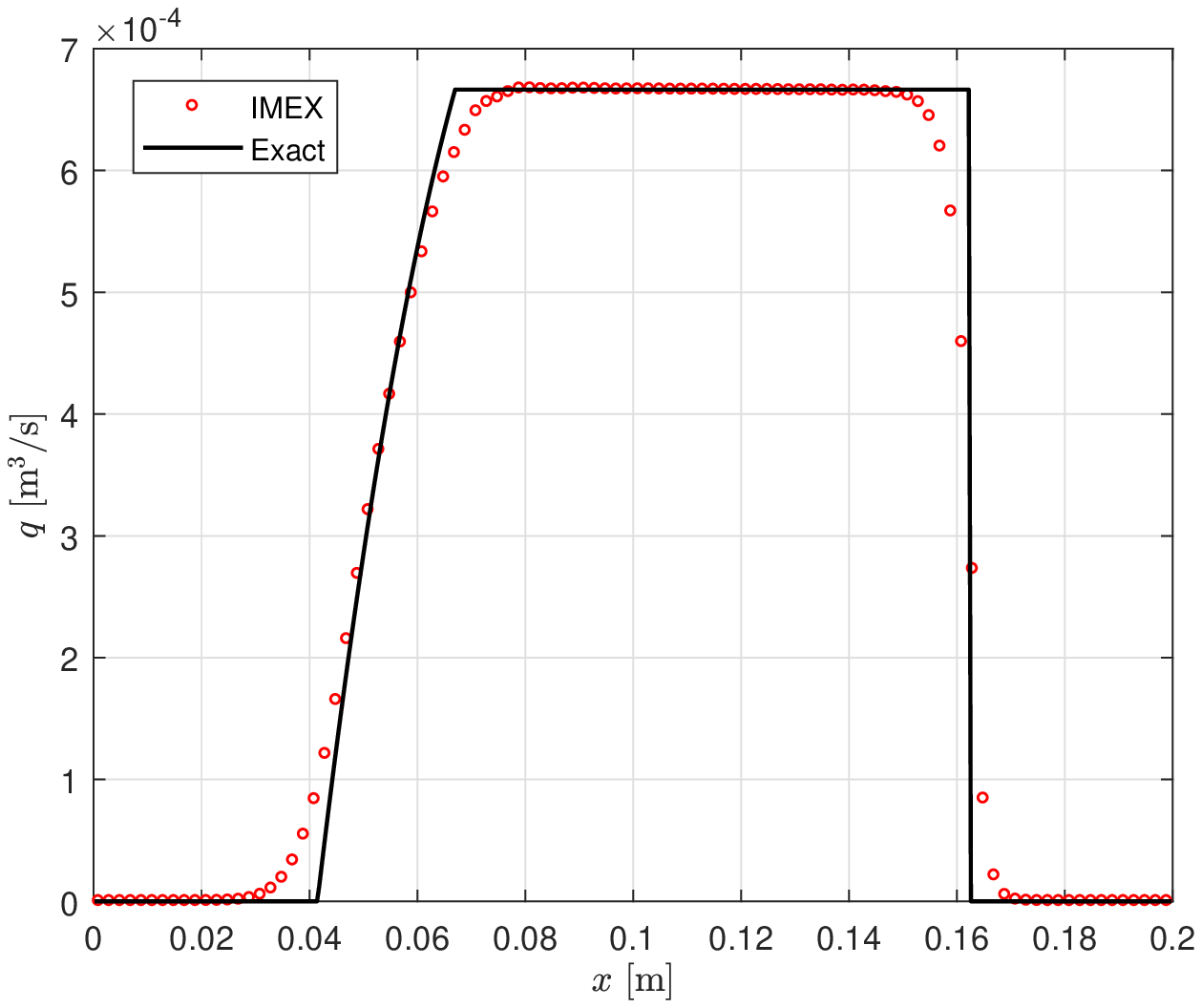}
\vspace*{-5mm}
\caption{}
\label{fig.RP4q}
\end{subfigure}
\begin{subfigure}{0.5\textwidth}
\centering
\includegraphics[width=1\linewidth]{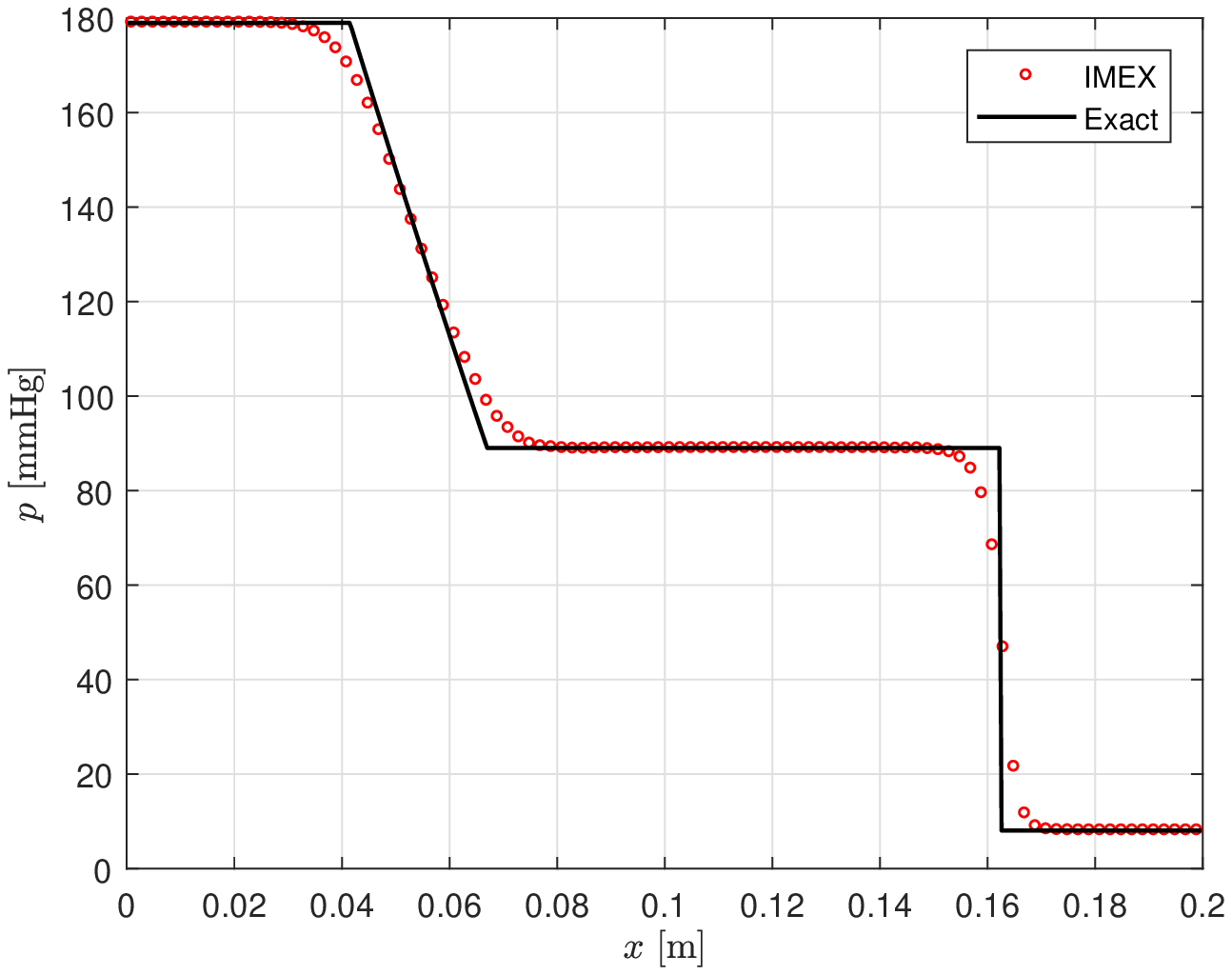}
\vspace*{-5mm}
\caption{}
\label{fig.RP4p}
\end{subfigure}
\begin{subfigure}{0.5\textwidth}
\centering
\includegraphics[width=1\linewidth]{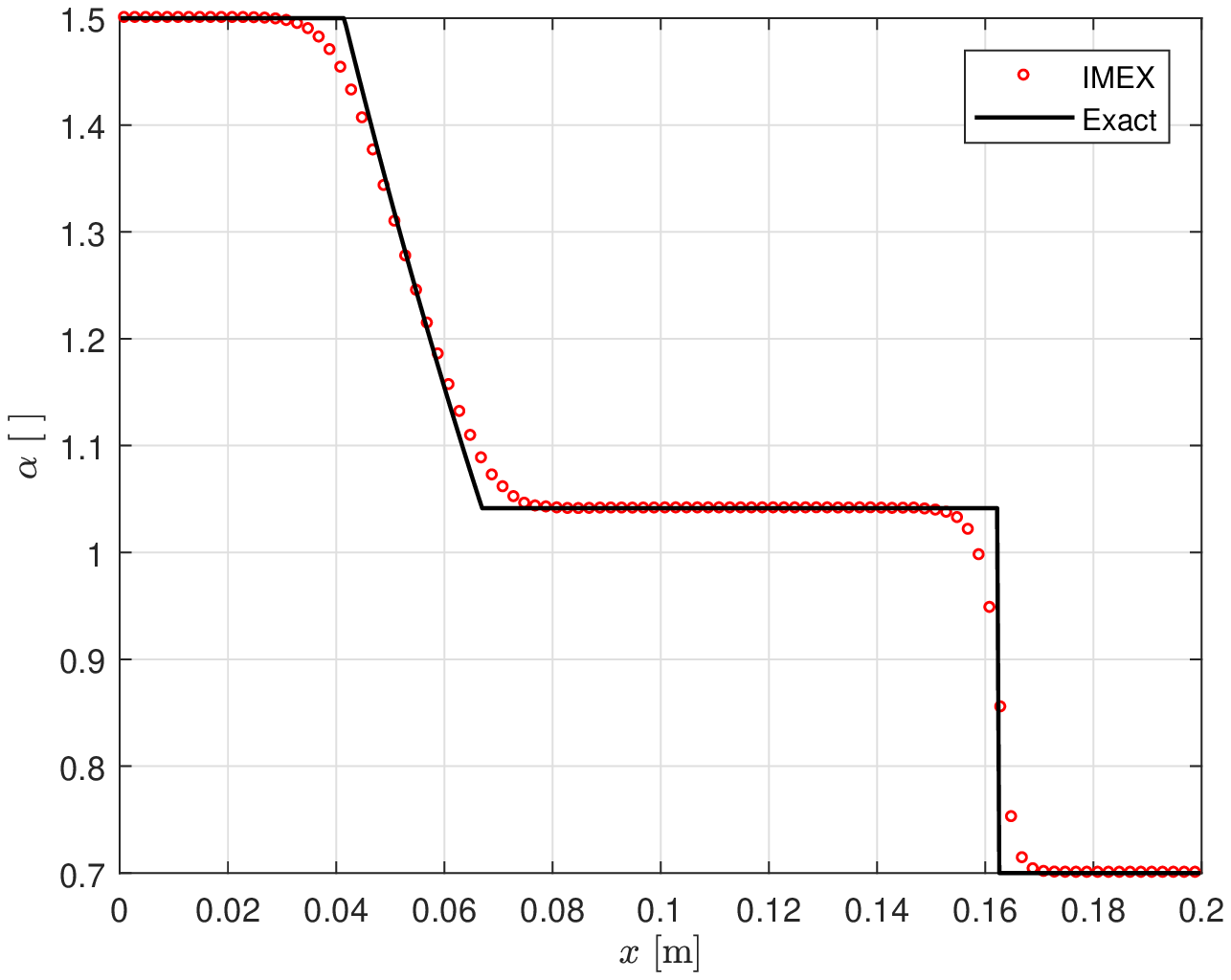}
\vspace*{-5mm}
\caption{}
\label{fig.RP4alpha}
\end{subfigure}
\begin{subfigure}{0.5\textwidth}
\centering
\includegraphics[width=1\linewidth]{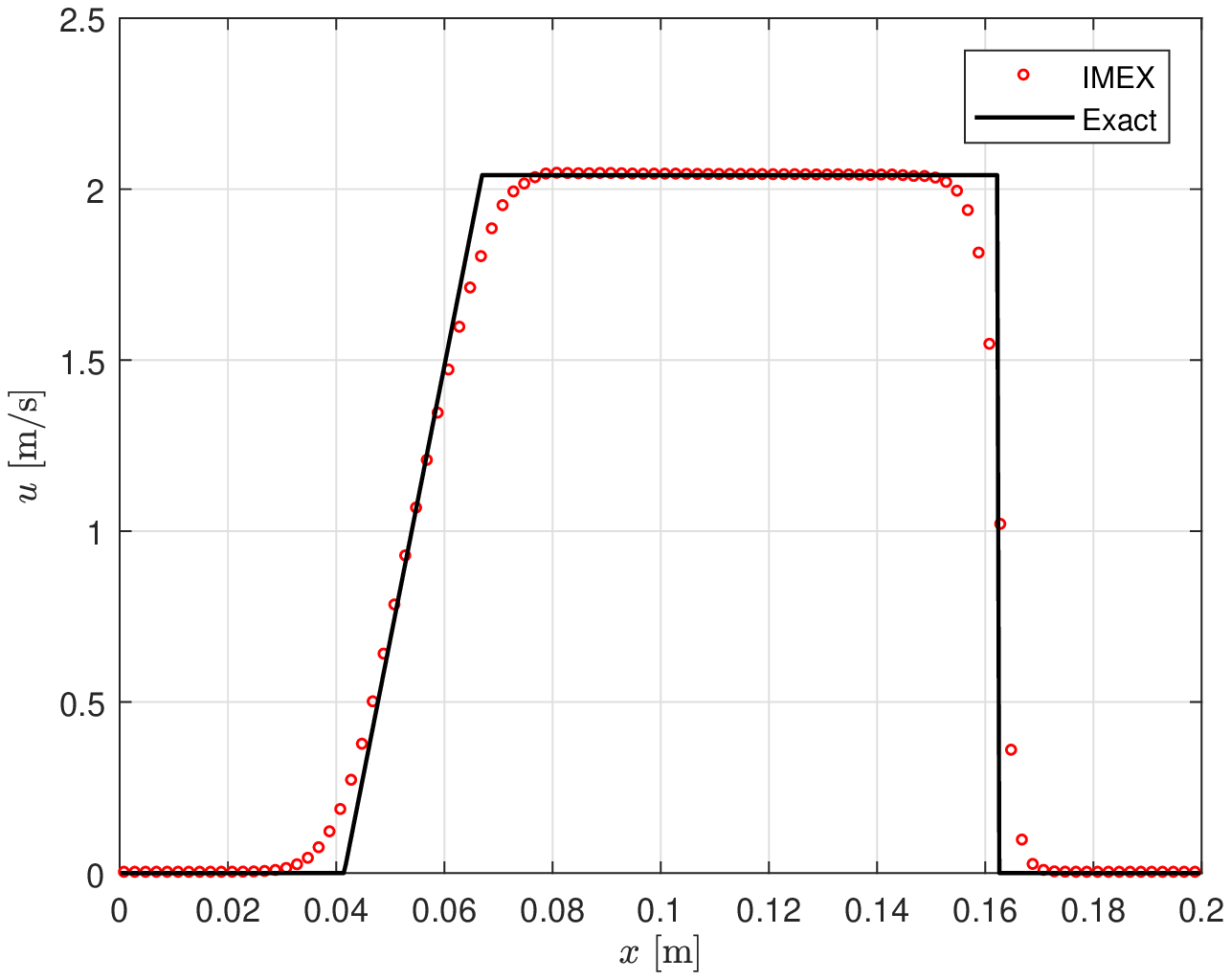}
\vspace*{-5mm}
\caption{}
\label{fig.RP4u}
\end{subfigure}
\caption{Results obtained in test RP4, at time \(t_{end}\) = 0.01 s, solving the augmented FSI system with the IMEX Runge-Kutta scheme in terms of (a) flow rate, (b) pressure, (c) non-dimensional cross-sectional area and (d) velocity, with respect to the exact solution.}
\label{fig.RP4}
\end{figure}

\begin{figure}[t!]
\begin{subfigure}{0.5\textwidth}
\centering
\includegraphics[width=1\linewidth]{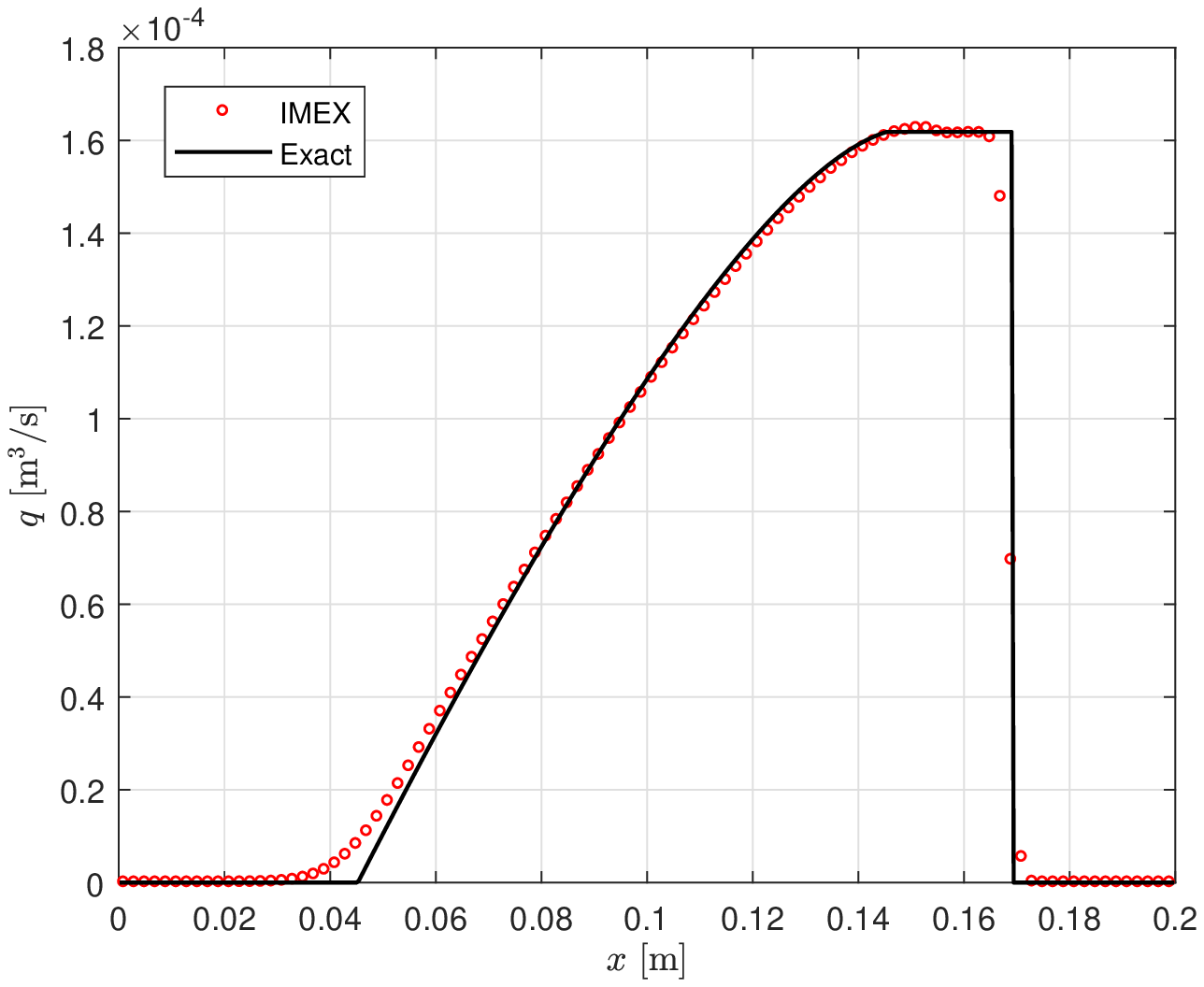}
\vspace*{-5mm}
\caption{}
\label{fig.RP5q}
\end{subfigure}
\begin{subfigure}{0.5\textwidth}
\centering
\includegraphics[width=1\linewidth]{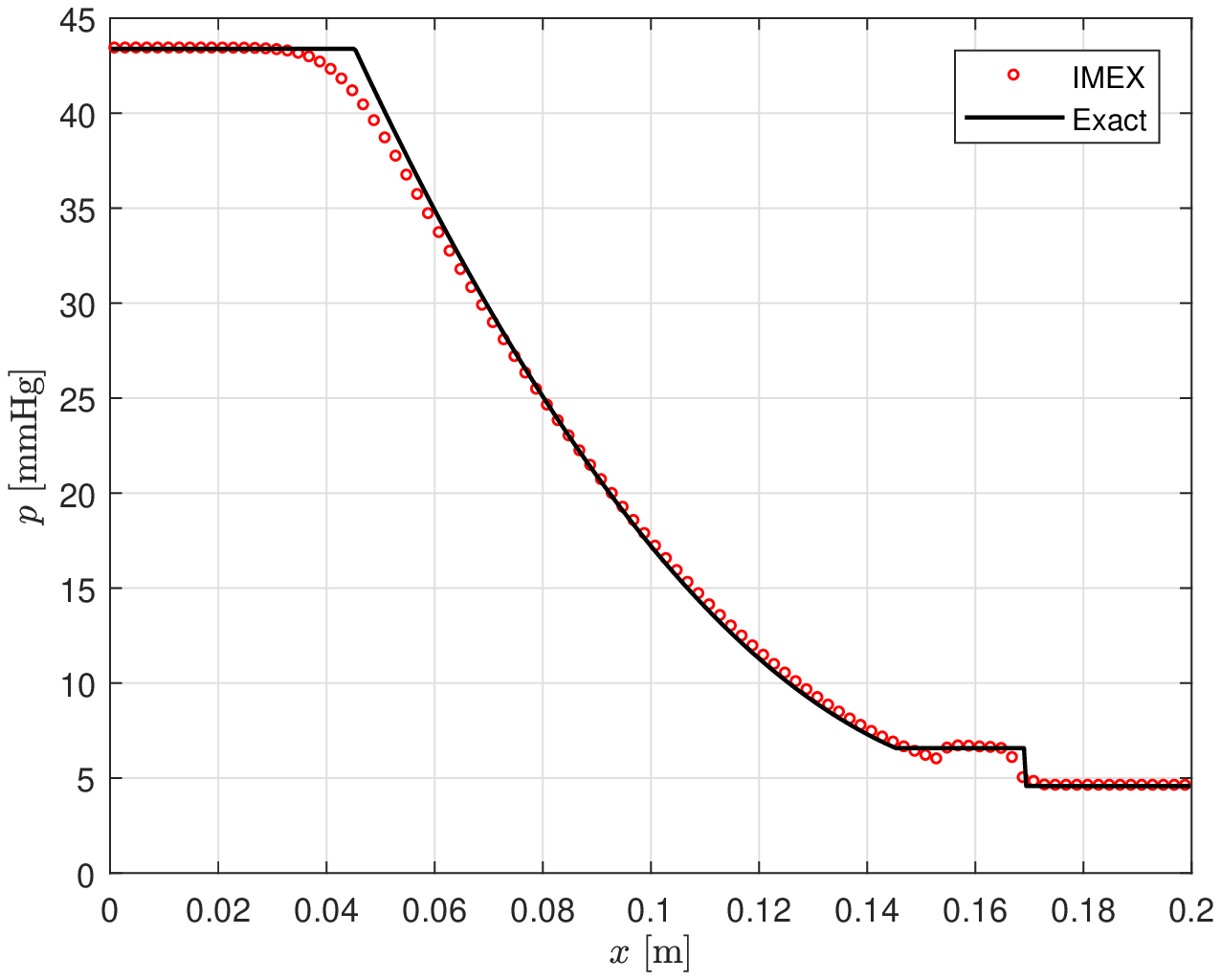}
\vspace*{-5mm}
\caption{}
\label{fig.RP5p}
\end{subfigure}
\begin{subfigure}{0.5\textwidth}
\centering
\includegraphics[width=1\linewidth]{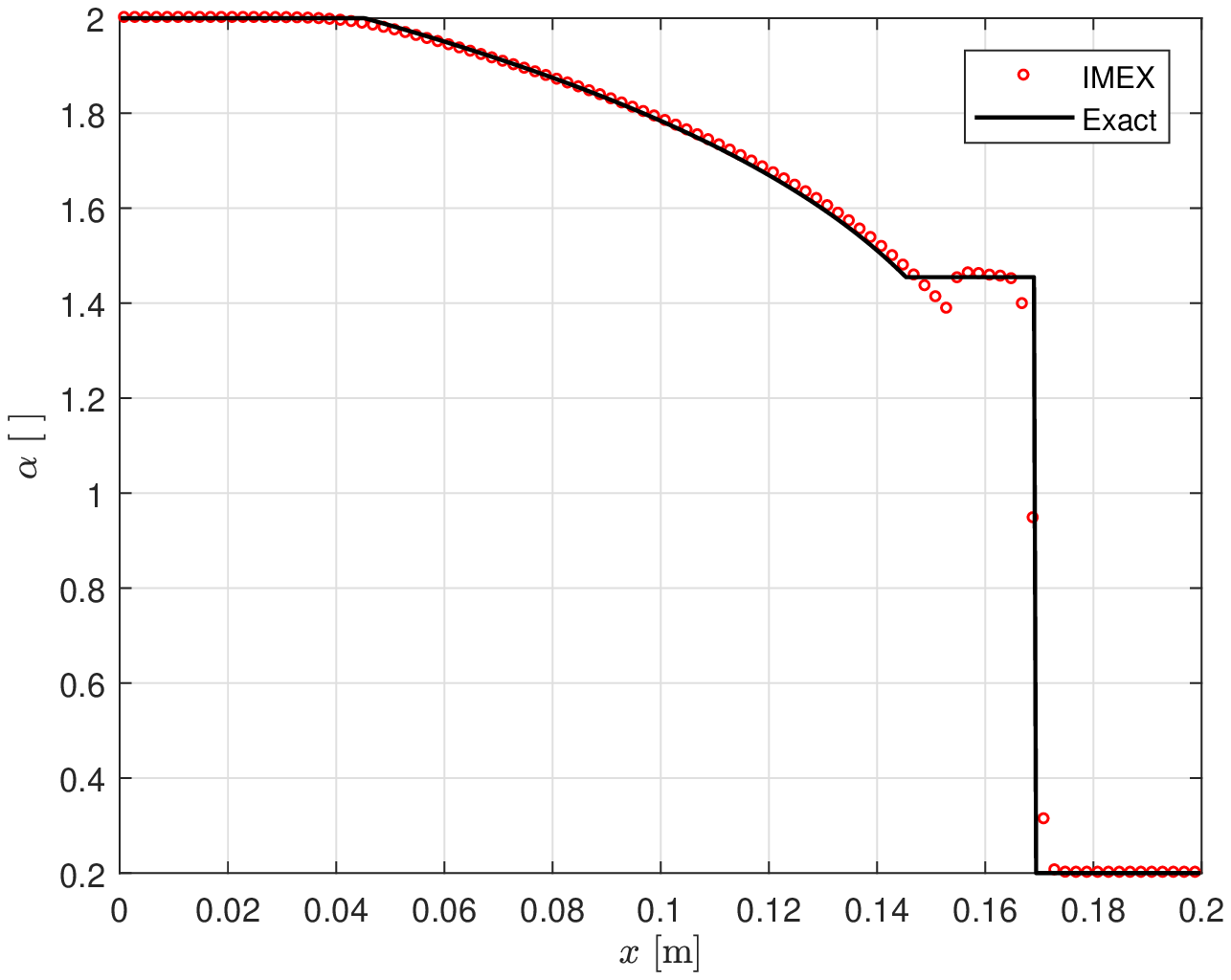}
\vspace*{-5mm}
\caption{}
\label{fig.RP5alpha}
\end{subfigure}
\begin{subfigure}{0.5\textwidth}
\centering
\includegraphics[width=1\linewidth]{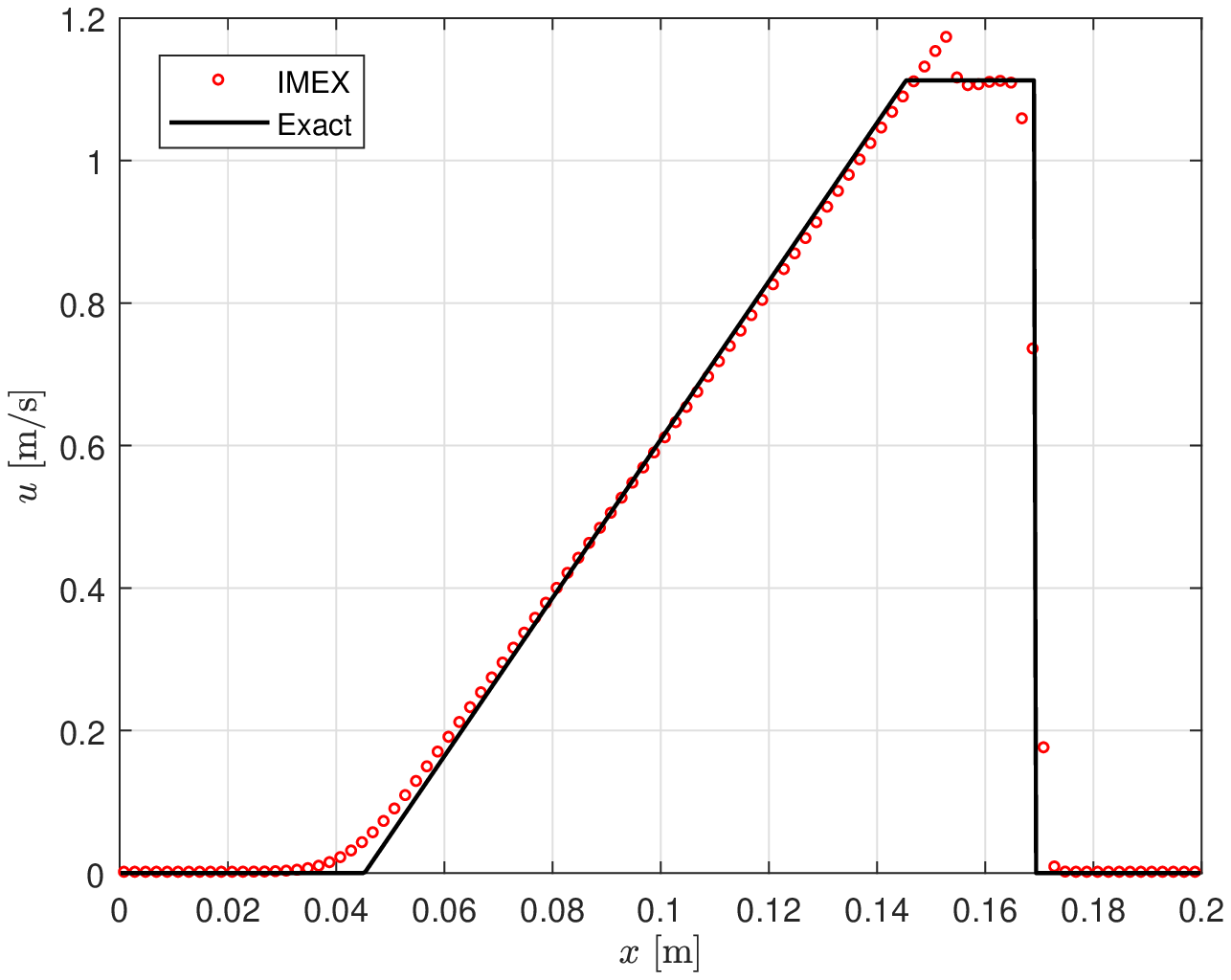}
\vspace*{-5mm}
\caption{}
\label{fig.RP5u}
\end{subfigure}
\caption{Results obtained in test RP5, at time \(t_{end}\) = 0.015 s, solving the augmented FSI system with the IMEX Runge-Kutta scheme in terms of (a) flow rate, (b) pressure, (c) non-dimensional cross-sectional area and (d) velocity, with respect to the exact solution.}
\label{fig.RP5}
\end{figure}
\section{Numerical results}
\label{section_numericalresults}
To check the accuracy and the robustness of the proposed model, for ranges of parameters found in the human body and reported in applications regarding cardiovascular mathematics, different types of test have been set up.\\
Initially, five Riemann problems (RP) have been selected, considering only an elastic behavior of the vessel wall, for which an exact solution is available \cite{toro2013}. The first of these RPs aims to verify the well-balancing of the scheme in a state of rest condition, while the others are related to unsteady problems in a tract of the aorta and in the internal jugular vein. Successively, the C-property of the scheme has been tested also in the case in which, on the left and on the right side of the discontinuity, smooth functions are defined, in the case of a generic artery and a generic vein. \\
Furthermore, given the fact that an exact solution of the problem does not exist when considering the viscoelasticity of the vessels, to validate also the contribution given by the viscoelastic source term in eq.~\eqref{source}, the Method of Manufactured Solutions (MMS) has been applied \cite{roache2002}. Thus, a problem for a modified non-linear system of equations that is a perturbation of the original one via a source term vector has been designed, applying it, again, to a generic artery and to a generic vein. Moreover, the accuracy of the scheme proposed has been verified using these latter problems. Results obtained with the IMEX Runge-Kutta scheme have also been compared to those obtained adopting the Strang splitting technique, for which a reduction of the expected order of accuracy has been observed.\\
An additional more realistic test case has been studied to assess the effects of viscoelasticity with respect to the simple elastic behavior of the vessel wall. Concerning the common carotid artery (CCA), a gaussian pulse wave has been prescribed at the inlet of the vessel, as presented for tests in \cite{acosta2015,sherwin2003a}.\\
In all the simulations presented in this work $\CFL$~=~0.9 with a constant number of cells in the domain \(N_x\)~=~100 (with the clear exception of the tests run for the accuracy analysis). For the geometric and mechanical parameters of the vessels we referred to \cite{muller2013,muller2014a,xiao2014}. The complete code, written in MATLAB (MathWorks Inc.) language, with the implemented test cases, is made available in the Mendeley Data repository associated to this article \cite{bertagliaDATA2019}. 
\subsection{Riemann problems}
The first Riemann problem, RP1, has been designed to verify the well-balancing, in a rest state, of the IMEX Runge-Kutta scheme here proposed for the resolution of the novel augmented FSI system~\eqref{completesyst}. Results of RP1 at $t_{end} = 0.01$ s (after 94 complete time iterations) confirm that the scheme is able to preserve the initial condition in the case of zero flow rate. Indeed, the $L^2$ norm evaluated for the evolution variables $A, Au$ and $p$ results, respectively: $3.83\times 10^{-20}, 0$ and $0$ (values in SI units). The problem presents an initial discontinuity in $A$, $A_0$, $E_0$ and $p_{ext}$.\\
The second Riemann problem, RP2, represents a hypothetical case of systolic pressure and peak flow arriving in a portion of the thoracic aorta. In this problem, the left side of the aorta, so the part that in the initial state was already reached by the systolic peak (represented in fig.~\ref{fig.RP2} by the shock wave on the right), is compressed, while on the right of the initial discontinuity the aorta is 10 times stiffer than the part on the left. This idealized configuration leads to a partial reflection of the incoming wave, which can be noticed in fig.~\ref{fig.RP2} by the presence of the shock wave on the left. From the same figure it is possible to observe that a good agreement between the exact solution and the numerical results is achieved for all the variables. \\
In the third Riemann problem, RP3, the effects of an idealized Valsalva maneuver on a portion of the internal jugular vein are depicted. The Valsalva maneuver consists in the practice of exhaling while closing all the airways, producing a sharp increase of the central venous pressure of the analyzed subject. Moreover, in this test it is considered that downstream, in the correspondence of the heart, there is an incompetent valve that causes a venous reflux towards the head (represented in fig.~\ref{fig.RP3} by the elastic jump traveling to the left). Even in this very challenging test case, in which there is also the presence of a rarefaction wave traveling to the right, the numerical results agree very well with the exact solution. These first three Riemann problems were reproduced with reference to \cite{muller2013}. \\
Riemann problems 4 (RP4) and 5 (RP5), presented in figs.~\ref{fig.RP4}~-~\ref{fig.RP5}, concern again a single tract of the thoracic aorta and a single tract of the internal jugular vein respectively. In this case a unique initial jump in the cross-sectional area (and thereby in the pressure, being evaluated through the tube law) is considered, to assess the effects of a solely geometrical discontinuity. Even in these tests, numerical results correctly capture the exact solution, confirming once more the suitability of the model to solve also unsteady problems, both in arteries and veins. It is here underlined that, in RP5, the small undershoot/overshoot of the numerical solution in correspondence of the tail of the rarefaction wave (figs.~\ref{fig.RP5alpha}~-~\ref{fig.RP5u}) depends on the slope limiter chosen. This irregularity, indeed, would not appear using the superbee limiter instead of the minmod. Nevertheless, it has been decided to present the results using a single slope limiter in a uniform manner, with the minmod being the most accurate among all the tests. Additionally, it is possible to observe, comparing the rarefaction wave in fig.~\ref{fig.RP5p} with the one in fig.~\ref{fig.RP5alpha}, the typical non-concave - non-convex behavior of veins \cite{spiller2017} given by the specific parameters adopted in the tube law presented in eq.~\eqref{elastictubelaw}. The complete set of initial data is listed, for each RP, in table~\ref{tab.RPdata}.
\begin{table}[b!]
\centering
\begin{tabular}{l | c c c c c c c}
\hline
\\[-1.1em]
	 Test &\(L\) [m] &\(h_0\) [mm] &\(u_0\) [m/s] &\(p_0\) [mmHg] &\(a\) [mm$^2$] &\(e_0\) [MPa] &\(p_{e}\) [mmHg] \\
\\[-1.1em]
\hline
	CP1 &0.10 &1.50 &0.00 &80.00 &1.00 &1.00 &80.00 \\
	CP2 &0.10 &0.30 &0.00 &10.00 &0.01 &0.10 &10.00 \\
\hline
\end{tabular}
\caption{Parameters used for the C-property tests: domain length $L$, vessel wall thickness $h_0$, reference velocity $u_0$, reference pressure $p_0$, reference cross-sectional area $a$, reference instantaneous Young modulus $e_0$, reference external pressure $p_e$.}
\label{tab.CPdata1}
\end{table}

\begin{table}[b!]
\centering
\begin{tabular}{c | c}
\hline
\\[-1.1em]
	Variable &IC \\
\\[-1.1em]
\hline
	\(u_L\)~[m/s] &$u_0$ \\
	\\[-1.1em]
	\(u_R\)~[m/s] &$u_0$ \\
	\\[-1.1em]
	\(p_L\)~[mmHg] &$p_0$ \\
	\\[-1.1em]
	\(p_R\)~[mmHg] &$p_0$ \\
	\\[-1.1em]
	\(A_{0,L}\)~[mm$^2$] &$a + \frac{a}{2}\sin\left(\frac{8\pi x}{L}\right)$ \\
	\\[-1em]
	\(A_{0,R}\)~[mm$^2$] &$2a + \frac{a}{2}\sin\left(\frac{8\pi x}{L}\right)$ \\
	\\[-1em]
	\(E_{0,L}\)~[MPa] &$e_0 + \frac{e_0}{2}\sin\left(\frac{8\pi x}{L}\right)$ \\
	\\[-1em]
	\(E_{0,R}\)~[MPa] &$2e_0 + \frac{e_0}{2}\sin\left(\frac{8\pi x}{L}\right)$ \\
	\\[-1em]
	\(p_{ext,L}\)~[mmHg] &$p_e + \frac{p_e}{2}\sin\left(\frac{8\pi x}{L}\right)$ \\
	\\[-1em]
	\(p_{ext,R}\)~[mmHg] &$2p_e + \frac{p_e}{2}\sin\left(\frac{8\pi x}{L}\right)$ \\
	\\[-1em]
	\(x_0\)~[m] &$0.5 \hspace{0.5mm} L$ \\
\hline
\end{tabular}
\caption{Initial conditions (IC) for the C-property tests CP1 and CP2. Subscripts \(L\) and \(R\) stand for the smooth initial values respectively on the left and on the right of the initial discontinuity, located in $x_0$. The initial condition of the area $A$ is evaluated solving the tube law presented in eq.~\eqref{elastictubelaw}. For parameters $u_0, p_0, a, e$ and $p_e$ refer to table \ref{tab.CPdata1}.}
\label{tab.CPdata2}
\end{table}
\subsection{C-property problems}
\begin{figure}[t!]
\begin{subfigure}{0.5\textwidth}
\centering
\includegraphics[width=1\linewidth]{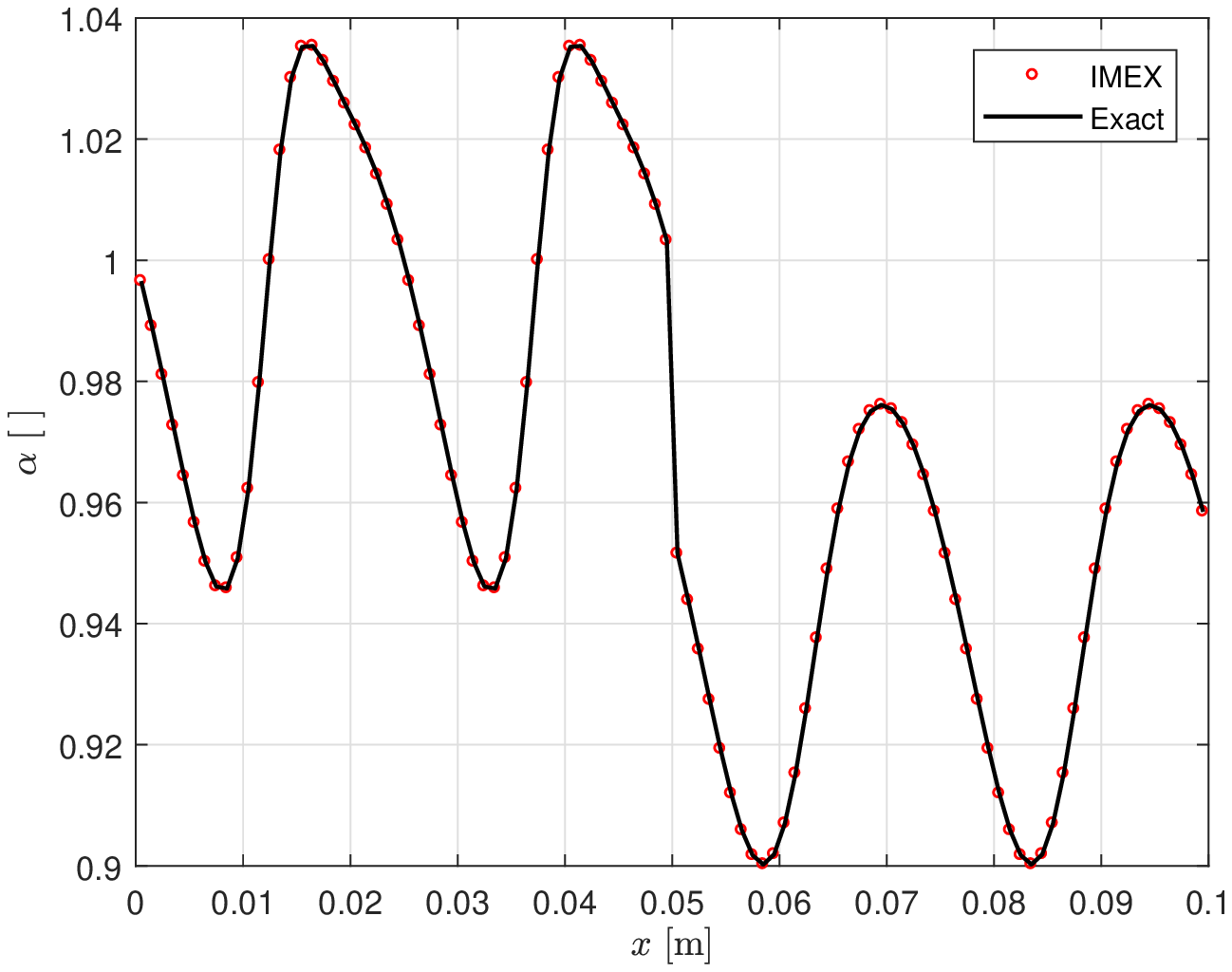}
\vspace*{-5mm}
\caption{}
\label{fig.CP1alpha}
\end{subfigure}
\begin{subfigure}{0.5\textwidth}
\centering
\includegraphics[width=1\linewidth]{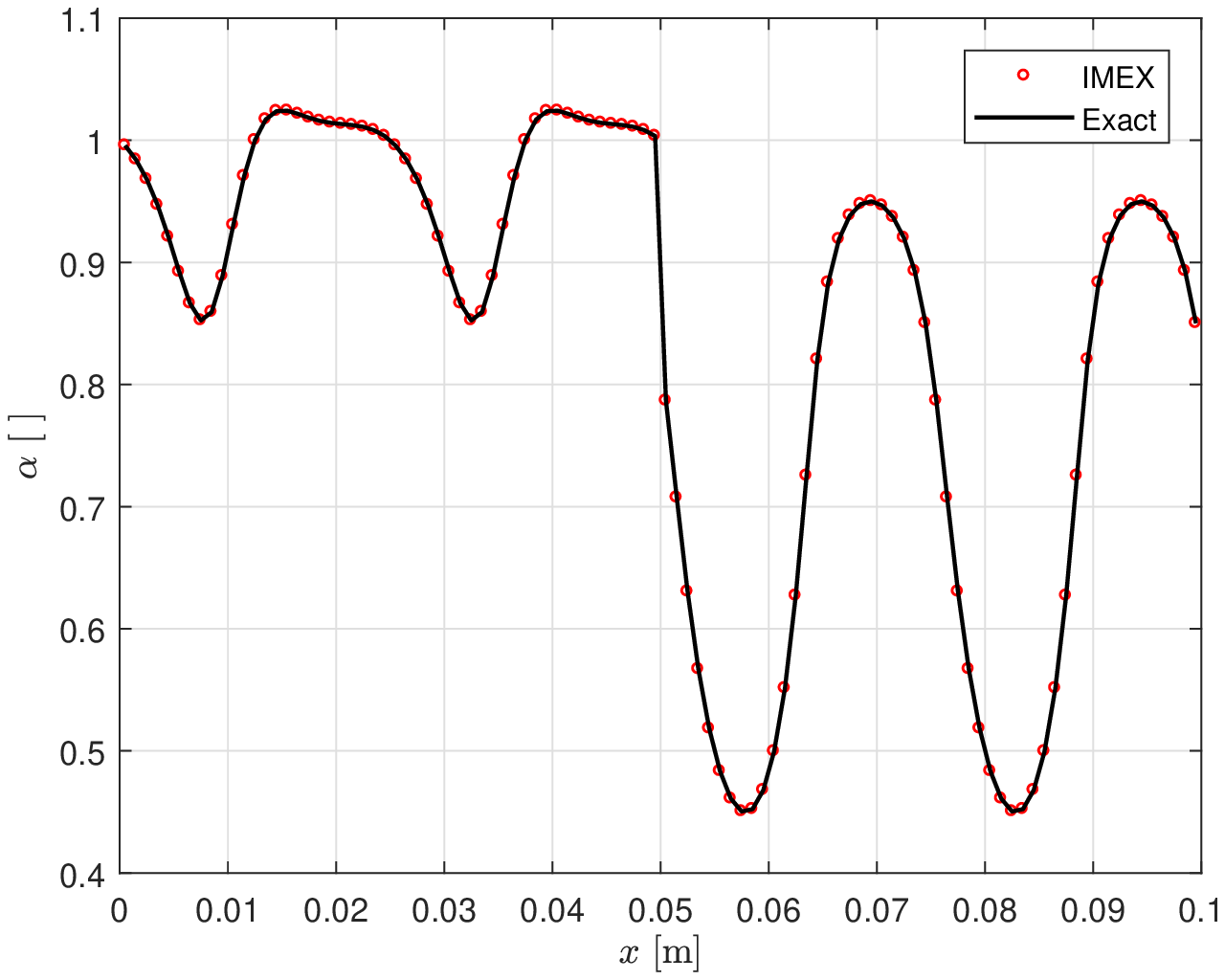}
\vspace*{-5mm}
\caption{}
\label{fig.CP2alpha}
\end{subfigure}
\caption{Results at time \(t_{end}\) = 0.25 s, in terms of non-dimensional cross-sectional area, obtained in problems CP1 (a) and CP2 (b), designed to test the C-property of the scheme solving the augmented FSI system, with respect to the motionless exact solution.}
\label{fig.CP}
\end{figure}

To test the C-property of the scheme even in the case in which not piece-wise constant but smooth functions are defined on the left and on the right side of a central initial discontinuity, two additional problems have been designed. The first test (CP1) takes into account a portion of a generic artery, while the second one (CP2) a portion of a generic vein, in both cases simply concerning an elastic behavior of the wall. Mechanical and geometrical reference data are given in table \ref{tab.CPdata1}, whereas initial conditions are specified in table \ref{tab.CPdata2}. The non-trivial initial conditions regard variables $A_0, E_0$ and $p_{ext}$, all influencing the initial condition of the area, evaluated involving the tube law. Results are graphically shown in figs.~\ref{fig.CP} for the non-dimensional cross-sectional area. Furthermore, errors are evaluated in terms of $L^2$ norm at time $t_{end} = 0.25$~s for both the tests. It is here specified that, with the chosen $t_{end}$, the number of complete time iterations is 4208 in CP1 and 1896 in CP2, being in fact fairly large. The $L^2$ norm for the evolution variables $A, Au$ and $p$ results, respectively: $1.18\times 10^{-20}, 0$ and $0$ in CP1; $1.34\times 10^{-22}, 0$ and $0$ in CP2 (values in SI units). By consequence, the well-balancing proof presented in section \ref{section_wellbalancing} results again confirmed.

\begin{figure}[ht!]
\begin{subfigure}{0.5\textwidth}
\centering
\includegraphics[width=1\linewidth]{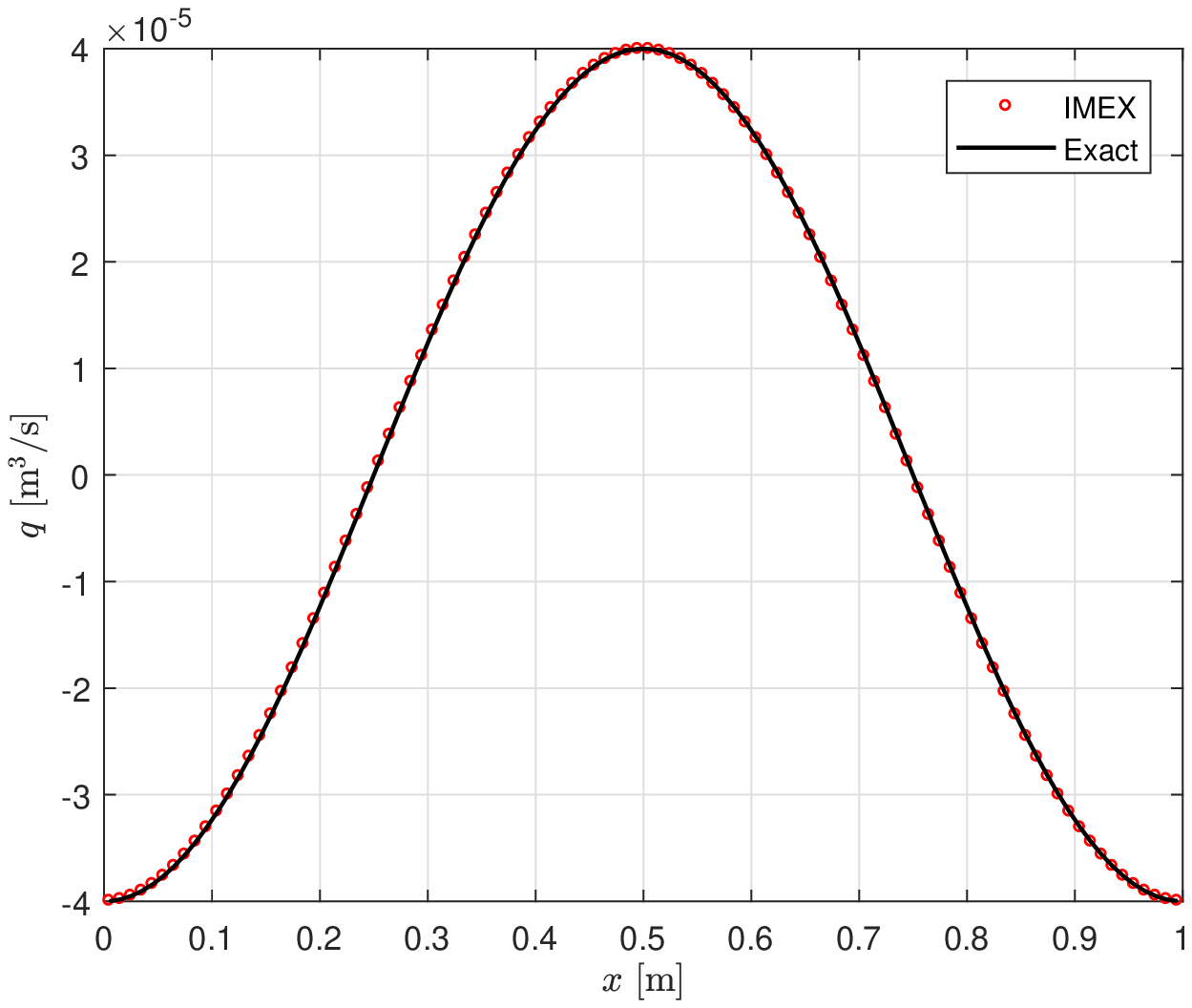}
\vspace*{-5mm}
\caption{}
\label{fig.VV1q}
\end{subfigure}
\begin{subfigure}{0.5\textwidth}
\centering
\includegraphics[width=1\linewidth]{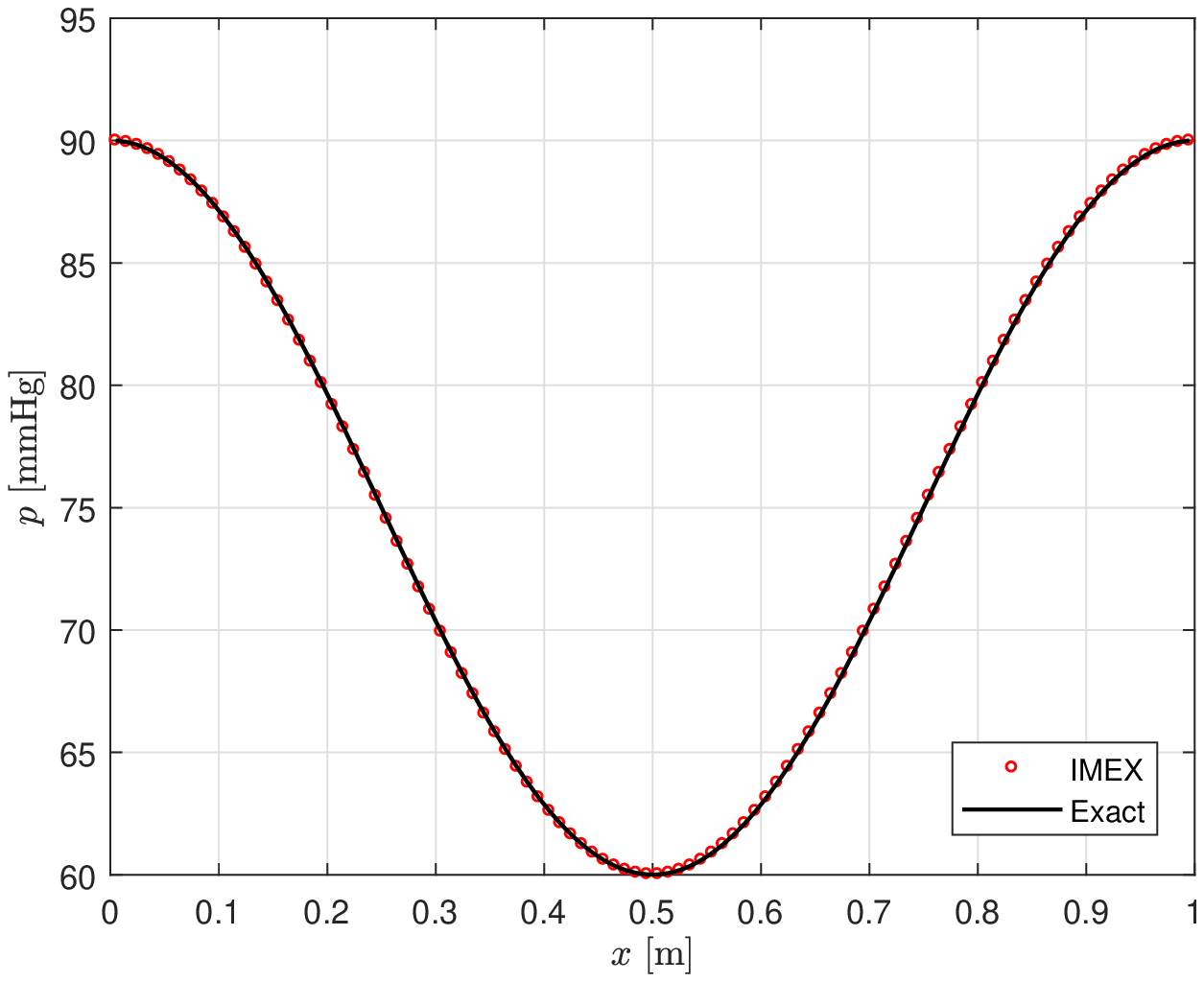}
\vspace*{-5mm}
\caption{}
\label{fig.VV1p}
\end{subfigure}
\begin{subfigure}{0.5\textwidth}
\centering
\includegraphics[width=1\linewidth]{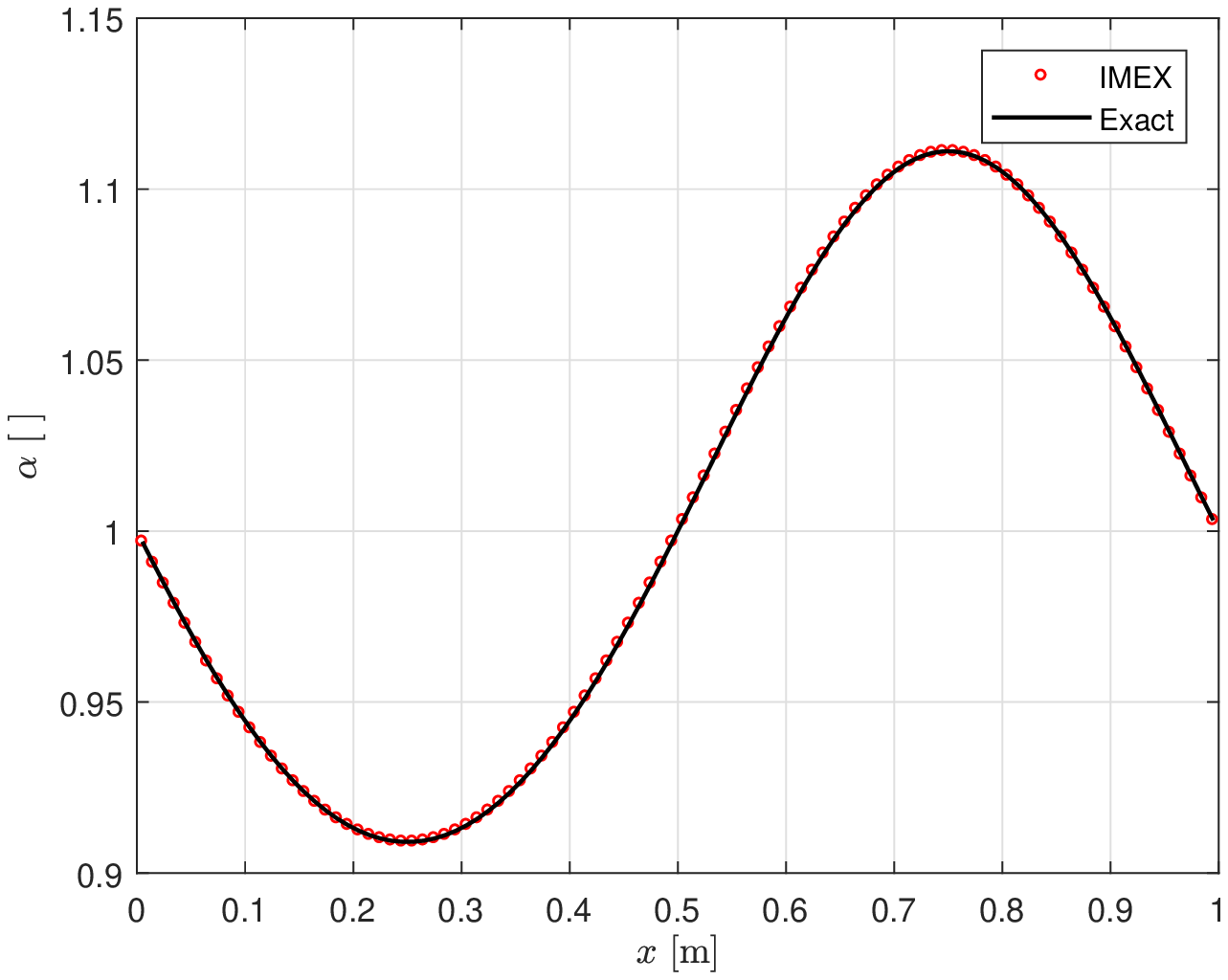}
\vspace*{-5mm}
\caption{}
\label{fig.VV1alpha}
\end{subfigure}
\begin{subfigure}{0.5\textwidth}
\centering
\includegraphics[width=1\linewidth]{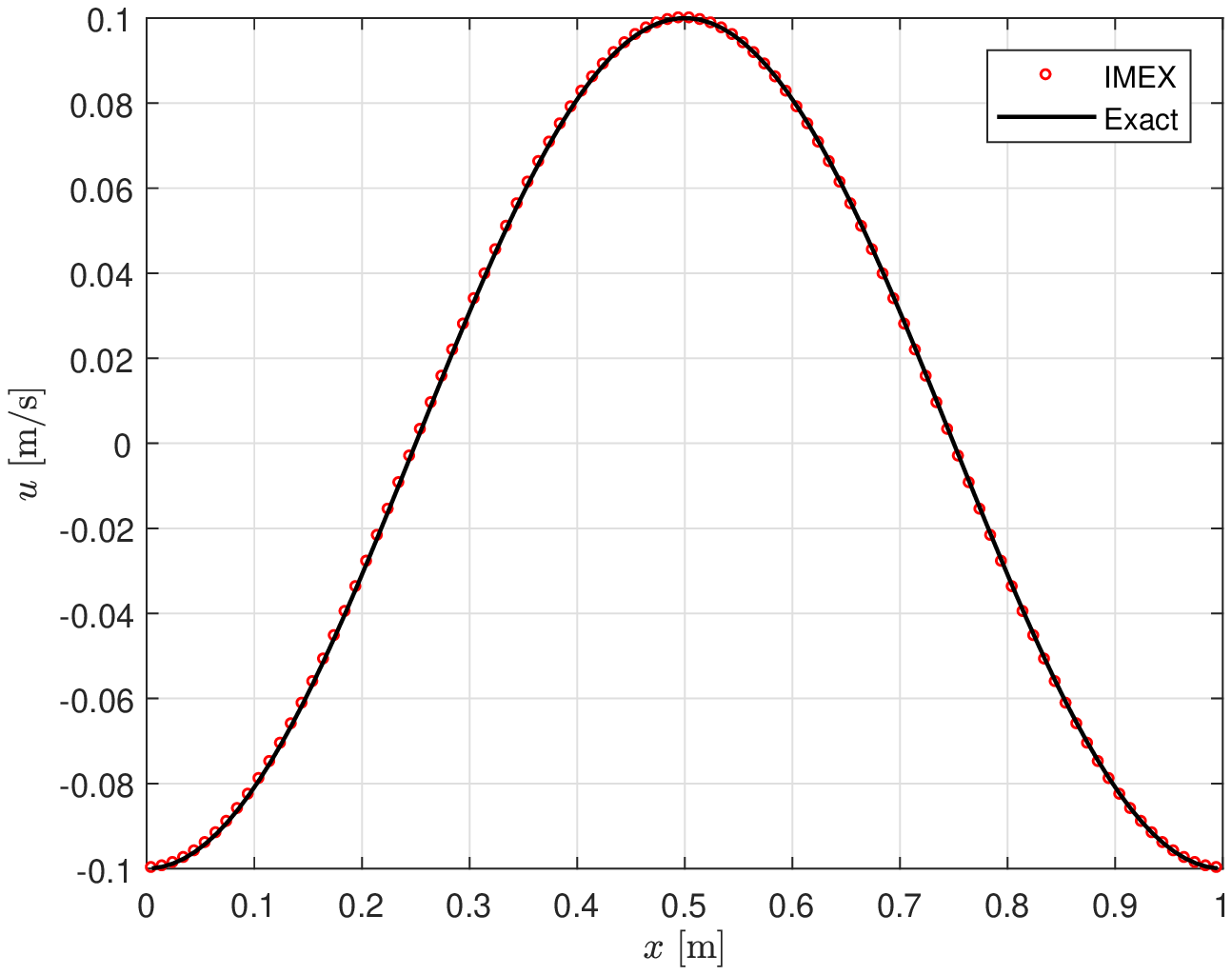}
\vspace*{-5mm}
\caption{}
\label{fig.VV1u}
\end{subfigure}
\caption{Results obtained in test VV1, built up to validate the viscoelastic contribution in a generic artery, at time \(t_{end}\) = 0.75 s, solving the augmented FSI system with the IMEX Runge-Kutta scheme, in terms of (a) flow rate, (b) pressure, (c) non-dimensional cross-sectional area and (d) velocity, with respect to the exact solution.}
\label{fig.viscovalidation_artery}
\end{figure}

\begin{figure}[ht!]
\begin{subfigure}{0.5\textwidth}
\centering
\includegraphics[width=1\linewidth]{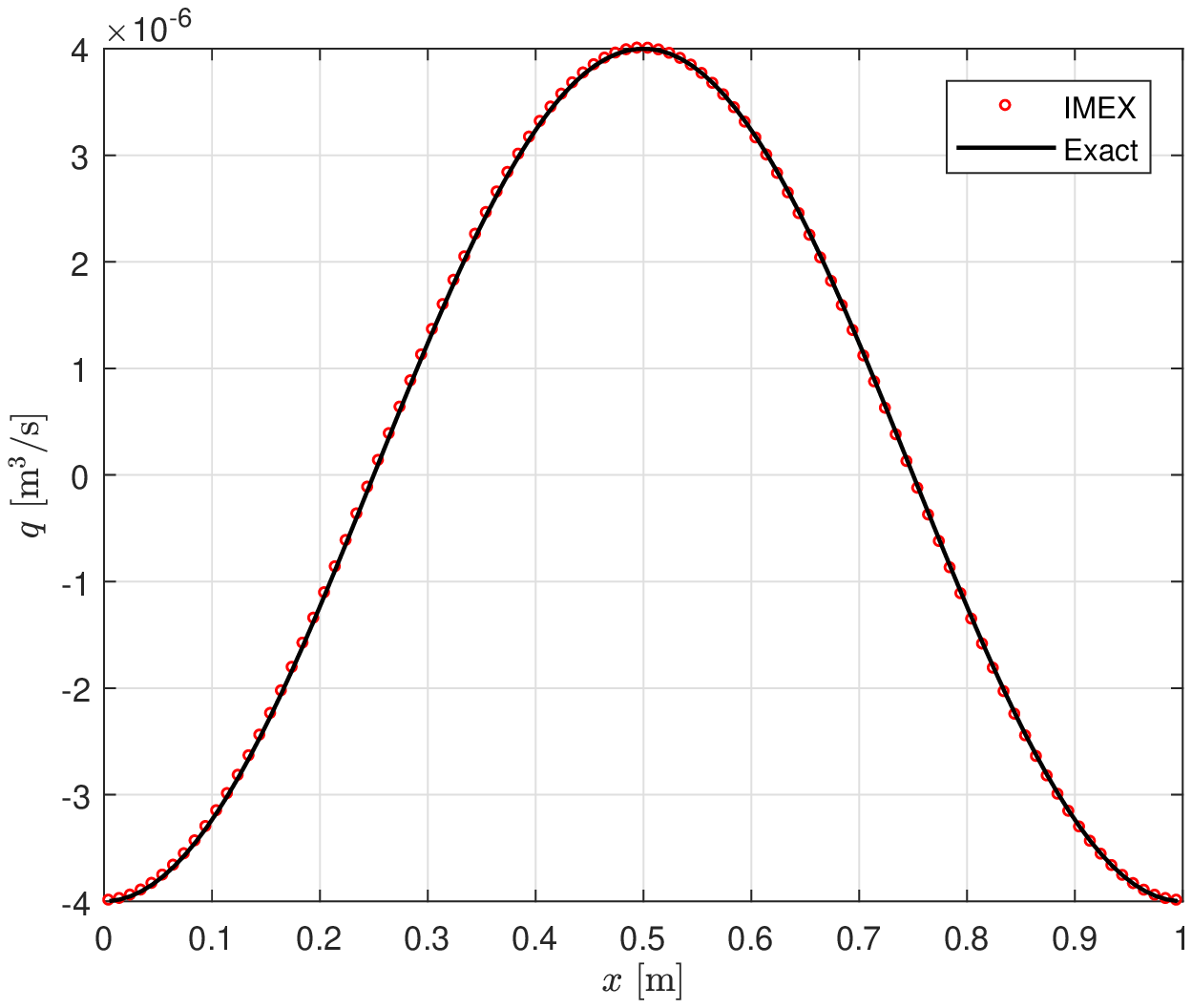}
\vspace*{-5mm}
\caption{}
\label{fig.VV2q}
\end{subfigure}
\begin{subfigure}{0.5\textwidth}
\centering
\includegraphics[width=1\linewidth]{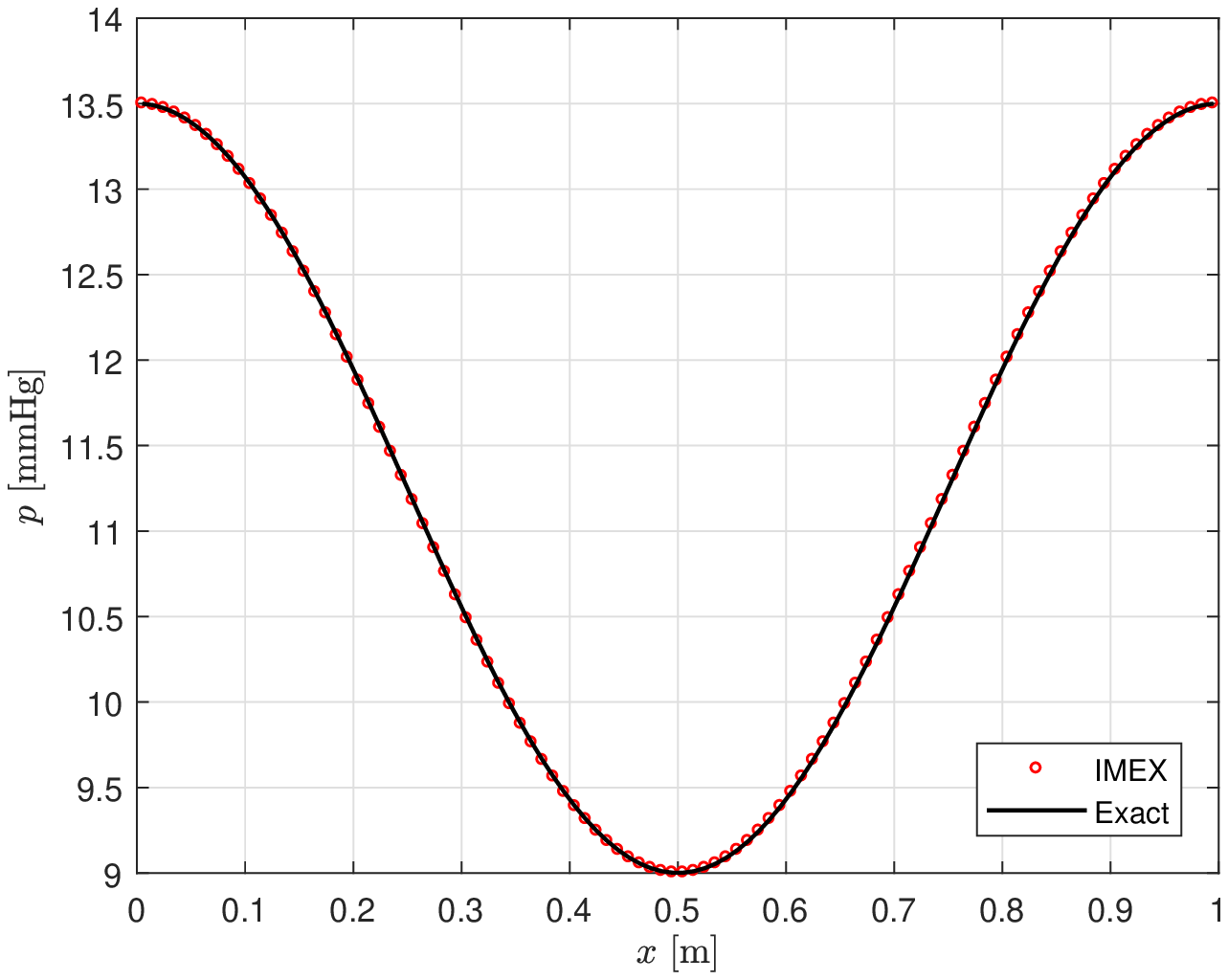}
\vspace*{-5mm}
\caption{}
\label{fig.VV2p}
\end{subfigure}
\begin{subfigure}{0.5\textwidth}
\centering
\includegraphics[width=1\linewidth]{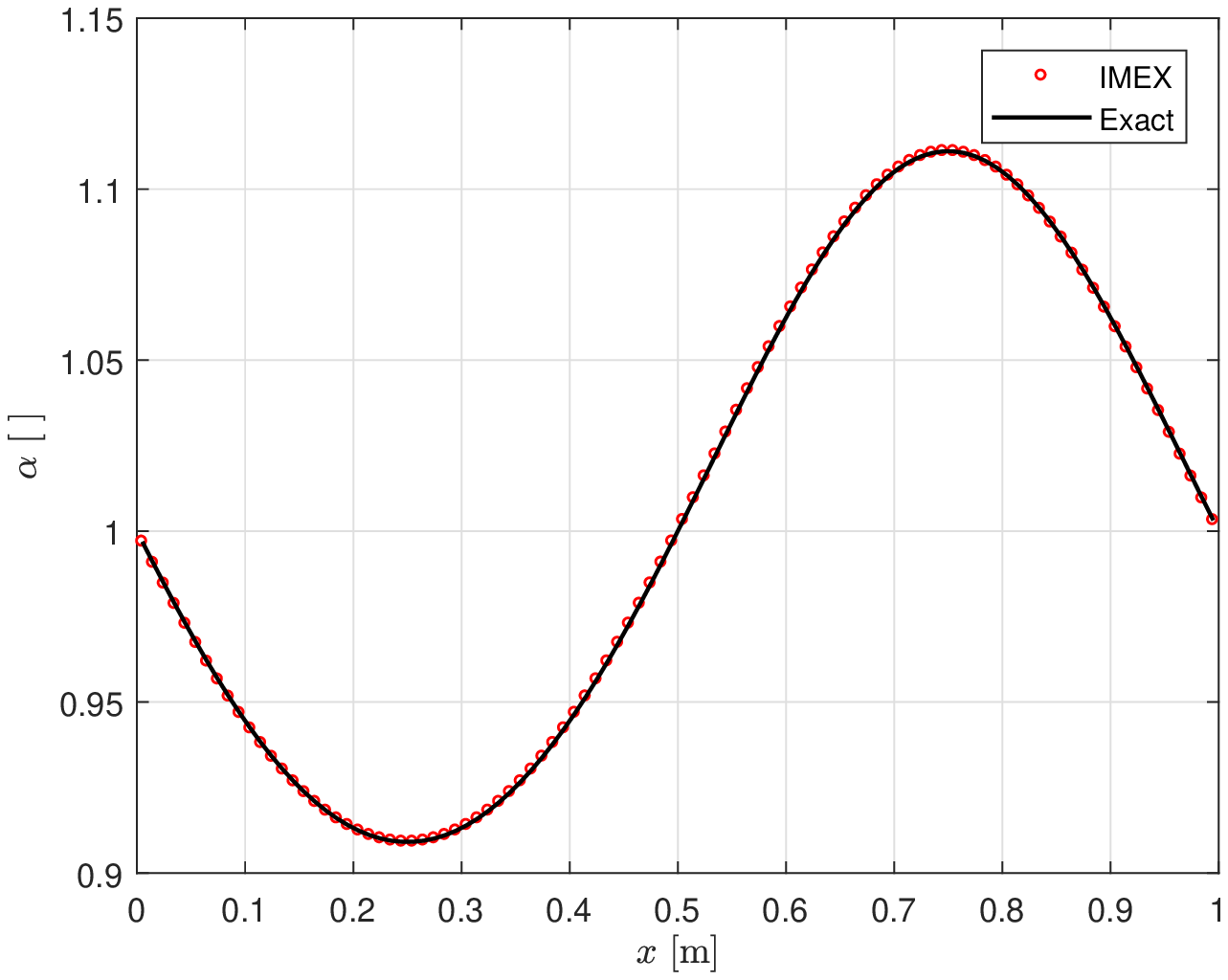}
\vspace*{-5mm}
\caption{}
\label{fig.VV2alpha}
\end{subfigure}
\begin{subfigure}{0.5\textwidth}
\centering
\includegraphics[width=1\linewidth]{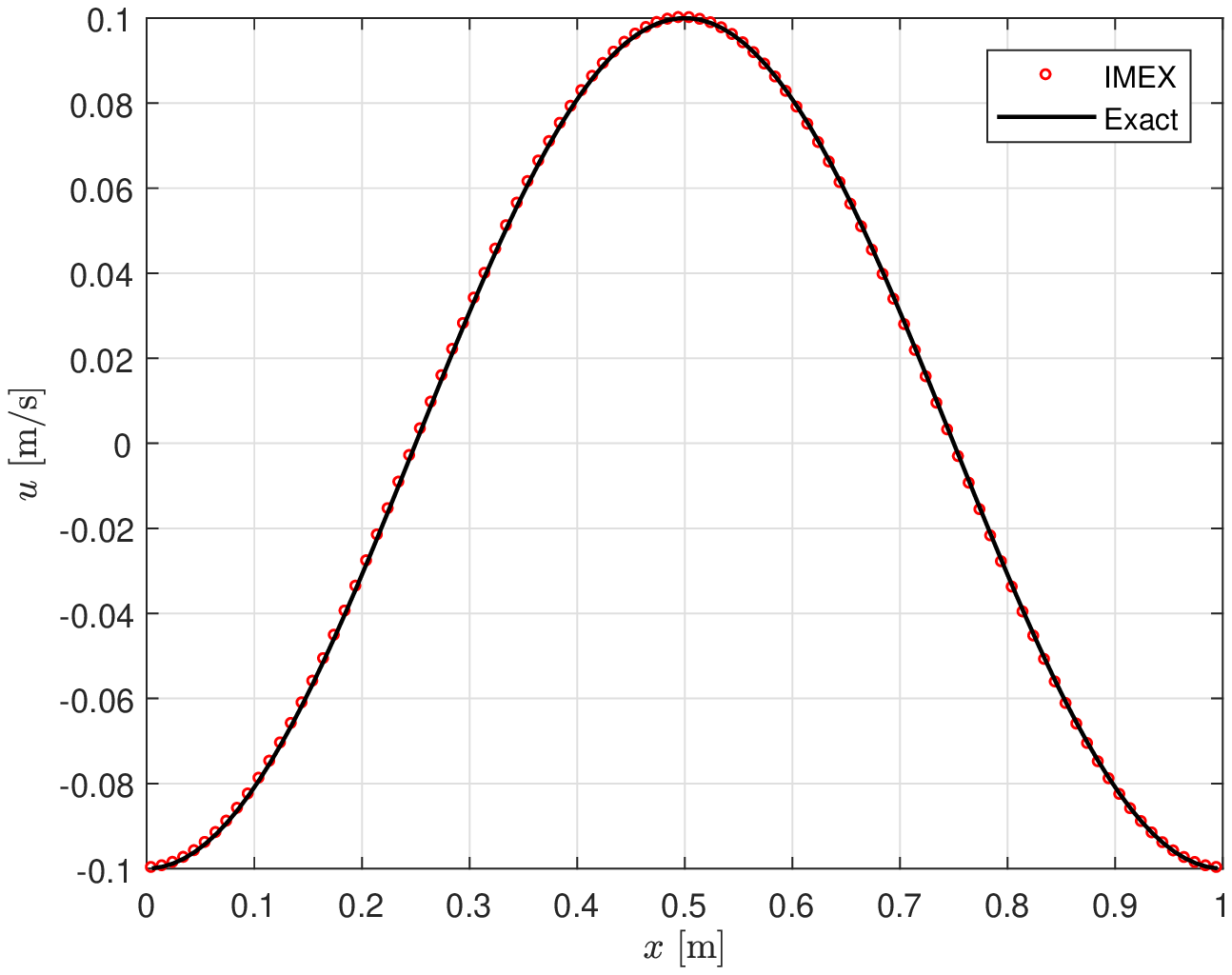}
\vspace*{-5mm}
\caption{}
\label{fig.VV2u}
\end{subfigure}
\caption{Results obtained in test VV2, built up to validate the viscoelastic contribution in a generic vein, at time \(t_{end}\) = 0.75 s, solving the augmented FSI system with the IMEX Runge-Kutta scheme, in terms of (a) flow rate, (b) pressure, (c) non-dimensional cross-sectional area and (d) velocity, with respect to the exact solution.}
\label{fig.viscovalidation_vein}
\end{figure}
\begin{table}[b!]
\centering
\begin{tabular}{l | c c c c c c c c c}
\hline
\\[-1.1em]
	 Test &$h_0$ [mm] &\(\tilde A\) [mm$^2$] &\(\tilde a\) [mm$^2$] &\(\tilde P\) [kPa] &\(\tilde p\) [kPa] &\(\tilde E_0\) [MPa] &\(\tilde e_0\) [MPa] &\(E_{\infty}\) [MPa] &\(\tau_r\) [ms]  \\
\\[-1.1em]
\hline
	VV1 &1.50 &4.00 &0.40 &10.00 &2.00 &2.00 &0.20 &1.60 &0.36 \\
	VV2 &0.30 &0.40 &0.04 &1.50  &0.30 &1.75 &0.10 &1.50 &0.06 \\
\hline
\end{tabular}
\caption{Parameters used for the tests built up to validate the viscoelastic contribution in a generic artery (VV1) and in a generic vein (VV2). Values chosen or calibrated in accordance with standard ranges found in the human body.}
\label{tab.VVdata}
\end{table}
\subsection{Problems for the validation of the viscoelastic contribution}
\label{section_numericalresults_VV}
To validate also the additional contribution given by viscoelasticity in the proposed model, we recurred to the Method of Manufactured Solutions \cite{roache2002}. Manufactured solutions are exact solutions to a set of governing equations that have been modified with forcing terms. Thus, starting from system \eqref{systqlinearform}, a smooth, exact solution of the perturbed system, which is a non-homogeneous non-linear system, is prescribed as follows:
	\[\boldsymbol{\hat Q}(x,t) =
\begin{pmatrix} 
  	\hat A(x,t) \\ \hat{Au}(x,t) \\ \hat p(x,t) \\ \hat A_0(x) \\ \hat E_0(x) \\ \hat p_{ext}(x)
\end{pmatrix}
=
\begin{pmatrix}
	\tilde A + \tilde{a} \hspace{0.5mm} \sin\left(\frac{2\pi}{L}x\right) \cos\left(\frac{2\pi}{T_0}t\right)\\ 
	\tilde{Au} - \tilde{a}\frac{L}{T_0} \hspace{0.5mm} \cos\left(\frac{2\pi}{L}x\right) \sin\left(\frac{2\pi}{T_0}t\right)\\ 
	\tilde P + \tilde{p} \hspace{0.5mm} \cos\left(\frac{2\pi}{L}x\right) \sin\left(\frac{2\pi}{T_0}t\right)\\ 
	\tilde A + \tilde{a} \hspace{0.5mm} \sin\left(\frac{2\pi}{L}x\right)\\ 
	\tilde E_0 + \tilde e_0 \hspace{0.5mm} \sin\left(\frac{2\pi}{L}x\right)\\ 
	\tilde P_{ext} + \tilde p_{ext} \hspace{0.5mm} \sin\left(\frac{2\pi}{L}x\right)
\end{pmatrix}\]
with $L = 1.0$~m, $T_0 = 1.0$~s, $\tilde{Au} = 0.0$~m$^3/$s, $\tilde P_{ext} = 0.0$~Pa and $\tilde p_{ext} = 50.0$~Pa. The rest of the parameters, characterised for each of the two tests, are listed in table \ref{tab.VVdata}. Introducing the prescribed solution in system \eqref{systqlinearform}, a non-homogeneous system with a residual source term $\boldsymbol{R}(x,t)$ (resulting because $\boldsymbol{\hat Q}$ is not the exact solution) is obtained:
\begin{equation}
\partial_t \boldsymbol{\hat Q} + \boldsymbol{A}(\boldsymbol{\hat Q})\partial_x \boldsymbol{\hat Q} - \boldsymbol{S}(\boldsymbol{\hat Q}) = \boldsymbol{R}(x,t) \hspace{0.5mm}.
\end{equation}
Knowing $\boldsymbol{R}(x,t)$, which can be analytically obtained with algebraic manipulations, it is possible to solve the system enriched by the additional source term:
\begin{equation}
\partial_t \boldsymbol{Q} + \boldsymbol{A}(\boldsymbol{Q})\partial_x \boldsymbol{Q} = \boldsymbol{S}(\boldsymbol{Q}) + \boldsymbol{R}(x,t) \hspace{0.5mm} .
\end{equation}
The numerical scheme must therefore reproduce $\boldsymbol{\hat Q}$ as unique solution. \\
The initial condition is fixed as $\boldsymbol{\hat Q}(x,0)$. For this type of test, periodic boundary conditions are defined, in accordance with the periodicity of the expected solution. Results are reported in fig.~\ref{fig.viscovalidation_artery} (test VV1, concerning an artery) and fig.~\ref{fig.viscovalidation_vein} (test VV2, concerning a vein) and confirm the validity of the model discussed in the present work also with respect to the proposed treatment of viscoelastic contributions of vessels.
\begin{table}[b!]
\centering
\begin{tabular}{l | c | c c c c c c c c}
\hline
\\[-1.1em]
	 Test &Variable &\(N_x\) &\(L^1\) &\(\mathcal{O}(L^1)\) &\(L^2\) &\(\mathcal{O}(L^2)\) &\(L^{\infty}\) &\(\mathcal{O}(L^{\infty})\) &\(t_{CPU}\) [s]  \\
\\[-1.1em]
\hline \\[-1.1em]

		 	 	&&9 & 4.70$\times 10^{-07}$ &-        & 5.19$\times 10^{-07}$ &-        &  7.52$\times 10^{-07}$ &-       &0.5 \\ 
				&&27 & 6.25$\times 10^{-08}$ &  1.84 & 6.97$\times 10^{-08}$ &  1.83 &  1.02$\times 10^{-07}$ &  1.82 &3.7 \\ 
		&$A$	&81 & 7.27$\times 10^{-09}$ &  1.96 & 8.09$\times 10^{-09}$ &  1.96 &  1.18$\times 10^{-08}$ &  1.96 &30.3 \\ 
				&&243 & 8.20$\times 10^{-10}$ &  1.99 & 9.12$\times 10^{-10}$ &  1.99 &  1.33$\times 10^{-9}$ &  1.99 &276.5 \\ 
				&&729 & 9.15$\times 10^{-11}$ &  2.00 & 1.02$\times 10^{-10}$ &  2.00 &  1.49$\times 10^{-10}$ &  2.00 &2478.7 \\ 
				\cline{2-10} \\[-1.1em]
			 	&&9 & 2.08$\times 10^{-06}$ &-         & 2.37$\times 10^{-06}$ &-        &  3.66$\times 10^{-06}$ &-        &0.5 \\ 
				&&27 & 1.71$\times 10^{-07}$ &  2.27 & 1.98$\times 10^{-07}$ &  2.26 &  3.34$\times 10^{-07}$ &  2.18 &3.7 \\ 
	VV1	&$Au$	&81 & 1.91$\times 10^{-08}$ &  1.99 & 2.21$\times 10^{-08}$ &  1.99 &  3.74$\times 10^{-08}$ &  1.99 &30.3 \\ 
				&&243 & 2.16$\times 10^{-09}$ &  1.98 & 2.50$\times 10^{-09}$ &  1.98 &  4.23$\times 10^{-09}$ &  1.99 &276.5 \\ 
				&&729 & 2.36$\times 10^{-10}$ &  2.02 & 2.73$\times 10^{-10}$ &  2.02 &  4.62$\times 10^{-10}$ &  2.02 &2478.7 \\ 
				\cline{2-10} \\[-1.1em]
			 	&&9 & 6.89$\times 10^{+01}$ &-        & 7.67$\times 10^{+01}$ &-        &  1.23$\times 10^{+02}$ &-        &0.5 \\ 
				&&27 & 8.20$\times 10^{+00}$ &  1.94 & 9.18$\times 10^{+00}$ &  1.93 &  1.45$\times 10^{+01}$ &  1.94 &3.7 \\ 
		&$p$	&81 & 9.03$\times 10^{-01}$ &  2.01 & 1.01$\times 10^{+00}$ &  2.01 &  1.59$\times 10^{+00}$ &  2.01 &30.3 \\ 
				&&243 & 1.02$\times 10^{-01}$ &  1.99 & 1.14$\times 10^{-01}$ &  1.99 &  1.79$\times 10^{-01}$ &  1.99 &276.5 \\ 
				&&729 & 1.10$\times 10^{-02}$ &  2.02 & 1.23$\times 10^{-02}$ &  2.02 &  1.94$\times 10^{-02}$ &  2.02 &2478.7 \\ 
\hline \\[-1.1em]
		 	 	&&9 & 2.49$\times 10^{-07}$ &-        & 2.83$\times 10^{-07}$ &-        &  4.04$\times 10^{-07}$ &- &0.1 \\ 
     			&&27 & 1.99$\times 10^{-08}$ &  2.30 & 2.27$\times 10^{-08}$ &  2.30 &  3.65$\times 10^{-08}$ &  2.19 &0.5 \\ 
		&$A$	&81 & 1.91$\times 10^{-09}$ &  2.14 & 2.22$\times 10^{-09}$ &  2.12 &  3.93$\times 10^{-09}$ &  2.03 &3.4 \\ 
				&&243 & 2.04$\times 10^{-10}$ &  2.03 & 2.38$\times 10^{-10}$ &  2.03 &  4.31$\times 10^{-10}$ &  2.01 &29.2 \\ 
    			&&729 & 2.25$\times 10^{-11}$ &  2.01 & 2.62$\times 10^{-11}$ &  2.01 &  4.77$\times 10^{-11}$ &  2.00 &257.5 \\ 
				\cline{2-10} \\[-1.1em]
			 	&&9 & 2.47$\times 10^{-07}$ &-        & 2.74$\times 10^{-07}$ &-        &  4.36$\times 10^{-07}$ &-  &0.1 \\ 
     			&&27 & 4.23$\times 10^{-08}$ &  1.61 & 4.74$\times 10^{-08}$ &  1.60 &  8.17$\times 10^{-08}$ &  1.52 &0.5 \\ 
	VV2	&$Au$	&81 & 5.11$\times 10^{-9}$ &  1.92 & 5.71$\times 10^{-09}$ &  1.92 &  9.59$\times 10^{-09}$ &  1.95 &3.4 \\ 
				&&243 & 5.82$\times 10^{-10}$ &  1.98 & 6.50$\times 10^{-10}$ &  1.98 &  1.09$\times 10^{-09}$ &  1.98 &29.2 \\ 
    			&&729 & 6.52$\times 10^{-11}$ &  1.99 & 7.28$\times 10^{-11}$ &  1.99 &  1.21$\times 10^{-10}$ &  1.99 &257.5 \\ 
				\cline{2-10} \\[-1.1em]
			 	&&9 & 5.98$\times 10^{+00}$ &-        & 6.68$\times 10^{+00}$ &-         &  1.18$\times 10^{+01}$ &-  &0.1 \\ 
     			&&27 & 4.95$\times 10^{-01}$ &  2.27 & 5.97$\times 10^{-01}$ &  2.20 &  1.30$\times 10^{+00}$ &  2.00 &0.5 \\ 
		&$p$	&81 & 5.20$\times 10^{-02}$ &  2.05 & 7.33$\times 10^{-02}$ &  1.91 &  1.76$\times 10^{-01}$ &  1.82 &3.4 \\ 
				&&243 & 5.54$\times 10^{-03}$ &  2.04 & 7.98$\times 10^{-03}$ &  2.02 &  1.95$\times 10^{-02}$ &  2.00 &29.2 \\ 
    			&&729 & 6.15$\times 10^{-04}$ &  2.00 & 9.04$\times 10^{-04}$ &  1.98 &  2.21$\times 10^{-03}$ &  1.98 &257.5 \\ 
\hline
\end{tabular}
\caption{Accuracy analysis results for the tests built up to validate the viscoelastic contribution in a generic artery (VV1) and in a generic vein (VV2). Errors are computed for variables $A, Au$ and $p$ in terms of norms $L^1, L^2$ and $L^{\infty}$, using the International System of Units. CPU times are reported for each simulation.}
\label{tab.VVaccuracy}
\end{table}
\subsection{Accuracy analysis}
\label{section_numericalresults_accuracy}
For the same test cases discussed in the previous section \ref{section_numericalresults_VV}, for which the prescribed solution is intended as the exact one, an accuracy analysis has been carried out. It can be seen, observing results reported in table \ref{tab.VVaccuracy}, that the expected second-order of accuracy is achieved for all the evolution variables. Given the set of parameters for each test case presented in table \ref{tab.VVdata}, following what stated in eq.~\eqref{eq:def.stiff}, it is worth to notice that, when considering a discretization with $\Delta x > 1.0\times 10^{-4}$~m, the two problems become stiff. Therefore, this analysis confirms that the chosen IMEX-SSP2 Runge-Kutta scheme is asymptotic accurate, preserving the expected order of accuracy in the stiff limit. However, it has been verified that resolving the same problems with a simple Strang splitting technique (which should provide second-order of accuracy if each step is at least second order accurate in space \cite{strang1968}) would lead to a reduction of the expected accuracy to first-order. The non-asymptotic behavior of this technique in the stiff limit was already discussed in literature for different contexts \cite{leveque1990,descombes2004,pareschi2005,duarte2011}. In the present paper this weakness is confirmed even for the specific application of blood flow modeling.
\begin{figure}[ht!]
\begin{subfigure}{0.5\textwidth}
\centering
\includegraphics[width=1\linewidth]{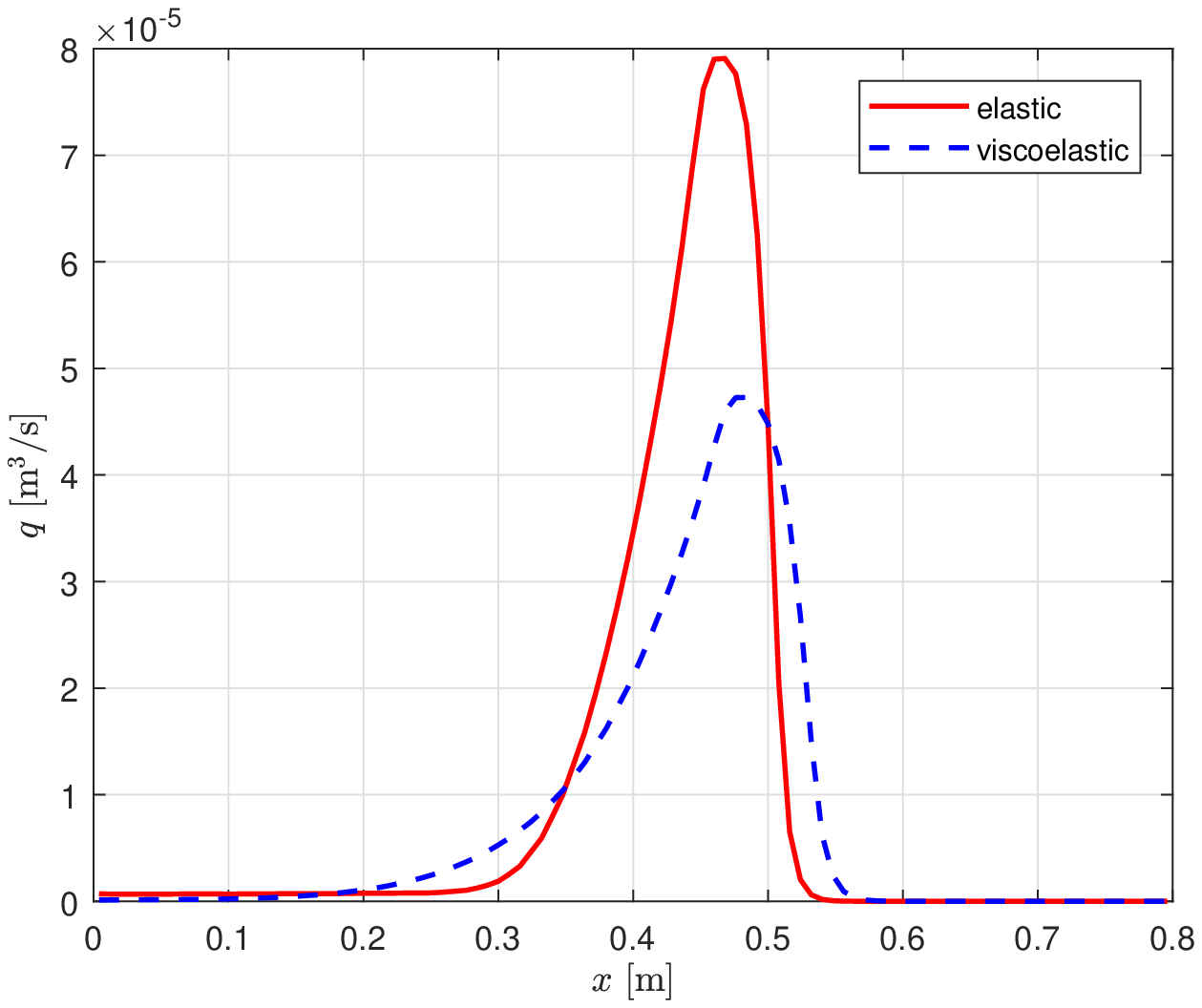}
\vspace*{-5mm}
\caption{}
\label{fig.PWq}
\end{subfigure}
\begin{subfigure}{0.5\textwidth}
\centering
\includegraphics[width=1\linewidth]{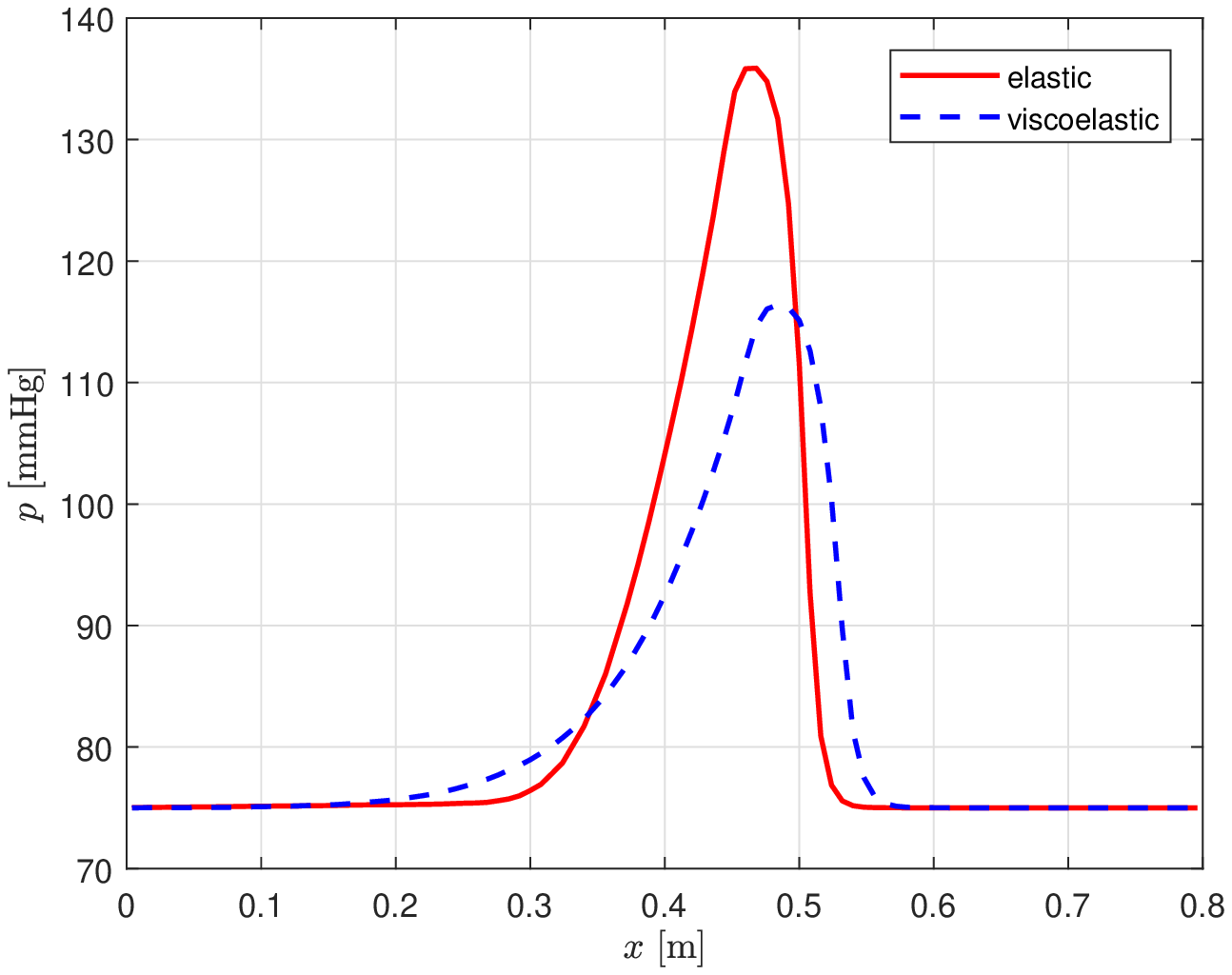}
\vspace*{-5mm}
\caption{}
\label{fig.PWp}
\end{subfigure}
\begin{subfigure}{0.5\textwidth}
\centering
\includegraphics[width=1\linewidth]{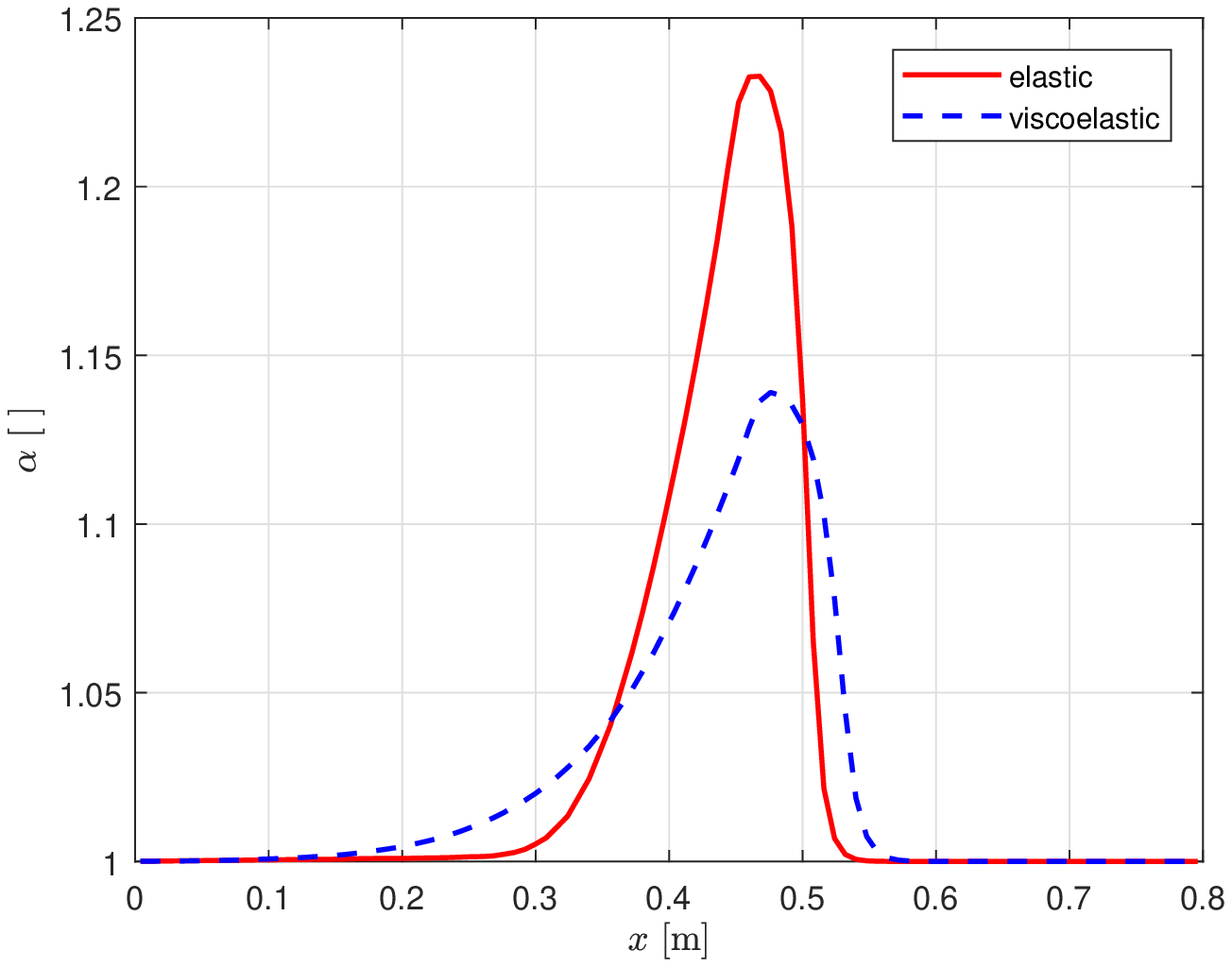}
\vspace*{-5mm}
\caption{}
\label{fig.PWalpha}
\end{subfigure}
\begin{subfigure}{0.5\textwidth}
\centering
\includegraphics[width=1\linewidth]{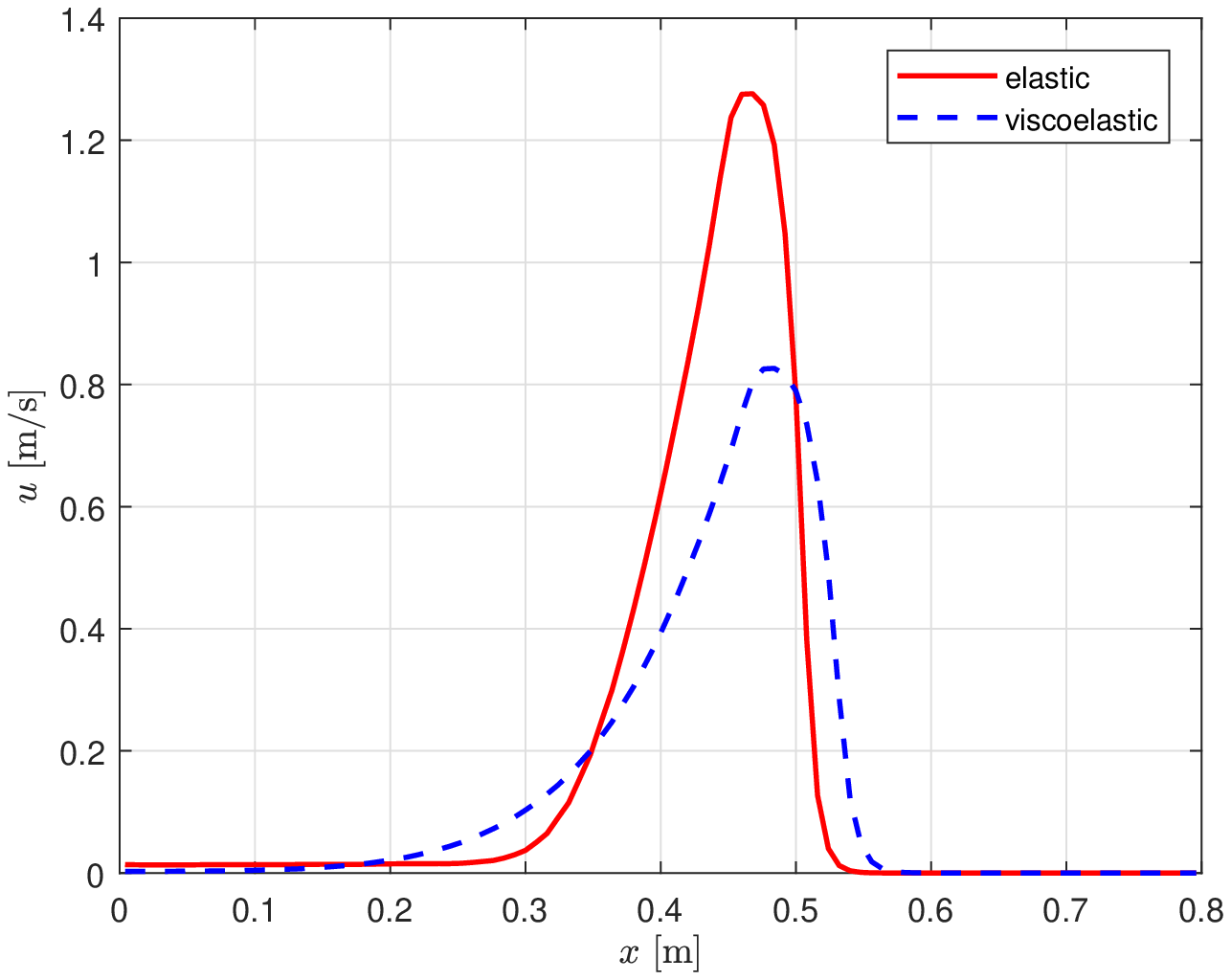}
\vspace*{-5mm}
\caption{}
\label{fig.PWu}
\end{subfigure}
\caption{Solution of the inlet pulse wave test, at time \(t_{end}\) = 0.08 s, solving the augmented FSI system with the IMEX Runge-Kutta scheme. Comparison of the results obtained adopting a simple elastic or a more realistic viscoelastic constitutive tube law, to characterize the mechanical behavior of the CCA wall, are compared.}
\label{fig.PW}
\end{figure}

\begin{figure}[ht!]
\centering
\includegraphics[width=0.6\linewidth]{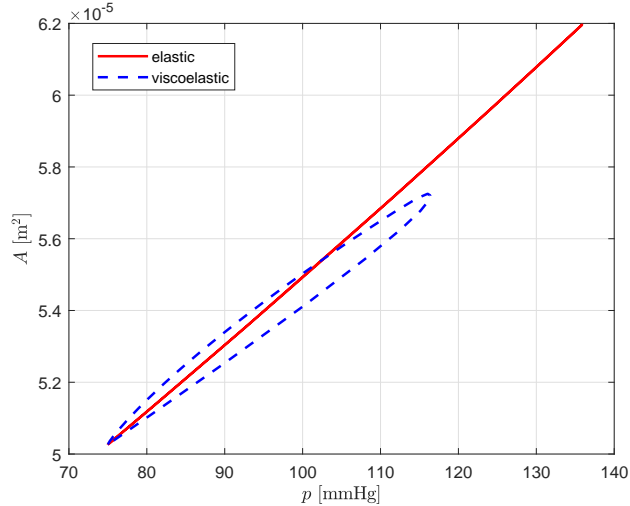}
\caption{Comparison of the $p$-$A$ curves obtained in test PW adopting a simple elastic tube law or a more realistic viscoelastic model, for which damping effects produce an hysteresis loop.}
\label{fig.PWhysteresis}
\end{figure}
\subsection{Inlet pulse wave problem}
An additional test case has been designed to assess the effects of viscoelasticity with respect to the simple elastic behavior of the vessel wall. In this pulse wave (PW) problem, a Gaussian-shape pulse pressure in time is prescribed at the inlet of the vessel, with function as in \cite{acosta2015}:
\begin{equation*}
p_{in}(t)= P_0 \hspace{0.5mm} e^{-\frac{(t-t_0)^2}{2\varsigma^2}} .
\end{equation*}
Inflow boundary conditions are defined prescribing the inlet pressure $p_{in}$ and applying the Riemann invariants $\Gamma_1$ and $\Gamma_2$ presented in section \ref{section_FSIsystem}. Then, considering a single tract of the common carotid artery (CCA), coefficients $P_0$, $t_0$ and $\varsigma$ are adapted to simulate an idealistic pulse wave passing trough the vessel. Thus, it is assumed $P_0 = 15.00$~kPa, $t_0 = 15.00$~ms and $\varsigma = 3.00$~ms. The rest of the parameters are set considering values available in literature \cite{alastruey2011,muller2014a,xiao2014} with regards to the CCA: equilibrium radius $R_0 = 4.00$~mm, wall thickness $h_0 = 0.15$~mm, external (equilibrium) pressure $p_{ext} = 75.00$~mmHg and reference celerity $c_0 = 5.92$~m/s. The instantaneous Young modulus is evaluated considering the reference celerity and then simply inverting eq.~\eqref{eq:cel}, resulting $E_0 = 0.25$~MPa. Concerning the remaining elastic and viscoelastic coefficients, the attribution of the viscosity coefficient $\eta$ is done through eq.~\eqref{gamma}, considering the relation between $\Gamma$ and $R_0$ defined in \cite{mynard2015}: $\Gamma = \sfrac{0.05}{R_0}+0.4$. The relaxation time thus obtained is $\tau_r = 7.90$~ms. Finally, the asymptotic Young modulus is fixed in order to reproduce a realistic energy loss (represented by the hysteresis loop) with respect to CCA pressure-diameter measurements available in literature \cite{nichols2011,salvi2012}, with an approach similar to the one proposed in \cite{alastruey2012a}. Hence, this sort of calibration has led to $E_{\infty} = 0.8\hspace{0.5mm} E_0 = 0.20$~MPa. The length of the domain was stretched with respect to standard lengths of the common carotid artery to conveniently plot the entire traveling pulse.\\
The relevance of taking into account viscoelastic damping effects appears evident when observing fig.~\ref{fig.PW}, in which numerical results over an elastic vessel wall are compared to those obtained applying the proposed viscoelastic model. The effective energy loss is also shown in fig.~\ref{fig.PWhysteresis}, with the occurrence of the hysteresis loop. 
\section{Conclusions}
\label{section_conclusions}
In the present work, an innovative augmented fluid-structure interaction system has been proposed for the blood flow modeling with regards to the viscoelastic effects of arterial and venous walls. The model has been validated through idealized time-dependent tests for situations close to reality, considered as a tool for checking the accuracy and the robustness of hemodynamics models. The selected IMEX Runge-Kutta scheme has been proved to preserve the expected second-order of accuracy also when dealing with stiff source terms, confirming the asymptotic-preserving property also in the blood flow context here discussed.\\
Given the satisfactory results obtained in all the test cases, it is believed that the proposed augmented FSI system of blood flow equations can be a valid instrument for modeling the human circulation, both in arteries and veins, representing a valuable resource for different real medical applications. A simple shift from elastic to viscoelastic characterization of the vessel wall  is ensured by the straightforward addition of a source term. Recurring to the augmented FSI system, in fact, would be mostly advantageous when taking into account the viscoelastic wall behavior of vessels: all the viscosity information would be enclosed within a source term, avoiding the presence of second order derivatives in the system. Moreover, in this manner the governing system of equations remains purely hyperbolic.\\
Finally, the impact of characterizing the mechanics of the vessel wall concerning viscoelastic effects and not only the elastic ones has been pointed out. In this context, the viscoelastic Standard Linear Solid Model proposed better describes the complex behavior of a viscoelastic material if compared with the Kelvin-Voigt model frequently adopted in the biofluid dynamics literature \cite{alastruey2011,montecinos2014,wang2014,mynard2015}, still maintaining ease of implementation and usage. 
\appendix
\section*{Appendix}
\label{section_appendix}
Analysing each of the $s=3$ Runge-Kutta steps necessary to apply the IMEX-SSP2(3,3,2) method chosen, the following is obtained:
\begin{equation*}
\begin{aligned}
	\boldsymbol{Q}^{(1)}_i &=  \boldsymbol{Q}^n_i + \Delta t a_{11} \boldsymbol{S}\left(\boldsymbol{Q}_i^{(1)}\right) \\
	\boldsymbol{Q}^{(2)}_i &=  \boldsymbol{Q}^n_i -  \frac{\Delta t}{\Delta x} \tilde{a}_{21} \left[ \left( \boldsymbol{F}_{i + \frac{1}{2}}^{(1)}  - \boldsymbol{F}_{i - \frac{1}{2}}^{(1)} \right)  + \left(\boldsymbol{D}_{i+\frac{1}{2}}^{(1)}  + \boldsymbol{D}_{i-\frac{1}{2}}^{(1)}\right) + \boldsymbol{B}\left( \boldsymbol{Q}_i^{(1)}\right) \Delta \boldsymbol{Q}_i^{(1)} \right] \\
	& + \Delta t \left[ a_{21} \boldsymbol{S}\left(\boldsymbol{Q}_i^{(1)}\right) + a_{22} \boldsymbol{S}\left(\boldsymbol{Q}_i^{(2)}\right) \right]  \\
	\boldsymbol{Q}^{(3)}_i &=  \boldsymbol{Q}^n_i -  \frac{\Delta t}{\Delta x} \Bigg\{ \tilde{a}_{31} \left[ \left( \boldsymbol{F}_{i + \frac{1}{2}}^{(1)}  - \boldsymbol{F}_{i - \frac{1}{2}}^{(1)} \right)  + \left(\boldsymbol{D}_{i+\frac{1}{2}}^{(1)} + \boldsymbol{D}_{i-\frac{1}{2}}^{(1)}\right) + \boldsymbol{B}\left( \boldsymbol{Q}_i^{(1)}\right) \Delta \boldsymbol{Q}_i^{(1)} \right] \\
	& + \tilde{a}_{32} \left[ \left( \boldsymbol{F}_{i + \frac{1}{2}}^{(2)}  - \boldsymbol{F}_{i - \frac{1}{2}}^{(2)} \right)  + \left(\boldsymbol{D}_{i+\frac{1}{2}}^{(2)}  + \boldsymbol{D}_{i-\frac{1}{2}}^{(2)}\right) + \boldsymbol{B}\left( \boldsymbol{Q}_i^{(2)}\right) \Delta \boldsymbol{Q}_i^{(2)} \right] \Bigg\} \\
	& + \Delta t \left[ a_{31} \boldsymbol{S}\left(\boldsymbol{Q}_i^{(1)}\right) + a_{32} \boldsymbol{S}\left(\boldsymbol{Q}_i^{(2)}\right) + a_{33} \boldsymbol{S}\left(\boldsymbol{Q}_i^{(3)}\right)\right] .
\end{aligned}
\label{eq.stepsIMEX}
\end{equation*}
In particular, analysing the structure of the system, it can be observed that each step can be analytically linearised, avoiding the adoption of a Newton-Raphson method (or similar ones) for the evaluation of the implicit part. Replacing the already explicit contributions of the convective part,
\begin{equation*}
\boldsymbol{EX}^{(1)} = \frac{\Delta t}{\Delta x} \tilde{a}_{21} \left[ \left( \boldsymbol{F}_{i + \frac{1}{2}}^{(1)}  - \boldsymbol{F}_{i - \frac{1}{2}}^{(1)} \right) + \left(\boldsymbol{D}_{i+\frac{1}{2}}^{(1)}  + \boldsymbol{D}_{i-\frac{1}{2}}^{(1)}\right) + \boldsymbol{B}\left( \boldsymbol{Q}_i^{(1)}\right) \Delta \boldsymbol{Q}_i^{(1)} \right] 
\end{equation*}
\begin{equation*}
\begin{aligned}
\boldsymbol{EX}^{(2)} &= \frac{\Delta t}{\Delta x} \Bigg\{ \tilde{a}_{31} \left[ \left( \boldsymbol{F}_{i + \frac{1}{2}}^{(1)}  - \boldsymbol{F}_{i - \frac{1}{2}}^{(1)} \right)  + \left(\boldsymbol{D}_{i+\frac{1}{2}}^{(1)} + \boldsymbol{D}_{i-\frac{1}{2}}^{(1)}\right) + \boldsymbol{B}\left( \boldsymbol{Q}_i^{(1)}\right) \Delta \boldsymbol{Q}_i^{(1)} \right] \\
	& + \tilde{a}_{32} \left[ \left( \boldsymbol{F}_{i + \frac{1}{2}}^{(2)}  - \boldsymbol{F}_{i - \frac{1}{2}}^{(2)} \right)  + \left(\boldsymbol{D}_{i+\frac{1}{2}}^{(2)}  + \boldsymbol{D}_{i-\frac{1}{2}}^{(2)}\right) + \boldsymbol{B}\left( \boldsymbol{Q}_i^{(2)}\right) \Delta \boldsymbol{Q}_i^{(2)} \right] \Bigg\}\end{aligned}
\end{equation*}
the final totally explicit formulation is here presented in details, for each of the 3 Runge-Kutta steps, only for the first three equations of system \eqref{completesyst}:
\begin{equation*}
\begin{aligned}
	A^{(1)}_i &=  A^n_i \\
	(Au)^{(1)}_i &=  (Au)^n_i \\
	p^{(1)}_i &=  \frac{\tau_r}{\tau_r + \Delta t a_{11}} p^n_i + \frac{\Delta t a_{11}}{\tau_r + \Delta t a_{11}} \left[ \frac{E_{\infty,i}}{E_{0,i}} \psi_{el}\left(A^{(1)}_i,A_{0,i},E_{0,i} \right) + p_{ext,i} \right] \\
	\\
	A^{(2)}_i &=  A^n_i - EX^{(1)}_{(A)} \\
	(Au)^{(2)}_i &=  (Au)^n_i - EX^{(1)}_{(Au)} \\
	p^{(2)}_i &= \frac{\tau_r}{\tau_r + \Delta t a_{22}} \Bigg\{ p^n_i - EX^{(1)}_{(p)} + \frac{\Delta t a_{21}}{\tau_r} \left[ \frac{E_{\infty,i}}{E_{0,i}} \psi_{el}\left(A^{(1)}_i,A_{0,i},E_{0,i} \right) - \left(p_i^{(1)} - p_{ext,i}\right) \right] \Bigg\} \\
	&+ \frac{\Delta t a_{22}}{\tau_r + \Delta t a_{22}} \left[ \frac{E_{\infty,i}}{E_{0,i}} \psi_{el}\left(A^{(2)}_i,A_{0,i},E_{0,i} \right) + p_{ext,i} \right]\\
	\\
	A^{(3)}_i &=  A^n_i - EX^{(2)}_{(A)} \\
	(Au)^{(3)}_i &=  (Au)^n_i - EX^{(2)}_{(Au)} \\
	p^{(3)}_i &= \frac{\tau_r}{\tau_r + \Delta t a_{33}} \Bigg\{ p^n_i - EX^{(2)}_{(p)} + \frac{\Delta t a_{31}}{\tau_r} \left[ \frac{E_{\infty,i}}{E_{0,i}} \psi_{el}\left(A^{(1)}_i,A_{0,i},E_{0,i} \right) - \left(p_i^{(1)} - p_{ext,i}\right) \right] \\
	&+ \frac{\Delta t a_{32}}{\tau_r} \left[ \frac{E_{\infty,i}}{E_{0,i}} \psi_{el}\left(A^{(2)}_i,A_{0,i},E_{0,i} \right) - \left(p_i^{(2)} - p_{ext,i}\right) \right] \Bigg\} \\
	&+ \frac{\Delta t a_{33}}{\tau_r + \Delta t a_{33}} \left[ \frac{E_{\infty,i}}{E_{0,i}} \psi_{el}\left(A^{(3)}_i,A_{0,i},E_{0,i} \right) + p_{ext,i} \right]
\end{aligned}
\label{eq.stepsIMEX_linear}
\end{equation*}
The time update of the variables finally results:
\begin{equation*}
\begin{aligned}
	\boldsymbol{Q}^{n+1}_i = \boldsymbol{Q}^n_i & -  \frac{\Delta t}{\Delta x} \Bigg\{ \tilde{\omega}_{1} \left[ \left( \boldsymbol{F}_{i + \frac{1}{2}}^{(1)}  - \boldsymbol{F}_{i - \frac{1}{2}}^{(1)} \right)  + \left(\boldsymbol{D}_{i+\frac{1}{2}}^{(1)}  + \boldsymbol{D}_{i-\frac{1}{2}}^{(1)}\right) + \boldsymbol{B}\left( \boldsymbol{Q}_i^{(1)}\right) \Delta \boldsymbol{Q}_i^{(1)} \right]  \\
	& + \tilde{\omega}_{2} \left[ \left( \boldsymbol{F}_{i + \frac{1}{2}}^{(2)}  - \boldsymbol{F}_{i - \frac{1}{2}}^{(2)} \right)  + \left(\boldsymbol{D}_{i+\frac{1}{2}}^{(2)}  + \boldsymbol{D}_{i-\frac{1}{2}}^{(2)}\right) + \boldsymbol{B}\left( \boldsymbol{Q}_i^{(2)}\right) \Delta \boldsymbol{Q}_i^{(2)} \right] \\
	& + \tilde{\omega}_{3} \left[ \left( \boldsymbol{F}_{i + \frac{1}{2}}^{(3)}  - \boldsymbol{F}_{i - \frac{1}{2}}^{(3)} \right)  + \left(\boldsymbol{D}_{i+\frac{1}{2}}^{(3)}  + \boldsymbol{D}_{i-\frac{1}{2}}^{(3)}\right) + \boldsymbol{B}\left( \boldsymbol{Q}_i^{(3)}\right) \Delta \boldsymbol{Q}_i^{(3)} \right] \Bigg\}\\
	& + \Delta t \left[ \omega_{1} \boldsymbol{S}\left(\boldsymbol{Q}_i^{(1)}\right)  + \omega_{2} \boldsymbol{S}\left(\boldsymbol{Q}_i^{(2)}\right) + \omega_{3} \boldsymbol{S}\left(\boldsymbol{Q}_i^{(3)}\right) \right] .\\
\end{aligned}
\label{eq.finalupdateIMEX}
\end{equation*}
\section*{Acknowledgements}
The authors warmly thank Prof. L. Pareschi (Department of Mathematics, Ferrara, Italy) for the very helpful discussions that contributed to the linearization of the IMEX algorithm of the present work.\par
This work was partially supported by MIUR (Ministero dell'Istruzione, dell'Universit\`a e della Ricerca) PRIN 2017 for the project \textit{``Innovative numerical methods for evolutionary partial differential equations and applications''}, code 2017KKJP4X.\\
The second author, Valerio Caleffi, was also funded by MIUR FFABR 2017.

\bibliographystyle{abbrv}

\begin{thebibliography}{10}

\bibitem{acosta2015}
S.~Acosta, C.~Puelz, B.~Rivi{\`{e}}re, D.~J. Penny, and C.~G. Rusin.
\newblock {Numerical method of characteristics for one-dimensional blood flow}.
\newblock {\em Journal of Computational Physics}, 294:96--109, 2015.

\bibitem{alastruey2011}
J.~Alastruey, A.~W. Khir, K.~S. Matthys, P.~Segers, S.~J. Sherwin, P.~R.
  Verdonck, K.~H. Parker, and J.~Peir{\'{o}}.
\newblock {Pulse wave propagation in a model human arterial network: Assessment
  of 1-D visco-elastic simulations against in vitro measurements}.
\newblock {\em Journal of Biomechanics}, 44(12):2250--2258, 2011.

\bibitem{alastruey2012a}
J.~Alastruey, T.~Passerini, L.~Formaggia, and J.~Peir{\'{o}}.
\newblock {Physical determining factors of the arterial pulse waveform:
  Theoretical analysis and calculation using the 1-D formulation}.
\newblock {\em Journal of Engineering Mathematics}, 77(1):19--37, 2012.

\bibitem{ambrosi2012}
D.~Ambrosi, A.~Quarteroni, and G.~Rozza.
\newblock {\em {Modeling of Physiological Flows}}.
\newblock Springer, 2012.

\bibitem{battista2015}
C.~Battista.
\newblock {\em {Parameter Estimation of Viscoelastic Wall Models in a
  One-Dimensional Circulatory Network}}.
\newblock PhD thesis, North Carolina State University, 2015.

\bibitem{bermudez1994}
A.~Bermudez and M.~E. Vazquez.
\newblock {Upwind methods for hyperbolic conservation laws with source terms}.
\newblock {\em Computers {\&} Fluids}, 23(8):1049--1071, 1994.

\bibitem{bertaglia2018}
G.~Bertaglia, M.~Ioriatti, A.~Valiani, M.~Dumbser, and V.~Caleffi.
\newblock {Numerical methods for hydraulic transients in visco-elastic pipes}.
\newblock {\em Journal of Fluids and Structures}, 81:230--254, 2018.

\bibitem{bertaglia2018a}
G.~Bertaglia, A.~Valiani, and V.~Caleffi.
\newblock {The augmented FSI system for blood flow in compliant vessels}.
\newblock In {\em Proc. of the 5th IAHR Europe Congress - New Challenges in
  Hydraulic Research and Engineering}, pages 153--154, Trento, Italy, 2018.
  
  \bibitem{bertagliaDATA2019}
G.~Bertaglia, V.~Caleffi and A.~Valiani.
\newblock {Data for: Modeling blood flow in viscoelastic vessels: the 1D augmented fluid-structure interaction system}.
\newblock {Mendeley Data}, v1, 2019.
\href{http://dx.doi.org/10.17632/zz4zw9kyz7.1}{http://dx.doi.org/10.17632/zz4zw9kyz7.1} 

\bibitem{bessems2008}
D.~Bessems, C.~G. Giannopapa, M.~C.~M. Rutten, and F.~N. van~de Vosse.
\newblock {Experimental validation of a time-domain-based wave propagation
  model of blood flow in viscoelastic vessels}.
\newblock {\em Journal of Biomechanics}, 41(2):284--291, 2008.

\bibitem{carpenter2001}
P.~W. Carpenter and T.~J. Pedley.
\newblock {Flow past Highly Compliant Boundaries and in Collapsible Tubes}.
\newblock In {\em Proceedings of the IUTAM Symposium}. Springer, 2001.

\bibitem{descombes2004}
S.~Descombes and M.~Massot.
\newblock {Operator splitting for nonlinear reaction-diffusion systems with an
  entropic structure: singular perturbation and order reduction}.
\newblock {\em Numerische Mathematik}, 97:667--698, 2004.

\bibitem{duarte2011}
M.~Duarte, M.~Massot, and S.~Descombes.
\newblock {Parareal operator splitting techniques for multi-scale reaction
  waves: numerical analysis and strategies}.
\newblock {\em Mathematical Modelling and Numerical Analysis}, 5(45):825--852,
  2011.

\bibitem{dumbser2011a}
M.~Dumbser and E.~F. Toro.
\newblock {A simple extension of the Osher Riemann solver to non-conservative
  hyperbolic systems}.
\newblock {\em Journal of Scientific Computing}, 48(1-3):70--88, 2011.

\bibitem{dumbser2011}
M.~Dumbser and E.~F. Toro.
\newblock {On universal Osher-type schemes for general nonlinear hyperbolic
  conservation laws}.
\newblock {\em Communications in Computational Physics}, 10(3):635--671, 2011.

\bibitem{formaggia2003}
L.~Formaggia, D.~Lamponi, and A.~Quarteroni.
\newblock {One-dimensional models for blood flow in arteries}.
\newblock {\em Journal of Engineering Mathematics}, 47(3-4):251--276, 2003.

\bibitem{formaggia2009}
L.~Formaggia, A.~Quarteroni, and A.~Veneziani.
\newblock {\em {Cardiovascular Mathematics: Modeling and simulation of the
  circulatory system}}.
\newblock Springer, 2009.

\bibitem{fung1997}
Y.~C. Fung.
\newblock {\em {Biomechanics: Circulation}}.
\newblock Springer, 2nd edition, 1997.

\bibitem{ghigo2016}
A.~R. Ghigo, X.~Wang, R.~Armentano, P.-Y. Lagr{\'{e}}e, and J.-M. Fullana.
\newblock {Linear and nonlinear viscoelastic arterial wall models: application
  on animals}.
\newblock {\em Journal of Biomechanical Engineering}, 139:011003, 2017.

\bibitem{holenstein1980}
R.~Holenstein, P.~Niederer, and M.~Anliker.
\newblock {A Viscoelastic Model for Use in Predicting Arterial Pulse Waves}.
\newblock {\em Journal of Biomechanical Engineering}, 102(4):318--325, 1980.

\bibitem{Lakes2009}
R.~Lakes.
\newblock {\em {Viscoelastic Materials}}.
\newblock Cambridge University Press, 2009.

\bibitem{leguy2019}
C.~Leguy.
\newblock {\em {Cardiovascular Computing: Methodologies and Clinical
  Applications. Chapter 11: Mathematical and Computational Modelling of Blood
  Pressure and Flow}}.
\newblock Springer, 2019.

\bibitem{leveque1990}
R.~J. LeVeque and H.~C. Yee.
\newblock {A Study of Numerical Methods for Hyperbolic Conservation Laws with
  Stiff Source Terms}.
\newblock {\em Journal of Computational Physics}, 86:187--210, 1990.

\bibitem{liang2018}
F.~Liang, D.~Guan, and J.~Alastruey.
\newblock {Determinant factors for arterial hemodynamics in hypertension:
  theoretical insights from a computational model-based study}.
\newblock {\em Journal of Biomechanical Engineering},
  140(3):031006--031006--14, 2018.

\bibitem{matthys2007}
K.~S. Matthys, J.~Alastruey, J.~Peir{\'{o}}, A.~W. Khir, P.~Segers, P.~R.
  Verdonck, K.~H. Parker, and S.~J. Sherwin.
\newblock {Pulse wave propagation in a model human arterial network: Assessment
  of 1-D numerical simulations against in vitro measurements}.
\newblock {\em Journal of Biomechanics}, 40(15):3476--3486, 2007.

\bibitem{mitsotakis2018a}
D.~Mitsotakis, D.~Dutykh, Q.~Li, and E.~Peach.
\newblock {On some model equations for pulsatile flow in viscoelastic vessels}.
\newblock {\em Wave motion}, 90:139--151, 2019.

\bibitem{montecinos2014}
G.~I. Montecinos, L.~O. M{\"{u}}ller, and E.~F. Toro.
\newblock {Hyperbolic reformulation of a 1D viscoelastic blood flow model and
  ADER finite volume schemes}.
\newblock {\em Journal of Computational Physics}, 266:101--123, 2014.

\bibitem{muller2019}
L.~O. M{\"{u}}ller, M.~Celant, E.~F. Toro, P.~J. Blanco, G.~Bertaglia,
  V.~Caleffi, and A.~Valiani.
\newblock {The Selfish-Brain Hypothesis as possible cause of arterial
  hypertension: a modelling study}.
\newblock In {\em 6th International Conference on Computational {\&}
  Mathematical Biomedical Engineering}, pages 592--595, Sendai City, Japan,
  2019.

\bibitem{muller2012}
L.~O. M{\"{u}}ller, G.~I. Montecinos, and E.~F. Toro.
\newblock {Some issues in modelling venous haemodynamics}.
\newblock In {\em Numerical Methods for Hyperbolic Equations: Theory and
  Applications. An international conference to honour Professor EF Toro}, pages
  347--354, 2013.

\bibitem{muller2013}
L.~O. M{\"{u}}ller and E.~F. Toro.
\newblock {Well-balanced high-order solver for blood flow in networks of
  vessels with variable properties}.
\newblock {\em International Journal for Numerical Methods in Biomedical
  Engineering}, 29(12):1388--1411, 2013.

\bibitem{muller2014a}
L.~O. M{\"{u}}ller and E.~F. Toro.
\newblock {A global multiscale mathematical model for the human circulation
  with emphasis on the venous system}.
\newblock {\em International Journal for Numerical Methods in Biomedical
  Engineering}, 30(7), 2014.

\bibitem{murillo2019}
J.~Murillo, A.~Navas-Montilla, and P.~Garc{\'{i}}a-Navarro.
\newblock {Formulation of exactly balanced solvers for blood flow in elastic
  vessels and their application to collapsed states}.
\newblock {\em Computers {\&} Fluids}, 186:74--98, 2019.

\bibitem{mynard2015}
J.~P. Mynard and J.~J. Smolich.
\newblock {One-Dimensional Haemodynamic Modeling and Wave Dynamics in the
  Entire Adult Circulation}.
\newblock {\em Annals of Biomedical Engineering}, 43(6):1443--1460, 2015.

\bibitem{nichols2011}
W.~W. Nichols, M.~F. O'Rourke, and C.~Vlachlopoulos.
\newblock {\em {McDonald's Blood Flow in Arteries}}.
\newblock Hodder Arnold, 6th edition, 2011.

\bibitem{Pares2006}
C.~Par{\'{e}}s.
\newblock {Numerical methods for nonconservative hyperbolic systems: a
  theoretical framework}.
\newblock {\em SIAM Journal on Numerical Analysis}, 44(1):300--321, 2006.

\bibitem{pareschi2005}
L.~Pareschi and G.~Russo.
\newblock {Implicit-explicit Runge-Kutta schemes and applications to hyperbolic
  systems with relaxation}.
\newblock {\em Journal of Scientific Computing}, 25(1):129--155, 2005.

\bibitem{vignon-clementel2011}
R.~Raghu, I.~E. Vignon-Clementel, C.~A. Figueroa, and C.~A. Taylor.
\newblock {Comparative Study of Viscoelastic Arterial Wall Models in Nonlinear
  One-Dimensional Finite Element Simulations of Blood Flow}.
\newblock {\em Journal of Biomechanical Engineering}, 133(8):081003, 2011.

\bibitem{reymond2009b}
P.~Reymond, F.~Merenda, F.~Perren, D.~Rufenacht, N.~Stergiopulos, and D.~Ru.
\newblock {Validation of a one-dimensional model of the systemic arterial
  tree}.
\newblock {\em AJP: Heart and Circulatory Physiology}, 297(1):H208--H222, 2009.

\bibitem{roache2002}
P.~J. Roache.
\newblock {Code Verification by the Method of Manufactured Solutions}.
\newblock {\em Journal of Fluids Engineering}, 124(1):4, 2002.

\bibitem{salvi2012}
P.~Salvi.
\newblock {\em {Pulse Waves: How vascular hemodynamics affects blood
  pressure}}.
\newblock Springer Verlag, 2012.

\bibitem{shapiro1977}
A.~H. Shapiro.
\newblock {Steady Flow in Collapsible Tubes}.
\newblock {\em Journal of Biomechanical Engineering}, 99:126--147, 1977.

\bibitem{sherwin2003a}
S.~J. Sherwin, L.~Formaggia, J.~Peir{\'{o}}, and V.~Franke.
\newblock {Computational modelling of 1D blood flow with variable mechanical
  properties and its application to the simulation of wave propagation in the
  human arterial system}.
\newblock {\em International Journal for Numerical Methods in Fluids},
  43:673--700, 2003.

\bibitem{spiller2017}
C.~Spiller, E.~F. Toro, M.~E. V{\'{a}}zquez-Cend{\'{o}}n, and C.~Contarino.
\newblock {On the exact solution of the Riemann problem for blood flow in human
  veins, including collapse}.
\newblock {\em Applied Mathematics and Computation}, 303:178--189, 2017.

\bibitem{strang1968}
G.~Strang.
\newblock {On the Construction and Comparison of Difference Schemes}.
\newblock {\em SIAM Journal on Numerical Analysis}, 5(3):506--517, 1968.

\bibitem{Toro2009}
E.~F. Toro.
\newblock {\em {Reimann Solvers and Numerical Methods for fluid dynamics}}.
\newblock Springer Verlag, 3rd edition, 2009.

\bibitem{toro2016a}
E.~F. Toro.
\newblock {Brain venous haemodynamics, neurological diseases and mathematical
  modelling. A review}.
\newblock {\em Applied Mathematics and Computation}, 272:542--579, 2016.

\bibitem{toro2013}
E.~F. Toro and A.~Siviglia.
\newblock {Flow in collapsible tubes with discontinuous mechanical properties:
  Mathematical model and exact Solutions}.
\newblock {\em Communications in Computational Physics}, 13(2):361--385, 2013.

\bibitem{tortora2013}
G.~J. Tortora and B.~Defrickson.
\newblock {\em {Principles of anatomy and phisiology}}.
\newblock John Wiley {\&} Sons, Inc., 12th edition, 2013.

\bibitem{valdez-jasso2009}
D.~Valdez-Jasso, M.~A. Haider, H.~T. Banks, D.~B. Santana, Y.~Z. German, R.~L.
  Armentano, and M.~S. Olufsen.
\newblock {Analysis of Viscoelastic Wall Properties in Ovine Arteries}.
\newblock {\em IEEE Transactions on Biomedical Engineering}, 56(2):210--219,
  2009.

\bibitem{wang2014}
X.~Wang, J.-M. Fullana, and P.-Y. Lagr{\'{e}}e.
\newblock {Verification and comparison of four numerical schemes for a 1D
  viscoelastic blood flow model}.
\newblock {\em Computer Methods in Biomechanics and Biomedical Engineering},
  23(16):37--41, 2014.

\bibitem{wang2016}
Z.~Wang, M.~J. Golob, and N.~C. Chesler.
\newblock {\em {Viscoelastic and Viscoplastic Materials. Chapter 7:
  Viscoelastic Properties of Cardiovascular Tissues}}.
\newblock InTech, 2016.

\bibitem{willemet2016}
M.~Willemet, S.~Vennin, and J.~Alastruey.
\newblock {Computational assessment of hemodynamics-based diagnostic tools
  using a database of virtual subjects: Application to three case studies}.
\newblock {\em Journal of Biomechanics}, 49(16):3908--3914, 2016.

\bibitem{wylie1978}
E.~Wylie and V.~Streeter.
\newblock {\em {Fluid Transients}}.
\newblock McGraw-Hill Inc., 1978.

\bibitem{xiao2014}
N.~Xiao, J.~Alastruey, and C.~A. Figueroa.
\newblock {A systematic comparison between 1-D and 3-D hemodynamics in
  compliant arterial models}.
\newblock {\em International Journal for Numerical Methods in Biomedical
  Engineering}, 30(2):204--231, 2014.

\end{thebibliography}



\end{document}